\documentclass[12pt]{article}

\usepackage{smacros5,sober,enumerate}
\usepackage{latexsym}
\usepackage{amsmath}
\usepackage{amssymb}
\usepackage{proof}

\usepackage{meaning}
\usepackage{bbm}
\usepackage{bm}
\allowdisplaybreaks

\usepackage{lscape}

\usepackage[dvipdfm]{graphicx}

\usepackage{blkarray}
\usepackage{multirow}

\usepackage{longtable}

\input xy
\xyoption{all}
\newcommand{\TR}[4]{{\sf Tr}^{#2}_{#3,#4}  \left( #1 \right) }
\newcommand{\Meanpre}[1]{\Mean{#1}^\dagger}

\newcommand{\abs}[1]{\mid\!{#1}\! \mid}

\newcommand{\Id}{\operatorname{Id}}

 \newcommand{\MLL}{{\sf MLL}}
  \newcommand{\LL}{{\sf LL}}
 \newcommand{\Int}{{\sf Int}}
  \newcommand{\MELL}{{\sf MELL}}
\newcommand{\MLLP}{{\sf MLLP}}

\newcommand{\MALL}{{\sf MALL}}
\newcommand{\LLP}{{\sf LLP}}

\renewcommand{\Rel}{{\sf Rel}}
\newcommand{\PInj}{{\sf PInj} }
\renewcommand{\Pfn}{{\sf Pfn} }

\newcommand{\Cplus}{\cC_{+}}
\newcommand{\Cminus}{\cC_{-}}

\newcommand{\down}[1]{\downarrow \! {#1} }
\newcommand{\up}[1]{\uparrow \! {#1} }

\newcount\PLv\newcount\PLw\newcount\PLx\newcount\PLy
\newdimen\PLyy\newdimen\PLX\newbox\PLdot \setbox\PLdot\hbox{\tiny.}
\def\scl{.08} 
\def\PLot#1{\PLx`#1\advance\PLx-42\PLy\PLx\PLv\PLx\divide\PLy9\PLw\PLy
\multiply\PLw9\advance\PLx-\PLw\advance\PLx-4\PLy-\PLy\advance\PLy4
\PLX=\the\PLx pt \advance\PLyy\the\PLy pt\wd\PLdot=\scl\PLX\raise\scl
\PLyy\copy\PLdot}
\def\draw#1{\ifx#1\end\let\next=\relax\else\PLot#1\let\next=\draw\fi\next}

\def\invamp{\hbox{\PLyy=70pt\draw
:::;DMV_gqppyyyyyooooxxxnnwvlutkjaWNE=%
5-./99:::CCCC:::99/..--544=EENWWaajjjkktttttttNNNVVVVVVVV\end
\hskip8pt}}

\newbox\iabox\setbox\iabox\invamp \def\Invamp{\copy\iabox}

\def\wp{\mathrel{\Invamp}} 

\newcount\smallPLv\newcount\smallPLw\newcount\smallPLx\newcount\smallPLy\newdimen\smallPLyy%
\newdimen\smallPLX\newbox\smallPLdot\setbox\smallPLdot\hbox{{\tiny.}}\def\smallscl{.062}
\def\smallPLot#1{\smallPLx`#1\advance\smallPLx-42\smallPLy\smallPLx\smallPLv\smallPLx%
\divide\smallPLy9\smallPLw\smallPLy\multiply\smallPLw9\advance\smallPLx-\smallPLw%
\advance\smallPLx-4\smallPLy-\smallPLy\advance\smallPLy4\smallPLX=\the\smallPLx pt%
\advance\smallPLyy\the\smallPLy pt\wd\smallPLdot=\smallscl\smallPLX%
\raise\smallscl\smallPLyy\copy\smallPLdot} 
\def\smalldraw#1{\ifx#1\end\let\next=\relax\else\smallPLot#1\let\next=\smalldraw\fi\next}

\begin{document}

\title{On Geometry of Interaction for Polarized Linear Logic}
\author{Masahiro Hamano
\thanks{
PRESTO, Japan Science and Technology Agency (JST), 
4-1-8 Honcho Kawaguchi, Saitama 332-0012, JAPAN.   ~
 {\tt hamano@jaist.ac.jp}
Research supported by a PRESTO grant from JST. 
}
\and
Philip Scott 
\thanks{
 Department of Mathematics and Statistics,  University of Ottawa, 
 585 King Edward, Ottawa, Ontario, K1N 6N5, CANADA.  
~{\tt phil@site.uottawa.ca } 
Research supported by an NSERC Discovery 
grant.
}}

\maketitle

\begin{abstract}
We present Geometry of Interaction (GoI) models
for Multiplicative Polarized Linear Logic, \MLLP,
which is the multiplicative fragment 
of Olivier Laurent's  Polarized Linear Logic.
This is done by uniformly adding {\em multipoints} to various categorical models  of GoI.
Multipoints are shown to play an essential role
in  semantically 
characterizing the dynamics of proof networks in     polarized
proof theory.  For example, they permit us to characterize the key
feature of polarization, {\em focusing}, as well as 
being fundamental to  our construction of
concrete polarized  GoI models.

Our approach to polarized GoI involves two independent studies, based
on different categorical perspectives of GoI. 

\vspace{1ex}

\begin{enumerate}[(i)]
\item Inspired by the work of Abramsky, Haghverdi, and Scott, a {\em polarized GoI situation} is defined in which  multipoints  are added to a traced monoidal category equipped with a 
reflexive object $U$. Using this framework, categorical
versions of Girard's Execution formula 
are  defined, as well as the GoI interpretation of  \MLLP\,  proofs.
Running the Execution formula is shown to characterize the focusing property
(and thus polarities) as well as  the dynamics of cut-elimination. 
\item The $\Int$ construction of Joyal-Street-Verity is another fundamental categorical structure
for modelling GoI.  Here, we investigate it in a multipointed
setting.  
 Our presentation yields
a compact version of Hamano-Scott's polarized categories, and thus denotational models
of \MLLP.   These arise from a contravariant  duality between monoidal categories of 
{\em positive} and {\em negative}  objects, along with an appropriate bimodule structure
(representing ``non-focused proofs")  between them.
\end{enumerate}

\vspace{1ex}

\noindent
Finally, as a special case of (ii) above, a compact model of $\MLLP$ is
 also presented based on $\Rel$ (the category of sets and relations) equipped with multi-points. 

\end{abstract}

\tableofcontents

\section{Introduction}

Linear Logic, introduced by Girard in 1987 \cite{Gi87}, originated from a profound analysis of the proof theory of traditional logic. In particular, linear logic involves a fine-grained study of how rules and connectives 
manipulate resources. This important development is by now quite familiar to researchers in many areas of logic and computer science.  What is somewhat less familiar is that shortly after the introduction of linear logic, Andreoli  \cite{and1,and2}  pointed out a different approach to the fundamental
connectives of linear logic; namely, to classify the connectives according to whether their introduction rules are    reversible  ({\em
negative}) or irreversible ({\em positive}) .   Positive connectives are the foundation of   
Andreoli's influential notion of {\em focusing} in 
proof search. The fundamental role of focusing in logic programming has been actively explored in numerous recent works  (for example, in papers of  D. Miller, K. Chaudhuri, et. al.   \cite{Dale,Dale2,Dale3,Chau, Chau2}).  

The Andreoli view also led to intrinsic studies of polarity and polarized logics, first taken up by Girard in  
\cite{Gir91,GirMean}, and systematically studied by O. Laurent in \cite{OLaur99,OLaur02}.   Such logics are also related to Girard's Ludics games
\cite{Girlocu} and other games-semantics models (cf.  the dialogue games and 
  recent dialogue categories with chiralities of  Melli\`es \cite{PAM2,PAM3}.)
In related categorical proof theory, we should mention the polarized categories and proof theory 
of  Cockett and Seely \cite{CS07}, which influenced  our own \cite{HamSc07}, as well as proof-theoretical papers of one of us \cite{HamTak08,HamTak10}.

In this paper we begin a study of the  dynamics of cut-elimination for the multiplicative
fragment $\MLLP$ of O. Laurent's polarized linear logic, using categorical versions  of Girard's Geometry of Interaction (GoI) program \cite{Gi89,Gi95}; here we follow the categorical GoI literature \cite{AHS02,HS06}.
 %
A common theme to the different parts of this paper is a fundamental new semantical idea first discussed 
in \cite{HamTak08}:  the addition of {\em multipoints}.  These multipoints have no syntactic counterpart but nevertheless provide a new understanding of the dynamics of cut-elimination in the presence of polarities.

\begin{itemize}
\item In Section \ref{polgoisit}, following the methods of Abramsky, Haghverdi, and Scott \cite{AHS02} we introduce {\em polarized GoI situations with
multipoints} as an appropriate but simple categorical setting for studying Girard's
Execution formula in the polarized setting.  
The version of the execution formula
we use is inspired by  the general categorical execution formula in
      Haghverdi-Scott \cite{HS06,HS11} for linear logic, but now
      extended to the multipointed setting.  For the polarized
      multiplicative system \MLLP, the execution formula becomes a {\em
      two-layered}  pair of execution formulas, one at the usual
      reflexive object level (as in \cite{HS06})
and a similar one at the multipoint-level.  The usual GoI properties for
      the dynamics of cut-elimination \cite{Gi89,HS06,HS11} turn out to
      be much stronger for \MLLP.  The execution formula(s) form   {\em
      full invariants} for normalization \footnote{in the sense that
      if $\pi \rightarrow_{*} \pi'$ by $\MLLP$ cut-elimination,
then ${\sf Ex}(\Mean{\pi}, \sigma) = {\sf Ex}(\Mean{\pi'}, \sigma')$}, which is well-known to fail in full linear 
logic (see \cite{AHS02} and Section \ref{polarex}  below). In fact the polarized execution formula(s)  satisfy a fundamental additional property. Namely, in Proposition 
\ref{converse_focus} below, we characterize  focusing, which is  intrinsic to \MLLP, as preservation of multipoints under the (polarized) execution  formulas.  Thus, in a precise sense, the execution formulas give rise to the polarities.

\item In the next  Sections \ref{polcomp-int} and \ref{rel-mp}, which are independent of
Section  \ref{polgoisit}, we consider general multipointed traced monoidal categories (TMCs) and study the \Int\, construction of
 Joyal, Street, and Verity \cite{JSV96} in this setting.  As is well-known \cite{AHS02, HS06},
 the \Int\, construction is an essential feature of all the different
      categorical approaches to GoI.  It yields a kind of  ``compact
      closure"   of a traced monoidal category; moreover, composition in
      the \Int\, category leads to  categorical versions of Girard's execution formula.
 As in Section \ref{polgoisit} above, we investigate  a two-layered $\Int$ construction,
the upper layer for general objects and the lower one   restricted to
      multipoints. 

In Section \ref{polcomp-int},  we  study a general categorical semantics for certain polarized linear logics, a simplified
 version of the bi-module duality framework in our paper \cite{HamSc07}. These are related to more general polarized categories  introduced for modelling polarities (see \cite{CS07,HamSc07,PAM2,PAM3}).  Our goal is to use GoI and the $\Int$\, construction to build compact polarized categorical models of \MLLP.
 
 In these sections multipoints  give a semantical framework for explaining the idea of bidirectional dataflow implicit in the \Int\, construction (see the discussion preceeding Proposition
  \ref{int-P-cat} below).
In this setting, multipoints satisfy a certain commutativity 
condition--corresponding to the focusing condition discussed in Section \ref{polgoisit}--which we show is compatible with  this construction.  As one expects, this yields an appropriate compact closed version of a denotational model for \MLLP.

Finally,  Section \ref{rel-mp} constructs a concrete instance of the \Int\, construction of Section \ref{polcomp-int}, when specialized to a multipointed version of \Rel.   It may be read independently of the previous section
and uses the relational calculus of Joyal-Street-Verity \cite{JSV96}.

\item 
In order to make the paper self-contained, the Appendices include some supplemental material.
In Appendix \ref{appdx1} we briefly recall the original {\em GoI situations} in the sense of 
Abramsky, Haghverdi and Scott  \cite{AHS02},  as well as the categorical approach to GoI
(for a survey, see \cite{HS11}).  In Appendix \ref{UDC} we recall Haghverdi's Unique Decomposition Categories
(UDC's) which provide a general framework (and matrix calculus) for ``particle-style" GoI, familiar from 
Girard's GoI 1 \cite{Gi89,HS06}.   The other Appendices include some detailed but routine proofs.


\end{itemize}

\section{Polarized Multiplicative Linear Logic $\MLLP$  }
\label{MLLP}
Following the work of Andreoli, polarities naturally arise within the proof
theory of linear logic, \LL. We can further divide the connectives 
according to whether their introduction rules
are reversible or not
\cite{GirMean,OLaur02}.
Those connectives which are reversible are called {\em
negative}; those which are not are called {\em positive}. 
Positive connectives are the foundation of   
Andreoli's influential  {\em focusing property} in 
proof search for linear logic  
\cite{and1,and2,Dale}.  Focusing is a property dual  
to reversibility. 

We recall Olivier Laurent's theory of {\em polarized
multiplicative linear logic},  \MLLP .  The theory
$\MLLP$ is a fragment (without structural rules)
of Laurent's full polarized linear logic $\LLP$ \cite{Girlocu}. 

 \vspace{1ex}

\begin{defn}[Polarized $\MLL$]{\em
The theory  $\MLLP$   is defined as
follows.

\vspace{1ex}

{\bf Syntax}: {\em Positive} and {\em Negative} formulas are
given by the following BNF notation:

$$\begin{array}{ccccccccccc}
P & ::= &  X          & \mid & P \otimes P & \mid &    \mbox{\bf 1} & \mid &    \down{N} \\
N & ::= & X^\perp & \mid & N \wp N      & \mid &    \bot 	      	   & \mid &      \up{P}
\end{array}$$

\noindent
Here $X$ is an atom, $\uparrow$ and $\downarrow$ are called {\em polarity shifting}
operations. Note that $\mbox{\bf 1}$ 
is the unit of $\otimes$
(and dually for  $ \bot$ 
 with respect to $ \wp$). 
In our categorical models introduced later, $\uparrow$ and $\downarrow$ will be  functorial
 operators  weaker than the traditional exponentials of linear logic.
\vspace{2ex}

{\bf Syntactic Negation:}
Following O. Laurent, we  adjoin to $\MLLP$ a syntactic strictly involutive
negation $( - )^\perp$, defined   by   general de Morgan duality.  Thus we extend the negation on
positive atoms to all formulas, as follows:
$ X^{\perp \perp} = X$ for atoms $X$,  and we assume  $\{\otimes, \wp\}$
are de Morgan duals, as in linear logic.  
Finally de Morgan duals for polarity changing operators are
$(\down{A})^\perp = \up{A^\perp}$ and
$(\up{A})^\perp = \down{A^\perp}$  for any formula $A$.  Positivity and negativity
of formulas may be defined as before, after cancelling any occurrences of
double-negations.

\vspace{1ex}

{\bf Rules of $\MLLP$}

\vspace{1ex}

\noindent
In the following rules, $M$ and $N$ (resp. ${\cal M}$ and ${\cal
 N}$) range  over  negative formulas  (resp. sequences of negative formulas)
and $P$ and $Q$ over positive formulas.
$\Gamma$ contains {\em at most one} positive formula.
We assume the rule of exchange, so sequents are closed under
permutation.

\vspace{2ex}

\begin{minipage}[t]{2in}
$\infer[\mbox{\footnotesize Axiom}]{\vdash N, N^\perp}{}$
\end{minipage}
\begin{minipage}[t]{2in}
$\infer[\otimes]{\vdash {\cal M}, {\cal N}, P \otimes Q}{
\vdash {\cal M}, P & \vdash {\cal N}, Q}$
\end{minipage}
\begin{minipage}[t]{2in}
$\infer[\wp]{\vdash \Gamma, N \wp M}{
\vdash \Gamma, N, M}$
\end{minipage}

\vspace{2ex}



\begin{minipage}[t]{2in}
$\infer[\down{}]{\vdash \down{N}, {\cal N}}{\vdash N, {\cal N}}$
\end{minipage}
 \begin{minipage}[t]{2in}
$\infer[\up{}]{\vdash \up{P}, {\cal N} }{\vdash P, {\cal N}}$
\end{minipage}
\begin{minipage}[t]{2in}
$\infer[\mbox{cut}]{\vdash \Gamma, {\cal M}}{\vdash \Gamma, N & \vdash
{\cal M}, N^{\perp}}$
\end{minipage}
\vspace{1ex}

  }

\end{defn}



\vspace{2ex}

\noindent
The following theorem is an important proof-theoretical
property of $\MLLP$, proved in \cite{OLaur99,OLaur02}:

\vspace{2ex}

\begin{prop}[Focalization Property]\label{fp} 
If \, $\vdash \Gamma$ is provable in $\MLLP$,
then the sequence $\Gamma$ contains at most
one positive formula. 
\end{prop}

\vspace{1ex}

A {\em focused sequent} is one of the form
$\vdash P, \cN$, while a {\em nonfocused sequent} has the form $\vdash \cN$, where
$\cN$ is a finite sequence of negative formulas and $P$ is positive.  Proposition \ref{fp}
says that every provable sequent of $\MLLP$ is either focused or
nonfocused.  We say a proof is focused if the sequent it proves is
focused.
We say  a proof {\em has the  focusing property} if it is focused.

\vspace{1ex}

\noindent
{\bf Notation:}  For the discussion of GoI and Cut-elimination, we use
 the Girard notation \cite{Gi89} for sequent calculus proofs:  a proof of
 the sequent $\vdash [\Delta] ,\Gamma$  denotes a  proof of the sequent $\vdash \Gamma$
 appended with the list $\Delta$ of all pairs of cut formulas $A,A^\perp$ used  in the proof.
Here   $|\Gamma| = n$ and $|\Delta| = 2m$, for some $m, n$.  This is further described in the
categorical GoI papers of Haghverdi-Scott (see Appendix \ref{appdx1} and Figure \ref{morgraph} there.)  

\vspace{2ex}

\noindent
{\bf Cut elimination for $\MLLP$:}
$\MLLP$ is the subsystem of polarized linear logic $\LLP$ \cite{OLaur02} without additive connectives and structural rules
(where $\downarrow$ and $\uparrow$ replace, respectively,  $!$ and $?$).
The cut elimination theorem for $\MLLP$ is 
obtained by restricting the one for
$\LLP$ (cf. Definition 5.27 of \cite{OLaur02}).
For this, the interpretation using polarized proof-nets (e.g. \cite{OLaur99})
is essential.  The crucial ingredient for our subsystem $\MLLP$
is the use of {\em boxes} to interpret the $\downarrow$-rules.
Each box has a principal door
and an auxiliary door on which occur, respectively,
the principal formula $\down{N}$ and the side formulas ${\cal M}$.

The crucial step for cut-elimination is the following  (cf. the
!/! case of Definition 5.27 of \cite{OLaur02}):  
reduction of a cut against a side formula of the $\downarrow$ rule.
 Here, a proof ending with a cut against a formula at an auxiliary door of
a box is reduced to that ending with the $\downarrow$-rule,
whose box is enlarged to contain all rules including the original cut.  This is illustrated
as follows:


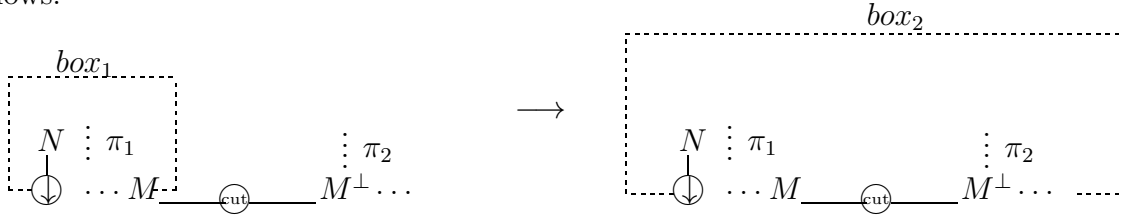
\begin{minipage}{0.4\hsize}
\begin{picture}(20,80)(60,30)\setlength{\unitlength}{0.25mm}
\put(100,78){$N$}
\put(105,75){\line(0,-1){10}}
\put(105,57){\circle{15}\makebox(0,0){\hspace{-4ex}$\downarrow$}}

\put(125,55){$\ldots$}

\put(125,70){$\infer*[\pi_1]{}{}$}

\put(148,52){$M$}

\put(165,50){\line(3,0){35}}
\put(205,52){\circle{15}\makebox(0,0){\hspace{-5ex} \tiny cut}}
\put(248,50){\line(-3,0){35}}

\multiput(97,57)(-5,0){3}{\line(-1,0){2}}
\multiput(85,57)(0,5){12}{\line(0,1){2}}
\multiput(85,117)(5,0){18}{\line(1,0){2}}
\multiput(162,57)(5,0){3}{\line(1,0){2}}
\multiput(174,57)(0,5){12}{\line(0,1){2}}

\put(110, 120){$box_1$}
\put(280,55){$\ldots$}
\put(250,52){$\infer*[\pi_2]{M^\bot}{}$}
\end{picture}
\end{minipage}
\begin{minipage}{0.1\hsize}
$\longrightarrow$
\end{minipage}
\begin{minipage}{0.4\hsize}
\begin{picture}(0,80)(60,30)\setlength{\unitlength}{0.25mm}
\put(100,78){$N$}
\put(105,75){\line(0,-1){10}}
\put(105,57){\circle{15}\makebox(0,0){\hspace{-4ex}$\downarrow$}}

\put(125,55){$\ldots$}

\put(125,70){$\infer*[\pi_1]{}{}$}

\put(148,52){$M$}
\put(165,50){\line(3,0){35}}
\put(205,52){\circle{15}\makebox(0,0){\hspace{-5ex} \tiny cut}}
\put(248,50){\line(-3,0){35}}
\put(250,52){$\infer*[\pi_2]{M^\bot}{}$}
\put(280,55){$\ldots$}
\multiput(97,55)(-5,0){5}{\line(-1,0){2}}
\multiput(72,55)(0,5){17}{\line(0,1){2}}
\multiput(72,140)(5,0){53}{\line(1,0){2}}
\multiput(312,55)(5,0){5}{\line(1,0){2}}
\multiput(337,55)(0,5){17}{\line(0,1){2}}
\put(200, 145){$box_2$}
\end{picture}
\end{minipage}

\vspace{1ex}

\begin{exam}[Extrusion of $\downarrow$-boxes through cut-elimination]
\label{cutexextru}{\em 
Consider the following three proofs, as instances of normalization: $\pi_1\succ\pi_2\succ\pi_3$.  

\begin{tabular}{ll}
\footnotesize $\pi_1 =$ & \footnotesize $\pi_2 =$ \\
\footnotesize
$
\infer[\mbox{\small cut}]{\vdash  \, [\uparrow\!
 X_1, \downarrow X_2^{\perp}, \uparrow\! X_2, \downarrow X_3^\perp], \,
\downarrow X_1^\perp,\uparrow\!
 X_3}{\infer[\mbox{\small cut}]{\vdash  \, [\uparrow\!
 X_1, \downarrow X_2^{\perp}], \,  \downarrow X_1^\perp, \uparrow\!
 X_2}{\infer[\downarrow]{\vdash \downarrow X_1^\perp,\uparrow\!
 X_1}{\infer{\vdash X_1^\perp,\uparrow \!X_1}{\vdash X_1^\perp,X_1}} &
 \infer*{\vdash  \downarrow X_2^{\perp},\uparrow\! X_2}{}} & 
\infer*{\vdash \downarrow X_3^\perp,\uparrow\! X_3}{} }
$ 
& 
\footnotesize
$
\infer[\mbox{\small cut}]{\vdash \, [\uparrow\!
 X_1, \downarrow X_2^{\perp}, \uparrow\! X_2, \downarrow X_3^\perp], \, \downarrow X_1^\perp,\uparrow\!
 X_3}{\infer[\downarrow]{\vdash \, [\uparrow\!
 X_1, \downarrow X_2^{\perp}], \, \downarrow X_1^\perp,\uparrow
 \!X_2}{\infer[\mbox{\small cut}]{\vdash \, [\uparrow\!
 X_1, \downarrow X_2^{\perp}], \, X_1^\perp,\uparrow
 \!X_2}{\infer{\vdash X_1^\perp,\uparrow\!X_1}{\vdash 
X_1^\perp,X_1}&\infer*{\vdash \downarrow X_2^\perp,\uparrow \!X_2}{}}} & 
\infer*{\vdash \downarrow X_3^\perp,\uparrow\! X_3}{}}
$
\end{tabular}

\bigskip

$$ \footnotesize
\raisebox{4ex}{$\pi_3$ \ \ = \ \ }
\infer[\downarrow]{\vdash \, [ \uparrow\! X_1, \downarrow X_2^{\perp}, \uparrow\! X_2, \downarrow X_3^\perp], \, \downarrow \!X_1^\perp,\uparrow\! X_3}
{\infer[\mbox{\small cut}]{\vdash \, [ \uparrow\! X_1, \downarrow X_2^{\perp},
 \uparrow\! X_2, \downarrow X_3^\perp], \, X_1^\perp,\uparrow
 \!X_3}{\infer[\mbox{\small cut}]{\vdash 
\, [\uparrow\!
 X_1, \downarrow X_2^{\perp}], \,
 X_1^\perp,\uparrow\! X_2}{\infer{\vdash X_1^\perp,\uparrow\! X_1}{\vdash X_1^\perp,X_1}&\infer*{\vdash \downarrow \!X_2^\perp,\uparrow\! X_2}{}} & \infer*{\vdash \downarrow\! X_3^\perp,\uparrow\! X_3}{}}}
$$


\begin{picture}(60,140)(40,30)\setlength{\unitlength}{0.3mm}
\put(105,89){\line(0,1){8}}
\put(100,78){$X_1^\perp$}
\put(155,89){\line(0,1){8}}
\put(150,78){$X_1$}
\put(105,97){\line(1,0){50}}
\put(105,75){\line(0,-1){10}}
\put(105,57){\circle{15}\makebox(0,0){\hspace{-5ex}$\downarrow_1$}}
\put(155,57){\circle{15}\makebox(0,0){\hspace{-5ex}$\uparrow_1$}}
\put(155,75){\line(0,-1){10}}

\put(160,50){\line(3,0){38}}
\put(205,50){\circle{15}\makebox(0,0){\hspace{-6ex} \scriptsize cut}}
\put(250,50){\line(-3,0){38}}

\multiput(97,57)(-5,0){3}{\line(-1,0){2}}
\multiput(85,57)(0,5){12}{\line(0,1){2}}
\multiput(85,117)(5,0){18}{\line(1,0){2}}
\multiput(162,57)(5,0){3}{\line(1,0){2}}
\multiput(174,57)(0,5){12}{\line(0,1){2}}

\put(110, 120){$box_1$}


\put(255,89){\line(0,1){8}}
\put(250,78){$X_2^\perp$}
\put(305,89){\line(0,1){8}}
\put(300,78){$X_2$}
\put(255,97){\line(1,0){50}}
\put(255,75){\line(0,-1){10}}
\put(255,57){\circle{15}\makebox(0,0){\hspace{-5ex}$\downarrow_2$}}
\put(305,57){\circle{15}\makebox(0,0){\hspace{-5ex}$\uparrow_2$}}
\put(305,75){\line(0,-1){10}}


\multiput(247,57)(-5,0){3}{\line(-1,0){2}}
\multiput(235,57)(0,5){12}{\line(0,1){2}}

\multiput(235,117)(5,0){18}{\line(1,0){2}}
\multiput(312,57)(5,0){3}{\line(1,0){2}}
\multiput(324,57)(0,5){12}{\line(0,1){2}}

\multiput(97,55)(-5,0){5}{\line(-1,0){2}}
\multiput(72,55)(0,5){19}{\line(0,1){2}}
\multiput(72,150)(5,0){53}{\line(1,0){2}}
\multiput(312,55)(5,0){5}{\line(1,0){2}}
\multiput(337,55)(0,5){19}{\line(0,1){2}}

\put(200, 155){$box_2$}

\put(405,89){\line(0,1){8}}
\put(400,78){$X_3^\perp$}
\put(455,89){\line(0,1){8}}
\put(450,78){$X_3$}
\put(405,97){\line(1,0){50}}
\put(405,75){\line(0,-1){10}}
\put(405,57){\circle{15}\makebox(0,0){\hspace{-5ex}$\downarrow_3$}}
\put(455,57){\circle{15}\makebox(0,0){\hspace{-5ex}$\uparrow_3$}}
\put(455,75){\line(0,-1){10}}

\put(310,50){\line(3,0){38}}
\put(355,50){\circle{15}\makebox(0,0){\hspace{-5ex}\scriptsize cut}}
\put(400,50){\line(-3,0){38}}


\multiput(397,57)(-5,0){3}{\line(-1,0){2}}
\multiput(385,57)(0,5){12}{\line(0,1){2}}

\multiput(385,117)(5,0){18}{\line(1,0){2}}
\multiput(462,57)(5,0){3}{\line(1,0){2}}
\multiput(474,57)(0,5){12}{\line(0,1){2}}

\multiput(97,50)(-5,0){10}{\line(-1,0){2}}
\multiput(52,50)(0,5){24}{\line(0,1){2}}
\multiput(52,170)(5,0){87}{\line(1,0){2}}
\multiput(462,50)(5,0){5}{\line(1,0){2}}
\multiput(487,50)(0,5){24}{\line(0,1){2}}
\put(330, 175){$box_3$}

\end{picture}

During normalization, boxes enter outer boxes and the scope of the outer
 box is extruded. That is,
for $i=1,2,3$, we represent proof $\pi_i$ by the proof-net obtained by choosing $box_i$
for the left-most $\downarrow$. The $box_i$ corresponds to the indicated $\downarrow$- rule in $\pi_i$.
Normalization gives rise to extrusions of $box_1$ 
into $box_2$ to include the middle box of $\pi_2$ and then
into $box_3$ to include both the middle and
right boxes of $\pi_3$.
}
\end{exam}

\section{Polarized GoI and GoI Situations}
\label{polgoisit}

{\em GoI situations} were first introduced in Abramsky, Haghverdi, and Scott  \cite{AHS02} as an algebraic
framework for Girard's GoI for full linear logic.   Later, Haghverdi and Scott \cite{HS06,HS11} gave a detailed categorical analysis of Girard's original GoI for \MELL\, in 
GoI situations associated to Unique Decomposition categories (UDCs), augmented with abstract (Hyland-Schalk) orthogonality relations.  For a quick summary of this categorical  GoI for the original (nonpolarized) setting, see  Appendix \ref{appdx1} (and its references) for GoI and Appendix \ref{UDC} for UDCs.  In particular, we will make use of the categorical version of the Execution Formula.

In  this section we develop an appropriate notion of {\em Polarized GoI Situation with  multipoints} for 
\MLLP.   Multipoints will enable us to give a detailed analysis of the dynamics and information flow in cut-elimination, via  polarized (2-layered) execution formulas.  It will also enable us to give a characterization of focusing. 
Our main result in this Section \ref{polgoisit} is to semantically characterize
the proof-theoretical notion of focusing
in terms of the polarized execution formulas. This goes as follows.
First, in Theorem \ref{invari},
the polarized execution formulas for focused sequents
are shown to satisfy an invariance property preserving the multipoints.
Conversely in Proposition \ref{converse_focus}, this invariance
property is shown to be sufficient to distinguish focused sequents.
Thus, in a precise sense, polarized GoI execution formulas give rise to
the polarities.

\vspace{1ex}

\noindent
{\bf Notation:}  In a category $\cC$, we denote identity arrows
$X\rightarrow X$ either as $\Id_X$ or just $X$, depending on context.
Let $\xymatrix@=0.1in
{Y \ar@<.8ex>[rr]^{g} &   & X \ar@<.8ex>[ll]^{f} }$ be a pair
of morphisms in  a category. 
 We write $g: Y\rhd X :f$ 
to mean $g \Comp f = \Id_X$ and we say {\em $X$ is a retract of $Y$ with
respect to $(g,f)$}.
This is often abbreviated by
$Y\rhd_{(g,f)} X$.
We say $X$ {\em is a retract of }$Y$ if there is a  pair $(g,f)$ as above such
that
$Y\rhd_{(g,f)}X$.
 

For simplicity, we assume all categories $\cC$ below are  locally small, so we can speak of hom-sets
(rather than hom-classes).

\subsection{Polarized GoI situations}

We introduce a polarized  analog of GoI situations from   \cite{AHS02} (see Appendix \ref{appdx1} for
a summary), which is suitable 
for polarized multiplicative linear logic.   The full exponential operators $T$ of GoI situations for
(nonpolarized) linear logic \LL~  are here replaced by much weaker,
functorial   {\em polarity shifters:} $\uparrow$ and $\downarrow$, as well as by introducing the critical notion of multipoints.
 
\bigskip

\begin{defn}[Polarized GoI situation] \label{GoIsitu}
{\em A {\em polarized GoI situation} is a tuple
$$({\cal C}, \otimes, I, s, U, 1, (\mathfrak{i}, \mathfrak{r}),
1 \stackrel{\alpha}{\longrightarrow} U,
(\mathfrak{i}_{\alpha},\mathfrak{r}_{\alpha}), 0)$$
where:
\begin{enumerate}
\item 
$({\cal C}, \otimes, I, s)$ is a traced symmetric monoidal category
      (with symmetry maps $s = \{s_{A,B}: A\otimes B\rightarrow B\otimes
      A\}$), and with unit object $I$, satisfying the usual identities
      (\cite{Mac Lane}).
(A {\em trace} is given by a family of
functions ${\sf Tr}_{X,Y}^Z : {\cal C}(X \otimes Z, Y \otimes Z)
\longrightarrow  {\cal C}(X, Y)$ subject to three 
naturalities:  (Natural in $X$), (Natural in $Y$), (Dinatural in $Z$),
and three axioms (Vanishing) (Superposing) and (Yanking).
We refer to \cite{JSV96, AHS02, HS06} for detailed treatment.)
{ In the monoidal structure, $X^m$ and $f^m$ denote respectively
 $m$-ary tensors  (also called {\em tensor foldings}) \,
$\overbrace{X \otimes \cdots \otimes X}^m$
and $\overbrace{f \otimes \cdots \otimes f}^m$. }


\item
$U$ is an object of ${\cal C}$
with a retraction  $U   \rhd_{(k,j)}     U \otimes U$, i.e.
$$\begin{array}{lr}
$\xymatrix{
U 
\ar@<1ex>[rr]^{k} &   & U\otimes U \ar@<1ex>[ll]^{j} }$
&
\mbox{such that
$k \Comp j = \Id_{U\otimes U}$.}
\end{array}$$

\item
$1$ is an object of ${\cal C}$  with
a  retraction $ 1\otimes 1\rhd_{(\mathfrak{i},\mathfrak{r})} 1 $ , i.e.
$$\begin{array}{lr}
$\xymatrix{
1\otimes 1
\ar@<1ex>[rr]^{\mathfrak{i}} &   & 1 \ar@<1ex>[ll]^{\mathfrak{r}} }$
& \mbox{
such that $\mathfrak{i} \Comp \mathfrak{r} = \Id_{1}$.}
\end{array}$$
{\em Note:} In general, $1\not = I$.

\item ({\em Distinguished point}) 
A morphism $1 \longrightarrow U$
is called a {\em point} of $U$.  A polarized GoI situation
has a distinguished point $1\stt{\alpha}U$ among the points of
$U$



\item ({\em Uniformity of Trace} ) \\
The trace of ${\cal C}$ is {\em uniform}
(cf. Simpson and Plotkin \cite{SimpPlot} and Hasegawa \cite{Hassei})
over points; i.e., 
every point    $p: 1 \longrightarrow U$
satisfies the following condition, which says points are trace invariant.

For any morphisms $f$ and $g$,
$$\xymatrix
@R=1pc
{
  X \otimes U  \ar[rr]^f    &     &  Y \otimes U \\
                                          &     &                       &   
\mbox{implies}  &  {\sf Tr}^{U}_{X,Y}(f)= {\sf Tr}^{1}_{X,Y}(g).                                 \\
X \otimes 1 \ar[uu]^{X \otimes  p}
 \ar[rr]_g   &    &  Y \otimes 1 \ar[uu]_{Y \otimes p}}
 $$



The final axioms $6$, $7$ and $9$ in the definition concern properties satisfied 
by the distinguished point 
$1 \stackrel{\alpha}{\longrightarrow} U$.

\item ({\em Lifting Property $U \otimes 1 \rhd_{(\mathfrak{i}_\alpha, \mathfrak{r}_\alpha)} U$
along  $\alpha$} ) \\
For the distinguished point $1 \stackrel{\alpha}{\longrightarrow} U$,
there exists a pair $(\mathfrak{i}_\alpha, \mathfrak{r}_\alpha)$ giving a
retraction $U \otimes 1 \rhd_{(\mathfrak{i}_\alpha, \mathfrak{r}_\alpha)}  U$
which lifts the retraction
$1\otimes 1 \rhd_{(\mathfrak{i},r)} 1$
along the point $\alpha$.
This means the following
diagram commutes  (in all possible ways) 
with $\mathfrak{i}_\alpha \Comp \mathfrak{r}_\alpha = \Id_{U}$
and $\mathfrak{i} \Comp r = \Id_{1}$:
$$
\xymatrix
      @R=1.2pc
{
U \otimes  1 \ar@<1ex>[rr]^{\mathfrak{i}_\alpha} &  &  \ar@<1ex>[ll]^{\mathfrak{r}_\alpha} U \\ \\
1 \otimes 1 \ar[uu]^{\alpha \otimes 1} \ar@<1ex>[rr]^{\mathfrak{i}}   &            &  
  \ar[uu]_{\alpha} \ar@<1ex>[ll]^{\mathfrak{r}}  1  \\
}
$$

\item
({\em Semi-invertibility of $\alpha$}) \\
The point
$\alpha : 1 \longrightarrow U$
is {\em semi-invertible}. That is, there exists
$ \alpha^{*} : U \longrightarrow 1$ such that
$\alpha^{*} \Comp \alpha=\Id_1 $.

\item ({\em $0$ morphisms}) \\     
The category $\cC$ has {\em zero morphisms}.  This means for every pair of objects
$X, Y\in \cC$, there is an assigned  map $0_{XY}: X\rightarrow Y$  such that the family  $\{0_{XY} \ | \ X, Y\in \cC\}$  satisfies:  for every $f: W\rightarrow X, g: Y\rightarrow Z$,
$$
\xymatrixcolsep{3pc}\xymatrix
{ 
W \ar[d]^f \ar[r]^{0_{WZ}}&Z\ar[d]^g\\ X	\ar[r]^ {0_{XY}}	&Y  
} 
$$
Note:  if $f$ or $g$ equals the identity, this amounts to the fact that any composition with
one factor zero is itself zero  (see also \cite{MA}). For simplicity,
$0$ denotes the zero morphism $0_{11}: 1\rightarrow 1$. 
Note that $f \otimes 0_{Y,Z}$ is not in general $0_{W \otimes Y, X \otimes Z}$
for any objects $X,Y,W,Z$.\footnote{Hence zero morphisms are absorbing with respect to composition, but not
with respect to tensor.}             

\item
({\em $(\mathfrak{i}_\alpha,\mathfrak{r}_\alpha)$ and $0$} ) \\
For every morphism
$f: V \otimes X \longrightarrow W \otimes X$ 
with $X \in \{ U, 1 \}$
and $0: 1 \longrightarrow
1$,
\begin{eqnarray*}
(\Id_W \otimes \mathfrak{i}_\alpha ) \Comp (f \otimes 0) \Comp
(\Id_V \otimes \mathfrak{r}_\alpha)  = f  &\mbox{and} &   
(\Id_W \otimes \mathfrak{i} ) \Comp ( f \otimes 0) \Comp
(\Id_V \otimes \mathfrak{r} ) = f 
\end{eqnarray*}
Diagrammatically, the following 
are the respective equations
when $X=U$ and $X=1$:

\begin{figure}[hbt]
\begin{minipage}{0.5\hsize} 
$\begin{picture}(180,85)(-60,-45)
\put(30,-15){\fbox{\rule[.5in]{.4in}{0in }}}
\put(12,-40){$1$}   \put(70,-40){$1$}

\put(30,-40){\vector(1,0){35}} \put(43,-50){$0$}

\put(12,18){$V$}    \put(70,18){$W$}
\put(41,0){$f$}
\put(12,-5){$U$} \put(70,-5){$U$}
\put(130, 0){$= \ \ f$}

\put(-15,-25){\vector(2,1){25}}  
\put(-15,-25){\vector(2,-1){25}}

\put(43,-30){$\otimes$}
\put(-30,-30){$U$}

\put(-5,-28){$\mathfrak{r}_{\alpha}$}
\put(85,-28){$\mathfrak{i}_{\alpha}$}
\put(12,-10){\vector(1,0){18}}
\put(66,12){\vector(1,0){25}}
\put(66,-10){\vector(1,0){10}}
\put(5,12){\vector(1,0){25}}


\put(78,-10){\vector(2,-1){25}}
\put(78,-40){\vector(2,1){25}}

\put(110,-30){$U$}
\end{picture}$
\end{minipage}
\begin{minipage}{0.1\hsize}
and 
\end{minipage}\hspace{-6ex}
\begin{minipage}{0.4\hsize} 
$\begin{picture}(180,85)(-50,-45)
\put(30,-15){\fbox{\rule[.5in]{.4in}{0in }}}
\put(12,-40){$1 $}   \put(70,-40){$1 $}

\put(30,-40){\vector(1,0){35}} \put(43,-50){$0$}

\put(12,18){$V$}    \put(70,18){$W$}
\put(41,0){$f$}
\put(12,-5){$1$} \put(70,-5){$1$}
\put(130, 0){$= \ \ f$}

\put(-15,-25){\vector(2,1){25}}  
\put(-15,-25){\vector(2,-1){25}}

\put(43,-30){$\otimes$}
\put(-30,-30){$1$}

\put(-5,-28){$\mathfrak{r}$}
\put(85,-28){$\mathfrak{i}$}
\put(12,-10){\vector(1,0){18}}
\put(66,12){\vector(1,0){25}}
\put(66,-10){\vector(1,0){10}}
\put(5,12){\vector(1,0){25}}

\put(78,-10){\vector(2,-1){25}}
\put(78,-40){\vector(2,1){25}}

\put(110,-30){$1$}
\end{picture}$\end{minipage}
\end{figure}
\end{enumerate}

}
\end{defn}
This ends the definition of a polarized
GoI situation.

\vspace{2ex}
 
\noindent
The following  Examples \ref{relexpl} and \ref{pfnexpl} of polarized GoI situations are 
built from the category  \Rel\ of sets and relations. 
$\Rel$ has two standard traced-monoidal structures
(see \cite{AHS02,JSV96} for details),
one with $\otimes = \times$ (cartesian product), the other $\otimes = +$ (disjoint union).
These two  categories are denoted  by $\Rel_\times$ and $\Rel_+$, respectively.
We don't
discuss $\Rel_\times$ in this paper, so  in what follows we often abbreviate $\Rel_+$ 
to  $\Rel$.   Below, we emphasize the additional polarized structure.

\vspace{2ex}

\begin{expl}[Polarized GoI situation $\Rel_+$]{\em
$$ ( \Rel, \otimes, I, s, \N,  1,  (\mathfrak{i},\mathfrak{r}), 
1 \stackrel{\alpha}{\longrightarrow} U,
(\mathfrak{i}_\alpha,\mathfrak{r}_\alpha), 0)$$
is a polarized GoI situation, denoted ${\sf Rel}_+$ (or just $\Rel$),   where we define:

\begin{itemize}
\item[-] The objects of $\Rel_+$  are sets and morphisms are (binary) relations between them.
\item[-] $\otimes$ is the disjoint union $+$, where $A+B : =  (\{1\}\times A)\cup (\{2\}\times B)$.
\item[-] $I:=\emptyset$, $U: = \N$, $1:=\{  * \}$. 
\item[-] The retraction $k: \N \rhd \N  +  \N  : j$
is a standard one often used in GoI  \cite{Gi89, AHS02,Hag00}:  

\qquad $j(1,n)=2n$ ,  $j(2,n)=2n+1$, and
$k(n)=  \left \{ \begin{array}{cc}
(1, \frac{n}{2}) & \mbox{$n$ even} \\
(2, \frac{n-1}{2}) & \mbox{$n$ odd} 
\end{array} \right. $

\item[-] $\mathfrak{r}: 1\rightarrow 1+1$ is the maximal relation, and $\mathfrak{i} : 1+1\rightarrow 1$ is its converse.

\item[-] 
$ \alpha : \{ * \} \longrightarrow \N$
is a {\em non-empty relation} that determines
a distinguished {\em singleton} subset $\{n_\alpha\}$ of $\N$.

\item[-] $\mathfrak{i}_\alpha: U + 1 \longrightarrow U$
is the relation whose restriction on $U$ (resp. on $1$) is $\Id_U$ (resp. $\alpha$)
and $\mathfrak{r}_\alpha$ is the converse relation of~$\mathfrak{i}_\alpha$.
\item[-] The zero morphism $0_{XY}$ is the empty relation, for all $X, Y$.
\end{itemize}

}
\label{relexpl}
\end{expl}

\begin{expl}[\Pfn and \PInj as degenerate polarized GoI situations]{\em
Starting with the category $\Rel$ above, we  consider the two major subcategories
{\sf Pfn}  (partial functions) and \PInj (partial
injective functions) from \cite{AHS02}, with the following choices of $\mathfrak{i}, r, \mathfrak{i}_{\alpha}, \mathfrak{r}_{\alpha}$:
for both {\sf Pfn} and \PInj, $\mathfrak{r}: 1\rightarrow 1+1$ is the left (or right) inclusion
and $\mathfrak{i}$ is its inverse.  $\mathfrak{r}_\alpha: U\rightarrow U+1$ is the left embedding. 
\begin{itemize} \itemsep=-2pt 
\item[-] {\sf Pfn}:   $\mathfrak{i}_{\alpha}: U + 1 \longrightarrow U$
is the total function whose restriction on $U$ (resp. on $1$) is $\Id_U$ (resp. $\alpha$).
\item[-] {\sf PInj}:
$\mathfrak{i}_{\alpha}: U + 1 \longrightarrow U$
is the partial injection 
determined by the identity on $U$.
\end{itemize}
We say these models are {\em degenerate} since in $\Rel$, $\mathfrak{r}[1]
:= \{  y \mid (*,y)  \in \mathfrak{r} \}  = 1 + 1$,
whereas in the other models  {\sf Pfn} and {\sf PInj}, $\mathfrak{r}[1] = 1$. This latter
property of $1$ causes the interpretation of polarized 
proofs in \MLLP \,to become degenerate (see Remark \ref{prfdegen} below.)
\label{pfnexpl}
}\end{expl}

\vspace{1ex}

We now make some remarks and observations about the axioms for  polarized GoI situations in   Definition \ref{GoIsitu} above.  We generalize Axioms {\bf 2} and {\bf 3} to $m$-ary tensors, and denote them by
Axioms {\bf 2'} and {\bf 3'} respectively.
We will apply these remarks in 
Proposition \ref{mplifting} below,
 where Axioms {\bf 6} and {\bf 9} are generalized,
 resulting in Axioms {\bf 6'} and {\bf 9'}, resp.

\begin{description}
\item[$\bullet$]  {\em Strengthening of Axiom 5 (Uniformity of Trace)}:    
\begin{prop}[Strong Uniformity of Trace:] \label{SUniP} 
{For any morphisms $$X_2 \otimes U \stt{f} Y_2 \otimes U \,
, \, X_1 \otimes 1 \stt{g} Y_1 \otimes 1 \, , \, X_1\stt{a}X_2 \, , \, Y_1\stt{b}Y_2 , $$
 and point $p: 1\rightarrow U$, we have:}
$$\xymatrix
@C=0.3in
@R=1pc
{
  X_2 \otimes U  \ar[rr]^f    &     &  Y_2 \otimes U 
& &  X_2 \ar[rr]^{
{\sf Tr}^{U}_{X_2,Y_2}(f)} &  & Y_2
\\ & & & \mbox{\em implies} &  & & &    \\
X_1 \otimes 1 \ar[uu]^{a \otimes  p}
 \ar[rr]_g   &    &  Y_1 \otimes 1 \ar[uu]_{b \otimes p}
 &  & X_1  \ar[uu]^{a} \ar[rr]_{ {\sf Tr}^{1}_{X_1,Y_1}(g) }
&    & Y_1 \ar[uu]_{b}
}$$
\end{prop}
\begin{prf}{}
Define
\begin{eqnarray*}
f'= f \Comp (a \otimes U) :  X_1 \otimes U \longrightarrow Y_2 \otimes
 U &  \mbox{and}   & 
g'= (b \otimes 1 ) \Comp g :   X_1 \otimes 1 \longrightarrow Y_2
 \otimes 1.
\end{eqnarray*}
Apply Axiom 5 to $f'$ and $g'$ with $X=X_1$ and $Y=Y_2$, then
\begin{eqnarray*}
{\sf Tr}^{U}_{X_1,Y_2}(f \, \Comp \, (a \otimes U)) & = &
{\sf Tr}^{1}_{X_1,Y_2}((b \otimes 1) \, \Comp \, g).
\end{eqnarray*}
By naturality, the L.H.S (resp. R.H.S) is equal to
${\sf Tr}^{U}_{X_2,Y_2}(f ) \, \Comp \, a $ 
(resp. $b \, \Comp \,
{\sf Tr}^{1}_{X_1,Y_1}( g)$), which proves the assertion.
\end{prf}
Observe, strong uniformity of trace implies
the original version (Axiom 5) by setting $a$ and $b$ to be the appropriate identity arrows.

\item[$\bullet$]  
{\em Generalizing Axiom 2 to the case of $m$-ary tensors.}


\begin{itemize}
\item[{Axiom \bf 2'.}] {\em For any natural numbers $m \geq 2$ 
there is a retraction  $  U   \rhd_{(k_m, j_m)}  U^m$}
$$\begin{array}{lr}
$\xymatrix{U 
\ar@<1ex>[rr]^{k_m} &   & U^m \ar@<1ex>[ll]^{j_m} }$
& 
\mbox{
such that $k_m \Comp j_m = \Id_{U^m}$.}
\end{array}$$
We define $j_m$ and $k_m$ as follows. This will be our
fixed choice  for the retraction structure for the rest
of the paper.
\begin{eqnarray}
j_m = j \, \Comp \, (j \otimes U) \, \Comp  \cdots \Comp \, 
(j \otimes U^{m-1}) &  & 
k_m = (k \otimes U^{m-1}) \,  \Comp \cdots \Comp \,  (k \otimes U) \,
 \Comp
\, k \label{jk}
\end{eqnarray}

\vspace{1ex}

{ When $m=1$, under the convention that $j_m=U=k_m$,
the retraction of 2' becomes the trivial identity.
Hence in what follows,
$(k_m, j_m)$ is used for any non-zero natural number $m$.
When $m=2$, we get the original Axiom 2. }
\end{itemize}

\item[$\bullet$] {\em  An $m$-fold tensor version
of Axiom 3 : ~~$1^m  \otimes 1^m \rhd_{(\mathfrak{i}^m,\mathfrak{r}^m)} 1^m  $ }. 

It is  straightforward to show the following:
\begin{itemize}
\item[{Axiom \bf 3'.}]
{\em $\mathfrak{i}^m$ has   right inverse $\mathfrak{r}^m$}, i.e. \ \ 
$\xymatrix{
1^m  \otimes 1^m
\ar@<1ex>[rr]^{\mathfrak{i}^m}
 &  &   1^m \ar@<1ex>[ll]^{\mathfrak{r}^m}
 }$  satisfying   $\mathfrak{i}^m \, \Comp \, \mathfrak{r}^m = \Id_{1^m}$, \\
where $\mathfrak{i}^m$ and $\mathfrak{r}^m$ are $m$-ary tensors of $\mathfrak{i}$ and $r$, respectively.
\end{itemize}
\end{description}

\smallskip

Finally, the reader may wonder why, in our definition of polarized GoI situation (Definition \ref{GoIsitu}), we introduced opposite directions
for the retractions on $U$ and on $1$.  This will be explained in   Appendix \ref{retract.remks} below.


\subsection{Multipoints in polarized GoI situations}
\label{multipts}
%
In this section we introduce the key notion of {\em multipoints}\,
for interpreting the weak exponentials $\uparrow$ and $\downarrow$ (polarity shifting) 
of \MLLP \, in polarized GoI situations.  We then generalize Axioms 6 and 9 of a polarized
GoI situation to the level of multipoints.
In the following Section \ref{goipfs}, we shall make use of multipoints to
translate  polarized formulas, and then extend this to a
 polarized GoI interpretation of \MLLP\, proofs.    This will be used
later (in Sections \ref{polarex}, \ref{mainthmsec1} below) to find
new invariants for cut-elimination, and to characterize focussing,
hence positivity and negativity, in polarized logics.

\smallskip

\begin{defn}[points and multipoints]\label{mpclass}~\\{\em
{\em Multipoints} in  a polarized GoI situation form a certain class (denoted $\mathbb{MP}$)  of
morphisms $1^m \longrightarrow U$ for natural
numbers $m$. {\em Points} are the subclass (denoted $\mathbb{P}$)  of multipoints in
 which $m=1$. 

First, $\mathbb{P}$ is constructed by the following BNF construction:
$$\mathbb{P} := \quad \alpha  \quad \arrowvert \quad
 j  \Comp (\mathbb{P} \otimes 0_{I,U} )
\quad \arrowvert \quad
 j  \Comp (0_{I,U} \otimes \mathbb{P} )
$$
\noindent That is,
\begin{itemize}
\item[1.] The distinguished point $\alpha: 1 \longrightarrow U$ is a point.
\item[2.]
If $\beta: 1 \longrightarrow U$ is a point, so are \\
$
\xymatrix{
1 \cong  1 \otimes I 
\ar[rr]^(0.5){\beta \otimes 0_{I,U}} &  & U \otimes U \ar[r]^j & U}
$
   and 
$
\xymatrix{
1 \cong  I \otimes 1 
\ar[rr]^(0.5){0_{I,U} \otimes \beta} & & U \otimes U \ar[r]^j & U}
$.
\end{itemize}
Second, $\mathbb{MP}$ is constructed from $\mathbb{P}$
and $j_m$ ranging over natural numbers $m \geq 2$, as follows:
$$\mathbb{MP} := \quad \mathbb{P} \quad \arrowvert \quad
 j_m \Comp \, \tau \, \Comp  \,
(\overbrace{\mathbb{P} \otimes \cdots \otimes \mathbb{P}}^m 
 )
\hspace{3ex}
\mbox{where $\tau$ ranges over the permutations of $U^m$. }
$$
That is, while all points are multipoints,
the second construction stipulates in addition that
\begin{itemize}
\item[3.] If $p_i: 1 \longrightarrow U$ 
are points for $i=1, \ldots, m$, and $\tau$ is a permutation of $U^m$, then
the following is a multipoint 
\begin{eqnarray}
 \xymatrixcolsep{3pc}
\xymatrix{
j \Comp \, \tau \, \Comp \bigotimes_{i=1}^m p_i : 
\hspace{2ex}
1^{m} \ar[rr]^(.7){\bigotimes_{i=1}^m p_i} &  & U^m \ar[r]^\tau & U^m  \ar[r]^{j_m} & U 
}.
\end{eqnarray}
\end{itemize}
}
\end{defn}

 \begin{rem}[Various contractions  $j_m \Comp \tau$ arising from permutations $\tau$.]  
{\em We make no assumptions on commutativity nor on associativity 
axioms for the monoidal $j$ and the comonoidal $k$.
Instead, we adopt a  minimal categorical setting for a GoI situation,
as suggested by the referee.
There are various ways of contracting $U^m$ to $U$, depending
on the choice of $U$ (to apply $j$ to) in each of the $m-1$ steps
of contraction.  Our specific choice of $j_m$ in equations (\ref{jk}) 
determines {\em one} such choice. Precomposing  each permutation
$\tau$ of $U^m$ with $j_m$ determines a different
choice of contraction.
Correspondingly,
the left inverse of $j_m \Comp \tau$ is given by the comonoidal
$\tau^{-1} \Comp k_m$, and the pair
gives a different retraction $U \rhd_{(j_m \Comp \tau, \, \tau^{-} \Comp  k_m)} U^m$.
Later in Section \ref{goipfs}, when we interpret every polarized
	  formula as a multipoint,  the permutation $\tau$
will be explicitly specified by the syntactic tree of the formula.}
\end{rem}

\smallskip


\begin{expl} \label{mprelexpl}
{\em 
In the $\Rel$ model
of Example \ref{relexpl} where the distinguished $\alpha$ is
taken to be a singleton subset $\{n_\alpha\}$ of $U = \N$,
a multipoint $1^m \longrightarrow U$ (generated by this $\alpha$) determines
a finite indexed family of cardinality $m$ of $U$. 
}
\end{expl}

\vspace{2ex}

Later we shall see how
polarized formulas can be interpreted in
a multipointed setting (Definition \ref{multipoint}
in Section \ref{goipfs} below). 
Together with a
two-layered GoI-interpretation of $\MLLP$ proofs
(Definition \ref{interprf}), 
this will turn out to be essential in our characterization of focusing
(see Theorem \ref{invari}).

\vspace{1ex}

We now prove generalizations of Axioms {\bf 6} and {\bf 9} of a polarized GoI situation 
to the level of multipoints,   
using the  $m$-fold setting of {\bf 3'}.   We call these {\bf Axioms} {\bf 6'} and {\bf 9'}
respectively.

\vspace{1ex}

\begin{prop}[Axiom 6': Lifting Property 
$U \otimes 1^m \rhd_{(\mathfrak{i}_{\bm{p}}, \mathfrak{r}_{\bm{p}})} U$
 along a multipoint $\bm{p}$] \label{mplifting} \
 
 \noindent
For any multipoint 
$\xymatrix{
\bm{p} :  1^m \ar[r] &  U
}$,
there exists a pair $(\mathfrak{i}_{
\bm{p}}, \mathfrak{r}_{\bm{p}})$ giving a
retraction
which lifts the retraction
$1^m\otimes 1^m \rhd_{(\mathfrak{i}^m,\mathfrak{r}^m)} 1^m$
along $\bm{p}$.
This means the following
diagram commutes (in all possible ways), with $\mathfrak{i}_{\bm{p}} \Comp \mathfrak{r}_{\bm{p}} = \Id_{U}$:
$$
\xymatrix
      @R=1.4pc{
U \otimes  1^m \ar@<1ex>[rr]^{\mathfrak{i}_{\bm{p}}} &  &  \ar@<1ex>[ll]^{\mathfrak{r}_{\bm{p}}} U \\ \\
1^m \otimes 1^m \ar[uu]^{\bm{p} \otimes 1^m} \ar@<1ex>[rr]^{\mathfrak{i}^m}   &            &  
  \ar[uu]_{\bm{p}} \ar@<1ex>[ll]^{\mathfrak{r}^m}  1^m  \\
}$$
\end{prop}
\begin{prf}{}
See Appendix \ref{REP}
\end{prf}
The next Proposition discusses the 
lifted retraction pair $(\mathfrak{i}_{\bm{p}}, \mathfrak{r}_{\bm{p}})$ of the above Axiom {\bf 6'}.

\vspace{1ex}

\begin{prop}
[Axiom 9':  On the lifted retraction pair $(\mathfrak{i}_{\bm{p}}, \mathfrak{r}_{\bm{p}})$] \


For any multipoint 
$\xymatrix{
\bm{p} :  1^m \ar[r] &  U
}$, any morphism 
$f: V \otimes  X \longrightarrow W \otimes X$
with $X \in \{U, 1^m \}$
and $0: 1^m \longrightarrow 1^m$,
\begin{eqnarray*}
(\operatorname{Id} \otimes \mathfrak{i}_{\bm{p}} ) \Comp (f \otimes 0) \Comp
(\operatorname{Id} \otimes \mathfrak{r}_{\bm{p}})  = f  &\mbox{and} &   
(\operatorname{Id} \otimes \mathfrak{i}^m ) \Comp ( f \otimes 0) \Comp
(\operatorname{Id} \otimes \mathfrak{r}^m ) = f 
\end{eqnarray*}
\end{prop}
This is illustrated in Figure \ref{figax9'} for the respective equations
when $X=U $ and $X=1^m$.
\begin{figure}[!h]
\begin{minipage}{0.5\hsize} 
$\begin{picture}(180,80)(-60,-50)
\put(30,-15){\fbox{\rule[.5in]{.4in}{0in }}}
\put(12,-40){$1^m $}   \put(70,-40){$1^m $}

\put(30,-40){\vector(1,0){35}} \put(43,-50){$0$}

\put(12,18){$V$}    \put(70,18){$W$}
\put(41,0){$f$}
\put(12,-5){$U$} \put(70,-5){$U$}
\put(130, 0){$= \ \ f$}

\put(-15,-25){\vector(2,1){25}}  
\put(-15,-25){\vector(2,-1){25}}

\put(43,-30){$\otimes$}
\put(-30,-30){$U$}

\put(-5,-28){$\mathfrak{r}_{\bm{p}}$}
\put(85,-28){$\mathfrak{i}_{\bm{p}}$}
\put(12,-10){\vector(1,0){18}}
\put(66,12){\vector(1,0){25}}
\put(66,-10){\vector(1,0){10}}
\put(5,12){\vector(1,0){25}}


\put(78,-10){\vector(2,-1){25}}
\put(78,-40){\vector(2,1){25}}

\put(110,-30){$U$}
\end{picture}$
\end{minipage}
\begin{minipage}{0.1\hsize}
and 
\end{minipage}\hspace{-6ex}
\begin{minipage}{0.4\hsize} 
$\begin{picture}(180,80)(-50,-50)
\put(30,-15){\fbox{\rule[.5in]{.4in}{0in }}}
\put(12,-40){$1^m$}   \put(70,-40){$1^m $}

\put(30,-40){\vector(1,0){35}} \put(43,-50){$0$}

\put(12,18){$V$}    \put(70,18){$W$}
\put(41,0){$f$}
\put(12,-5){$1^m$} \put(70,-5){$1^m$}
\put(130, 0){$= \ \ f$}

\put(-15,-25){\vector(2,1){25}}  
\put(-15,-25){\vector(2,-1){25}}

\put(43,-30){$\otimes$}
\put(-30,-30){$1^m$}

\put(-5,-28){$\mathfrak{r}^m$}
\put(85,-28){$\mathfrak{i}^m$}
\put(12,-10){\vector(1,0){18}}
\put(66,12){\vector(1,0){25}}
\put(66,-10){\vector(1,0){10}}
\put(5,12){\vector(1,0){25}}

\put(78,-10){\vector(2,-1){25}}
\put(78,-40){\vector(2,1){25}}

\put(110,-30){$1^m$}
\end{picture}$
\end{minipage}
\caption{Axiom {\bf 9'}}
\label{figax9'}
\end{figure}
 \begin{prf}{}
Straightforward induction on the construction of $\bm{p}$.
\end{prf}



\vspace{2ex}
 
The above definition of multipoints
  is compatible with Axiom
{\bf 5} (Uniformity of Trace) in Definition \ref{GoIsitu},
in the sense that  uniformity of trace generalizes to multipoints
(see Axiom {\bf 5'} in Proposition \ref{uniftracemulti} below.) 

\vspace{1ex}

\begin{lem}[Invariance of traces under conjugate actions
of the retractions]  \label{leminvunderconj} 
\

\noindent
For any non-zero natural number $m$
and a multipoint $\bm{p}: 1^m \longrightarrow U$,
the retractions $(k_m, j_m)$ and
$(\mathfrak{i}_{\bm{p}}, \mathfrak{r}_{\bm{p}})$
respectively act by conjugation 
 on  morphisms $f
: X \otimes U^m \longrightarrow Y \otimes U^m$
and $g: X \otimes U \longrightarrow Y \otimes U$ as follows:
\begin{eqnarray*}
f^{(k_m, j_m)} &   := &  (X \otimes j_m) \Comp f \Comp (X \otimes k_m) 
\quad : X \otimes U \longrightarrow Y \otimes U \\
g^{(\mathfrak{i}_{\bm{p}},\mathfrak{r}_{\bm{p}})}  &  :=  & (X \otimes \mathfrak{r}_{\bm{p}}) \Comp g \Comp 
(X \otimes \mathfrak{i}_{\bm{p}}) \quad
: X \otimes U \otimes 1^m \longrightarrow Y \otimes U \otimes
1^m
\end{eqnarray*}
Then the following invariant equations hold:
\begin{eqnarray}
\TR{f}{U^m}{X}{Y} & = & 
\TR{f^{(k_m, j_m)}}{U}{X}{Y} \label{2'} \\
\TR{g}{U}{X}{Y} & = & 
\TR{g^{(\mathfrak{i}_{\bm{p}}, \mathfrak{r}_{\bm{p}})}}{U \otimes 1^m}{X}{Y}
 \label{5'} 
\end{eqnarray}
\end{lem}
The two equations guarantee the invariance of
taking traces along $U$ (instead of $U^m$)
and along $U \otimes 1^m$ (instead of $U \otimes 1$).
\begin{prf}{}
See Appendix \ref{REP}
\end{prf}
 
 \vspace{1ex}
 
 \noindent

 \noindent
The following  Axiom {\bf 5'} generalizes Axiom {\bf 5}:
\begin{prop} [Axiom {\bf 5'} (Uniformity of Trace on multipoints)]
\label{uniftracemulti} ~\\
Every multipoint $\bm{p}: 1^m \longrightarrow U$
satisfies the following condition: for any morphisms $f$ and $g$,
$$\xymatrix
@C=0.3in
{
  X \otimes U \ar[r]^{X \otimes k_m}
 & X \otimes U^m  
\ar[rr]^{f}  
 &  & 
Y \otimes U^m  \ar[r]^{Y \otimes j_m}
    &  Y \otimes U \\
&                                           &     &                       &   
\\
 & X \otimes 1^m \ar[uul]^{X \otimes  \bm{p}}
 \ar[rr]_g  &   &    Y \otimes 1^m \ar[uur]_{Y \otimes \bm{p}}}
 $$
commuting implies
$$\begin{array}{ccc}
\TR{g}{1^m}{X}{Y} &  = \TR{f^{(k_m, j_m)}}{U}{X}{Y} 
&  = \TR{f}{U^m}{X}{Y}        \\
           \end{array}.$$
           \end{prop}
{\bf Note: } 

\noindent
(i) The composition of
the top horizontal arrows above is $f^{(k_m, j_m)}$ of Lemma \ref{leminvunderconj}.
 
 \noindent
(ii)  Axiom {\bf 5} is a special case, by setting $m=1$ and using   
  our convention that $j_1=U=k_1$.

\begin{prf}{}

We first note that the second equation of the assertion
is by (\ref{2'}) of Lemma \ref{leminvunderconj}.

We prove the assertion by
induction on $m\geq 1$; i.e.,   by induction on the construction of a
 multipoint $\bm{p}: 1^m \longrightarrow U$

\smallskip

\noindent (Base Case): $m = 1$. The assertion is the original  Axiom $5$ of Definition \ref{GoIsitu}.

\smallskip

\noindent (Induction Case for $m+1$) \\
The given commutative diagram with $m+1$ factors as follows from
the construction of $\bm{p}$ by Definition \ref{mpclass}
for some permutation $\tau$ on $U^m$. 
$$
\xymatrix{
 X \otimes U  \ar[r]^{X \otimes k}
& X \otimes U \otimes U \ar@<1.5ex>[r]^{X \otimes U \otimes 
(\tau^{-} \Comp \, k_m)}
& X \otimes U \otimes U^{m}  \ar[rr]^{f}  & & 
Y \otimes U \otimes U^{m} \ar@<1.5ex>[r]^{Y \otimes U \otimes (j_m \Comp
\, \tau)} & 
Y \otimes U \otimes U \ar[r]^{Y \otimes j} & 
Y \otimes U \\ \\
&
 &
X \otimes 1 \otimes 1^{m} 
\ar[uull]^(.7){X \otimes \bm{p}}
\ar[uul]_(.6){X \otimes \bm{p'} \otimes p_{m+1}   }
\ar[uu]_{ X \otimes \otimes_{i=1}^{m+1} p_i}
\ar[rr]_g
    &  & 
 Y \otimes 1 \otimes 1^m
\ar[uu]^{ Y \otimes \otimes_{i=1}^{m+1} p_i}
\ar[uur]^(.6){Y \otimes \bm{p'} \otimes p_{m+1}  }
\ar[uurr]_(.7){Y \otimes \bm{p}} 
}$$
\begin{eqnarray*}
\mbox{where} \quad \bm{p}= j_{m+1} \Comp \, \tau
\Comp \otimes_{i=1}^{m+1} p_i & \mbox{and} &
\bm{p'}= j_{m} \Comp \, \tau \Comp \otimes_{i=1}^{m} p_i.
\end{eqnarray*}

Suppose the outer trapezium commutes. 
We note that the inner trapezium commutes
by both precomposing $X \otimes U \stackrel{X \otimes j}{\longleftarrow} X
\otimes U \otimes U$ and postcomposing
$Y \otimes U \otimes U \stackrel{Y \otimes k}{\longleftarrow} Y
\otimes U$ respectively on the top left-most and right-most horizontal arrows, 
because $k \Comp j=\Id_{U \otimes U}$ and since the left and right triangles commute.

Then applying the I.H. to this inner trapezium, we have:
$$
\xymatrix
@C=0.3in
@R=1.2pc
{
 X \otimes U  
\ar@<1.5ex>[rrr]^{
\TR{f^{(j_m \Comp
 \, \tau, \, \,\tau^{-} \Comp \, k_m)}}{U}{X \otimes U}{Y \otimes U}
}
&  &  & 
Y \otimes U \\  & &  \\
X \otimes 1 \ar[uu]^{X \otimes p_{m+1}}
 \ar[rrr]_{ {\sf Tr}^{1^m}_{X \otimes 1, Y \otimes 1}(g) }
   &  &  & 
Y \otimes 1 \ar[uu]_{Y \otimes p_{m+1}}
}
$$
The upper horizontal arrow is equal to;
\begin{eqnarray*}
 \TR{f^{(j_m \Comp
 \, \tau, \, \,\tau^{-} \Comp \, k_m)}}{U}{X \otimes U}{Y \otimes U}
& =
\TR{f^{(j_m , \, k_m)}}{U}{X \otimes U}{Y \otimes U}
& =
\TR{f}{U^m}{X \otimes U}{Y \otimes U}
\end{eqnarray*}
The first equation is by dinaturality on $\tau$,  cancelling with
its inverse $\tau^{-}$, and the second equation is by (\ref{2'}).

Then applying the original Axiom $5$ (Uniformity of Trace) to the above
 square, we  obtain
\begin{eqnarray*}
\TR{
\TR{f}{U^m}{X \otimes U}{Y \otimes U}
}{U}{X}{Y}
  & = &
\TR{
\TR{g}{1^m}{X \otimes 1}{Y \otimes 1}
}{1}{X}{Y}
\end{eqnarray*}
By Vanishing applied to both sides of the equation, 
we have 
$\TR{f}{U^{m+1}}{X}{Y} = 
\TR{g}{1^{m+1}}{X}{Y}$
\end{prf}

As in the case for points in Proposition \ref{SUniP},
a stronger version of  {\bf 5'} can be derived for multipoints:

\smallskip

\begin{cor}[Strong Uniformity of Trace on
multipoints] \label{stronguniform}~\\
Every multipoint $\bm{p}: 1^m \longrightarrow U$
satisfies the following condition:
for any morphisms $f$, $g$, $a$, $b$,

\begin{minipage}{0.55\hsize}
$\xymatrix
@C=0.25in
{
  X_2 \otimes U \ar[r]^{X \otimes k_m}
 & X_2 \otimes U^m  
\ar[rr]^{f}  
 &  & 
Y_2 \otimes U^m  \ar[r]^{Y \otimes j_m}
    &  Y_2 \otimes U \\
&                                           &     &                       &   
\\
 & X_1 \otimes 1^m \ar[uul]^{a \otimes  \bm{p}}
 \ar[rr]_g  &   &    Y_1 \otimes 1^m \ar[uur]_{b \otimes 
\bm{p}}}
$
\end{minipage}
\begin{minipage}{0.15\hsize}
\mbox{implies}
\end{minipage}
\begin{minipage}{0.4\hsize}
$\xymatrixcolsep{3pc}\xymatrix{
  X_2 \ar[rr]^{
\TR{f}{U^m}{X_2}{Y_2}=}_{
\TR{f^{(k_m, j_m)}}{U}{X_2}{Y_2}
} &  & Y_2
\\   & & &    \\
 X_1  \ar[uu]^{a} \ar[rr]_{ \TR{g}{1^m}{X_1}{Y_1} }
&    & Y_1 \ar[uu]_{b}
}
$
\end{minipage}
\end{cor}
\begin{prf}{}
Same as Proposition \ref{uniftracemulti}, using naturality.
\end{prf}




\subsection{The GoI interpretation of $\MLLP$ proofs}
\label{goipfs}

We now define one of the central notions of this paper:  the GoI interpretation of $\MLLP$
 proofs in polarized GoI situations.  We shall begin with a detailed
 discussion of how to interpret multipoints of polarized formulas.  We then present  a categorical approach
to GoI in the polarized case, influenced by the   categorical approach to ordinary GoI of Haghverdi and
 Scott\cite{HS06,HS11} in (ordinary) GoI situations,  as summarized in Appendix \ref{appdx1} below. 

\smallskip

\begin{defn}[Multipoints associated with Formulas] \label{multipoint}
{\em Given a polarized GoI situation and a polarized $\MLLP$ formula $A$,
we will inductively construct  below a morphism ${\sf mp}(A)$ together with its domain $\mathbbm{1}_A$ 
and codomain $U_A$, where $U_A := U$.
\begin{eqnarray*}
{\sf mp}(A) :   \mathbbm{1}_A    & \longrightarrow  U_A   .
\end{eqnarray*}
 More generally, for a sequence ${\cal M}=A_1, \ldots , A_n$ of polarized formulas,
we will construct a morphism $${\sf mp}({\cal M}) %
: \mathbbm{1}_{\cal M} \longrightarrow U_{\cal M}
:= U_{A_1} \otimes \cdots \otimes U_{A_n}.$$
from the constructed domain $\mathbbm{1}_{\cal M}$ to the codomain $ U_{\cal M}$.
The arrows ${\sf mp}(A)$ (resp. ${\sf mp}({\cal M})$) defined below are called {\em the multipoint
 associated with formula $A$ (resp. with sequence} ${\cal M}$).
}\end{defn}

\vspace{1ex}

\noindent
{\bf Construction of multipoints}.  

First, to each positive (resp. negative) formula $P$ (resp. $N$) of \MLLP,
we associate  an object $\mathbbm{1}_P$ (resp. $\mathbbm{1}_N$),
which is  a tensor product of $1$s, defined inductively as follows:
$$
\begin{array}{lllll}
\mathbbm{1}_X :=1 & &
\mathbbm{1}_{P \otimes Q} :=
\mathbbm{1}_P \otimes \mathbbm{1}_Q
 & &  \mathbbm{1}_{\down{N}}:= 1 \\
\mathbbm{1}_{X^\perp} :=1 & &  
\mathbbm{1}_{N \wp M} := \mathbbm{1}_N \otimes \mathbbm{1}_M
 & & \mathbbm{1}_{\up{P}}:= 1 \\
\end{array}$$
For a sequence ${\cal M}=A_1, \ldots , A_n$ of polarized formulas,
$\mathbbm{1}_{\cal M}:=\mathbbm{1}_{A_1} \otimes \cdots \otimes
\mathbbm{1}_{A_n}$

\smallskip

Second, 
with each positive (resp. negative) formula $P$ (resp. $N$),
we associate three objects $\bm{U}_P$, $P^\flat$ and $P^\sharp$
(resp. $\bm{U}_N$, $N^\flat$ and $N^\sharp$)
so that 
\begin{eqnarray}
\bm{U}_A \cong A^\sharp \otimes A^\flat \label{FUD}
\end{eqnarray}
 inductively as follows:
$$
\begin{array}{lllll}
\makebox{\boldmath $U$}_X
:=U & &  \bm{U}_{P \otimes Q} := 
\bm{U}_P \otimes \bm{U}_Q & &  \bm{U}_{\down{N}}:= U \otimes
 \bm{U}_N \\
X^\flat :=I & &  (P \otimes Q)^\flat := P^\flat \otimes Q^\flat & & (\down{N})^\flat:= \bm{U}_N \\
 X^\sharp :=U & & (P \otimes Q)^\sharp := P^\sharp \otimes Q^\sharp & & (\down{N})^\sharp:= U  \\ \\
\bm{U}_{X^\perp} :=U & & 
\bm{U}_{N \wp M} := \bm{U}_N \otimes \bm{U}_M & &  
\bm{U}_{\up{P}}:= \bm{U}_P \otimes U  \\
(X^\perp)^\flat :=I & &  (N \wp M)^\flat := N^\flat \otimes M^\flat & & (\up{P})^\flat:= \bm{U}_P \\
 (X^\perp)^\sharp :=U & & (N \wp M)^\sharp := N^\sharp \otimes M^\sharp & & (\up{P})^\sharp:= U
\end{array}$$
All these objects are isomorphic to tensor products of $U$s. 
For a sequence ${\cal M}=A_1, \ldots , A_n$ of polarized formulas,
we define the object $\bm{U}_{\cal M}:=
\bm{U}_{A_1} \otimes \cdots \otimes \bm{U}_{A_n}$
and for $\star \in \{D,U \}$,
we define the  object
${\cal M}^\star :=A_1^\star \otimes \cdots \otimes A_n^\star$. \\

Note that the following hold for any polarized formula $A$.
\begin{itemize} \itemsep=-2pt 
\item[-]
If $A$ is a literal, $U_A =\bm{U}_A$. Otherwise,
\begin{eqnarray} \label{*retract}
U_A \quad \rhd_{(\tau \, \Comp \, k_{n-1}, \, \, j_{n-1} \, \Comp \, \tau)} \quad \bm{U}_A 
\end{eqnarray}
where $n$
 is the number logical connectives in $A$ so that
$\bm{U}_A = U^{n}$, as in
Axiom {\bf 2'}, and $\tau$ is a permutation of $U^{n}$.

\item[-]
There exists a natural number $m$ such that
\begin{eqnarray} \label{AU1A}
A^\sharp \cong U^m & \mbox{and} & \mathbbm{1}_A \cong 1^m
\end{eqnarray}
\end{itemize}

\noindent
Finally,
using (\ref{FUD}), (\ref{*retract}) and (\ref{AU1A}),
we define ${\sf mp}(A)$ by the following composition (we 
give a simpler, equivalent definition in Remark \ref{multidefequiv} below):
\begin{eqnarray}
{\sf mp}(A): \xymatrixcolsep{4pc} \xymatrix{
\mathbbm{1}_{A} \cong \mathbbm{1}_{A} \otimes I
\ar[r]^(.6){\alpha^m \otimes 0_{I, A^\flat}} 
& 
A^\sharp \otimes A^\flat \ar[r]^{\tau} 
& U^{n}  \ar[r]^(.6){j_{n-1}} & U_A,
}\label{factormp} \end{eqnarray}
where the permutation $\tau$ on the tensor folding
$\bm{U}_A$ is determined by the syntax tree of the polarized formula
$A$.
Finally, in the case of a sequence $\cM = A_1,\ldots,A_n$,
\begin{eqnarray}{\sf mp}({\cal M})
:={\sf mp}(A_1) \otimes \cdots \otimes {\sf mp}(A_n) 
: \mathbbm{1}_{\cal M} \longrightarrow U_{\cal M}.
\end{eqnarray}

\smallskip

\begin{rem}
\label{multidefequiv}
{\em   Multipoints
$ {\sf mp}(A)$ can be {\em inductively} defined as follows, where the
first, second and third constructions correspond respectively to 1, 2
and 3 of Definition \ref{mpclass}:}

\begin{enumerate} 
\item[-]
${\sf mp}(X):=\alpha$ and ${\sf mp}(X^\perp):=\alpha$ 

\item[-]  
${\sf mp}(\down{N}) := j \Comp (\alpha \otimes 0_{I,U_N})$
and 
${\sf mp}(\up{P}) := j \Comp (0_{I,U_P} \otimes \alpha )$.



\item[-]
${\sf mp}(P \otimes Q) := j \Comp (
{\sf mp}(P)  \otimes {\sf mp}(Q))$ and
${\sf mp}(N \wp M) := j \Comp (
{\sf mp}(N)  \otimes {\sf mp}(M))$.


\end{enumerate}

\noindent{\em
We observe that ${\sf mp}(A)$ so inductively defined uniquely factors
as (\ref{factormp}), in which the permutation $\tau$
is determined by the syntax tree of the $\MLLP$ formula $A$.}
\smallskip

{\em
In what follows, we make the convention that
$U_\downarrow$ (resp. $U_\uparrow$)
denotes the codomain $U$ of 
$\alpha$ in the above construction of
${\sf mp}(\down{N})$
(resp. of ${\sf mp}(\up{P})$).
That is, $U_\downarrow$ (resp. $U_\uparrow$)
denotes $U$ of $(\down{N})^\sharp$
(resp. of $(\up{P})^\sharp$).
}

\end{rem}

\smallskip

\vspace{1ex}

\begin{exam}\label{Relex}{ \em 
In the $\Rel$ polarized GoI situation,
\begin{itemize} \itemsep=-2pt 
\item[-] $\mathbbm{1}_{\up{X}}=1 = \{ n_\alpha \}$
and ${\sf mp}(\up{X}) : 1 \cong
I + 1
\stackrel{0 \otimes \alpha}{\longrightarrow}  
 U_X  +  U_{\uparrow} 
\stackrel{j}{\longrightarrow} U_{\up{X}}$
is a singleton subset consisting of
the element $j((2,n_\alpha))=2n_\alpha +1
\in \N$.
\item[-]
$\mathbbm{1}_{Y^\perp \wp X^\perp}=1^2= \{  n_\alpha \} + \{
	n_\alpha \}$
and 
${\sf mp}(Y^\perp \wp X^\perp) : 1^2 \stackrel{\alpha^2}{\longrightarrow}
	U_{Y^\perp} + U_{X^\perp}
\stackrel{j}{\longrightarrow} U_{Y^\perp \wp X^\perp}$
is a subset of cardinality $2$
consisting of the elements $j((1,n_\alpha))=2n_\alpha$ and $j((2,n_\alpha))=2n_\alpha +1$ in
$\N$.
\end{itemize}
}\end{exam}


\begin{defn}[retractions $(\mathfrak{i}_A, \mathfrak{r}_A)$ 
 ]\label{rA!A}{\em
For a polarized formula $A$, we define two morphisms $\mathfrak{r}_A$ and $\mathfrak{i}_A$
to give a retraction
$$\begin{array}{lr}
 A^\sharp \otimes  \mathbbm{1}_A  \rhd_{(\mathfrak{i}_A,\mathfrak{r}_A)} A^\sharp 
& \mbox{so that $\mathfrak{i}_A \Comp \mathfrak{r}_A = A^\sharp$.}
\end{array}$$
These are defined inductively on $A$ as follows: \\[1ex]
\smallskip
$\mathfrak{r}_X$ and $\mathfrak{r}_{X^\perp}$ are $\mathfrak{r}_\alpha$,
$\mathfrak{r}_{P \otimes Q} := \mathfrak{r}_{P} \otimes \mathfrak{r}_{Q}$,
$\mathfrak{r}_{N \wp M} := \mathfrak{r}_{N} \otimes \mathfrak{r}_{M}$,
$\mathfrak{r}_{\down{N}}$ and $\mathfrak{r}_{\up{P}}$ are $\mathfrak{r}_\alpha$, \\
$\mathfrak{i}_X$ and $\mathfrak{i}_{X^\perp}$ are $\mathfrak{i}_\alpha$, $\mathfrak{i}_{P \otimes Q} := \mathfrak{i}_{P} \otimes \mathfrak{i}_{Q}$,
$\mathfrak{i}_{N \wp M} := \mathfrak{i}_{N} \otimes \mathfrak{i}_{M}$,
$\mathfrak{i}_{\down{N}}$ and $\mathfrak{r}_{\up{N}}$ are $\mathfrak{i}_\alpha$.  

\vspace{1ex}

\noindent
For a sequence ${\cal M}=A_1, \ldots , A_n$ of polarized formulas,
define $\mathfrak{r}_{\cal M}$ (resp. $\mathfrak{i}_{\cal M}$) to be
$\mathfrak{r}_{A_1} \otimes \cdots \otimes \mathfrak{r}_{A_n}$
(resp. $\mathfrak{i}_{A_1} \otimes \cdots \otimes \mathfrak{i}_{A_n}$).
}\end{defn}

\bigskip

Then the lifting property (Axiom 6' of Proposition \ref{mplifting})
has the following variant:
\begin{prop}[Lifting property $A^\sharp \otimes \mathbbm{1}_A 
\rhd_{(\mathfrak{i}_A, \mathfrak{r}_A) } A^\sharp$ along $\alpha^m$] \label{varlifting}\

\noindent
Suppose $m$ is a natural number satisfying  
(\ref{AU1A}).
Then $(\mathfrak{i}_A, \mathfrak{r}_A)$ gives a retraction which lifts the retraction
$\mathbbm{1} \otimes \mathbbm{1} \rhd_{(\mathfrak{i}^m, \mathfrak{r}^m)} \mathbbm{1}$
along $\alpha^m$.
This means the following diagram commutes with 
$\mathfrak{i}_A \Comp \mathfrak{r}_A = \Id_{A^\sharp}$.
\end{prop}
$$
\xymatrix
@R=1.4pc{
A^\sharp \otimes  \mathbbm{1}_A \ar@<1ex>[rr]^{\mathfrak{i}_A} &  &
 \ar@<1ex>[ll]^{\mathfrak{r}_{A}} 
A^\sharp \\ \\
\mathbbm{1}_A \otimes \mathbbm{1}_A \ar[uu]^{\alpha^m \otimes 1^m} \ar@<1ex>[rr]^{\mathfrak{i}^m}   &            &  
  \ar[uu]_{\alpha^m} \ar@<1ex>[ll]^{\mathfrak{r}^m}  \mathbbm{1}_A  \\
}$$
\begin{prf}{}
Straightforward by noting that
composing the right ( resp. left)
vertical arrow 
with the second and third
maps of (\ref{factormp}) (resp. $\otimes 1^m $ of these maps)
gives rise to Axiom 6'.
\end{prf}


In what follows, we introduce the GoI interpretation for \MLLP.   The original GoI situations in Abramsky, et. al. \cite{AHS02} 
  (summarized in
Appendix \ref{appdx1} below) form a very basic framework for interpreting 
GoI.  For example, their exponential structure, which   is sufficient for defining linear combinatory 
algebras on $End_{\cC}(U)$, for a reflexive object $U$, 
  does not include the more elaborate categorical 
structure of the exponentials (e.g.  cocommutative coalgebras,
comonoids, etc.) in genuine models of linear logic \cite{PAM}.

Recall that  the GoI interpretation
of an \MLL \, proof $\pi$ of the sequent $\vdash \! [\Delta] ,\Gamma$  in an (ordinary) GoI situation $\cC$ yields an endomorphism 
$\Mean{\pi}\in End_{\cC}(U^{2m+n})$, for the reflexive object $U$.  We now introduce the analog for the polarized case
of  \MLLP.

Let us sketch the framework, before going into details.  Consider   a {\em polarized} GoI situation $\cC$,
with a reflexive object $U$ and an object $1$. 
The {\em polarized 
GoI Interpretation}   in Definition \ref{interprf} below will yield  {\em a pair } of endomorphisms, 
 $(\Mean{\pi}, f_\pi)\in End_{\cC}(U^{2m+n})\times  End_{\cC}(1^{2m' +n'})$.  
 
 We think of the two endomorphisms as layers:  an ``upper" and a ``lower" GoI interpretation.
 
 \begin{itemize}
 
\item[-] The {\em upper} interpretation, $\Mean{\pi}\in End_{\cC}(U^{2m+n})$,
is on the level of the reflexive object $U$. 
It is analogous to the non-polarized GoI interpretation, using the
	 polarized retraction structure
$U \rhd U \otimes U$ 
{for  coding ``untyped" GoI, by folding tensors of $U$'s into a single $U$}.
Here $U^{2m+n}$ comes from
	 $U_{\Delta,\Gamma}$ of Definition \ref{multipoint};
 $n$ (resp. $2m$) is the number of formulas in $\Gamma$ (resp. $\Delta$).
At this level, the polarity will be handled by the retraction
$U \otimes 1 \rhd U$.
{Hence, both the retraction structures
3 and 6 of Definition \ref{GoIsitu}
are used.}

  \item[-] The {\em lower} interpretation, $f_\pi \in End_{\cC}(1^{2m'+n'})$, is a similar GoI
formula to $\Mean{\pi}$, but without assuming any reflexivity on $1$.
It is  defined on the level of multipoints.
Here $1^{2m'+n'}$ comes from
	 $\mathbbm{1}_{\Delta,\Gamma}$ of Definition \ref{multipoint};
$n'$ (resp. $2m'$) is the sum,
over formulas in $\Gamma$ (resp. $\Delta$),
of 
the number of $\MLLP$ logical connectives (including literals) not bounded by
polarity shifting operations.
At this level, the polarity will be handled by the retraction
$1 \otimes 1 \rhd 1$.
{ Hence,
only  retraction 3
of Definition \ref{GoIsitu} is used.}

\end{itemize}

\begin{defn}[The two-layered Polarized GoI interpretation of $\MLLP$ proofs]  \label{interprf}

\noindent
{\em
An {\sf MLLP} proof $\pi$ of $\vdash [\Delta], \Gamma$
is interpreted by two endomorphisms
%

\begin{eqnarray*}
\Mean{\pi}  \in    {\cal C} ( U_{\Delta, \Gamma}, U_{\Delta, \Gamma} )
&  \mbox{and} &  
f_{\pi}         \in   {\cal C} (\mathbbm{1}_{\Delta, \Gamma}, \mathbbm{1}_{\Delta, \Gamma})
\end{eqnarray*}
%
%
}\end{defn}
We see the polarized view as a {\em two-layered
interpretation:}  an {\em upper} layer $\Mean{\pi}$ at the level of
 reflexive objects $U$, a {\em lower} layer $f_\pi$ at the level of multipoints
 $1$.

We define simultaneously 
\footnote{{\bf Notation:}  In Appendix \ref{appdx1} for traditional GoI, (e.g., see  Figure \ref{morgraph}), we illustrate
proofs of sequents $\vdash [\Delta], \Gamma$ as I/O boxes with labelled wires as interface.  In what follows below, for typographical reasons, we often omit the wires and just write I/O labels for the interface (e.g. see
the Cut-Rule below.)  \\
 On the I/O box, interface formulas are ordered (from top to bottom) as follows:
first,  the unique  focused (positive) formula (if it exists), then the negative formulas,
and finally the sequence $\Delta$ of cut formulas.}
$\Mean{\pi}$ and
$f_{\pi}$
by induction on $\pi$. \\

\noindent1. ({\em Axiom}): \quad
$\pi$ \,  is \, $\vdash P^\bot, P$. \\
Remember that $U_P=U_{P^\perp}=U$ and 
$\mathbbm{1}_{P} \cong \mathbbm{1}_{P^\bot}  : \cong 1^n$
for a certain natural number $n$.
\begin{eqnarray*} 
\Mean{\pi}:= s_{U,U}: &   U_P \otimes U_{P^\bot}  \longrightarrow 
  U_P \otimes U_{P^\bot}   \\
f_\pi   := s_{1^n,1^n}:  &  
\mathbbm{1}_{P} \otimes \mathbbm{1}_{P^\perp} 
  \longrightarrow \mathbbm{1}_{P} \otimes \mathbbm{1}_{P^\perp} 
\end{eqnarray*}
where
$\xymatrix
@C=0.1in 
@R=0.25pc   
{ 
& U_{P}  \ar[ddrr]  & 
&  U_{P}  \\
s_{U,U}=&              &  &             \\
& U_{P^\bot}   \ar[uurr]       &    &  U_{P^\bot}
}$
$\xymatrix
@C=0.1in 
@R=0.25pc
{  & \mathbbm{1}_{P} \ar[ddrr]  &
 &  \mathbbm{1}_{P} \\
\mbox{and~} s_{1^n,1^n}= &             &  &             \\
& \mathbbm{1}_{P^\bot}   \ar[uurr]       &    &  \mathbbm{1}_{P^\bot}
}$

Each arrow consisting of the crossing
$s_{1^n,1^n}$ denotes
the permutation (between factors of the tensor foldings
$\mathbbm{1}_P$ and
$\mathbbm{1}_{P^\bot}$) which is induced by De Morgan duality 
between polarized formulas $P$ and $P^\bot$.

\vspace{3ex}

\noindent 2. ({\em Cut}):

\hspace{1.2in}
$\pi$ \, is 

\vspace{-10ex} 

$$
\infer[cut]{
\vdash [\Delta', \Delta'', P, P^\bot], {\cal N}, \Xi}{
\infer*[\pi']{\vdash [\Delta'], P, {\cal N}}{}
&
\infer*[\pi'']{\vdash [\Delta''], P^\perp, \Xi}{}
}$$

We define
\begin{align*}
\Mean{\pi} :=  \tau^{-} (\Mean{\pi'} \otimes \Mean{\pi''}) \, \,
 \tau 
& & 
f_\pi      :=  \tau^{-} (f_{\pi'} \otimes f_{\pi''}) \, \, \tau
\end{align*}
where $\tau$ denotes the indicated exchange
for the conclusions and the cut-formulas.  Here $\tau$ and its inverse
$\tau^{-}$ are
simply permutations of the interface (denoted by $U_{(~)}$ 
 and $\mathbbm{1}_{(~)}$).
See the following:

\begin{figure}[htbp]
\begin{minipage}{0.5\hsize}
$$\begin{picture}(20,100)(20,-60) 

\put(30,0){\fbox{\rule[.5in]{.5in}{0in}}}

\put(30,-60){\fbox{\rule[.5in]{.5in}{0in}}}

\put(-10,-60){\fbox{\rule[1.35in]{.1in}{0in}}}

\put(100,-60){\fbox{\rule[1.35in]{.1in}{0in}}}

\put(10,30){$U_P$}      \put(80,30){$U_P$}
\put(10,15){$U_{\cal N}$}           \put(80,15){$U_{\cal N}$}
\put(10,0){$U_{\Delta'}$}     \put(80,0){$U_{\Delta'}$} 
     \put(42, 15){$\Mean{\pi'}$}
     \put(48, -15){$\otimes$}
       \put(42, -45){$\Mean{\pi''}$}
     \put(-5, -15){$\tau$}     \put(102, -15){$\tau^{-}$}
     \put(-70,-15){$\Mean{\pi}:=$}
\put(10,-30){$U_\Xi$}       \put(80,-30){$U_\Xi$}
\put(7,-45){$U_{P^\bot}$}     \put(77,-45){$U_{P^\bot}$}
\put(10,-60){$U_{\Delta''}$}     \put(80,-60){$U_{\Delta''}$}

\put(-30,30){$U_\Xi$}            \put(120,30){$U_\Xi$}   
\put(-30,10){$U_{\cal N}$}        \put(120,10){$U_{\cal N}$} 
\put(-30,-10){$U_P$}     \put(120,-10){$U_P$}

\put(-30,-25){$U_{P^\bot}$}        \put(120,-25){$U_{P^\bot}$}     
\put(-30,-45){$U_{\Delta'}$}               \put(120,-45){$U_{\Delta'}$}    
\put(-30,-60){$U_{\Delta''}$}       \put(120,-60){$U_{\Delta''}$}   
\end{picture}$$
\end{minipage}
 \begin{minipage}{0.5\hsize}
$$\begin{picture}(20,100)(20,-60)

\put(30,0){\fbox{\rule[.5in]{.5in}{0in}}}

\put(30,-60){\fbox{\rule[.5in]{.5in}{0in}}}

\put(-10,-60){\fbox{\rule[1.35in]{.1in}{0in}}}

\put(100,-60){\fbox{\rule[1.35in]{.1in}{0in}}}
\put(10,30){$\mathbbm{1}_P$}      \put(80,30){$\mathbbm{1}_P$}
\put(10,15){$\mathbbm{1}_{\cal N}$}           \put(80,15){$\mathbbm{1}_{\cal N}$}
\put(10,0){$\mathbbm{1}_{\Delta'}$}     \put(80,0){$\mathbbm{1}_{\Delta'}$} 
     \put(42, 15){$f_{\pi'}$}
     \put(48, -15){$\otimes$}
       \put(42, -45){$f_{\pi''}$}
     \put(-5, -15){$\tau$}     \put(102, -15){$\tau^{-}$}
     \put(-70,-15){$f_{\pi}:=$}
\put(10,-30){$\mathbbm{1}_{\Xi}$}       \put(80,-30){$\mathbbm{1}_{\Xi}$}
\put(10,-45){$\mathbbm{1}_{P^\bot}$}     \put(80,-45){$\mathbbm{1}_{P^\bot}$}
\put(10,-60){$\mathbbm{1}_{\Delta''}$}     \put(80,-60){$\mathbbm{1}_{\Delta''}$}

\put(-30,30){$\mathbbm{1}_{\Xi}$}            \put(120,30){$\mathbbm{1}_{\Xi}$}   
\put(-30,10){$\mathbbm{1}_{\cal N}$}        \put(120,10){$\mathbbm{1}_{\cal N}$} 
\put(-30,-10){$\mathbbm{1}_{P}$}     \put(120,-10){$\mathbbm{1}_{P}$}

\put(-30,-25){$\mathbbm{1}_{P^\bot}$} \put(120,-25){$\mathbbm{1}_{P^\bot}$}     
\put(-30,-45){$\mathbbm{1}_{\Delta'}$}    \put(120,-45){$\mathbbm{1}_{\Delta'}$}    
\put(-30,-60){$\mathbbm{1}_{\Delta'' }$}       \put(120,-60){$\mathbbm{1}_{\Delta'' }$}   
\end{picture}$$
\end{minipage}
\end{figure}

\noindent
In the following cases, we define
$\Meanpre{\pi} \in {\cal C} ( \bm{U}_{\Delta, \Gamma},
\bm{U}_{\Delta, \Gamma})$
so that $\Mean{\pi}= j_\ell \Comp \Meanpre{\pi} \Comp k_\ell$,
where $j_\ell$ and $k_\ell$ are the retractions
for $U_{\Delta,\Gamma} \rhd_{(k_\ell,j_\ell)} \bm{U}_{\Delta,\Gamma}$
in equation (\ref{*retract}) of the construction of multipoints after 
Definition \ref{multipoint}.
Note that $\ell$ is a number of logical connectives of formulas
contained in $\Gamma, \Delta$.
 


\smallskip

\noindent 3.  ({\em Linear Connectives}):
For a $\wp$-rule, the interpretation remains
that of
the premise proof.  
For a $\otimes$-rule, the interpretations are
the same
as those of the cut-rule defined
      above. To be precise:  
\smallskip

\noindent ($\wp$ rule) We define
\begin{align*}
\Meanpre{\pi} =   
\Meanpre{\pi'} &  & \mbox{and} & & 
f_\pi =  f_{\pi'}
\end{align*}
\noindent ($\otimes$ rule) We define
\begin{align*}
\Meanpre{\pi} = 
\tau^{-}  \Meanpre{\pi'}  \tau & & \mbox{and} &
   & 
f_\pi  =  \tau^{-}  f_{\pi'}  \tau, 
\end{align*}
where $\tau$ for $\Meanpre{\pi}$ (resp. for $f_\pi$)
is $\bm{U}_P \otimes \bm{U}_Q \otimes \bm{U}_{\Gamma'} \otimes \bm{U}_{\Gamma''} 
\otimes \bm{U}_{\Delta'} \otimes \bm{U}_{\Delta''}
\longrightarrow
\bm{U}_P \otimes \bm{U}_{\Gamma'} \otimes \bm{U}_{\Delta'} 
\otimes 
\bm{U}_Q
\otimes
\bm{U}_{\Gamma''} 
\otimes
\bm{U}_{\Delta''} 
$
(resp. 
$ \mathbbm{1}_P
\otimes \mathbbm{1}_Q \otimes
\mathbbm{1}_{\Gamma'} \otimes \mathbbm{1}_{\Gamma''} \otimes \mathbbm{1}_{\Delta'} \otimes \mathbbm{1}_{\Delta''}
\longrightarrow
\mathbbm{1}_P
\otimes  \mathbbm{1}_{\Gamma'} \otimes \mathbbm{1}_{\Delta'} \otimes 
\mathbbm{1}_Q \otimes
\mathbbm{1}_{\Gamma''} 
\otimes
\mathbbm{1}_{\Delta''} $).

\bigskip

Note that in the above Cases 1 - 3, $\Meanpre{\pi}$ and $f_\pi$
are defined in the same way: replace $\bm{U}$'s by $\mathbbm{1}$'s. 
However this is no longer true 
in the following Cases 4 and 5
(cf. the paragraph above Definition \ref{interprf}). 

\noindent 4. ({\em Polarity Changing $\uparrow$}):
\vspace{5ex}

\hspace{1.75in} $\pi$ \, is  

\vspace{-8ex}
$$\infer*[\pi']{\infer[\uparrow]{\vdash [\Delta], {\cal M}, \up{P}}{
\vdash [\Delta], {\cal M},P}}{}$$

We define
$$
\begin{array}{ccl}
\Meanpre{\pi} & =  &  
  \Meanpre{\pi'} \otimes 0_{U,U}  
  \quad \mbox{with $0_{U,U}: U_\uparrow \longrightarrow
U_\uparrow .$} \\
f_\pi  & = &  
(
(0_{\mathbbm{1}_P, I} \otimes \mathbbm{1}_{\cal M} \otimes \mathbbm{1}_{\Delta})
\, \, \Comp \, \,  
f_{\pi'}
\, \, \Comp \, \, 
(0_{I, \mathbbm{1}_P,} \otimes \mathbbm{1}_{\cal M} \otimes \mathbbm{1}_{\Delta})
)
\otimes 
0_{1,1}  
 \quad  \mbox{with
$0_{1,1}: 1 \longrightarrow 1 .$
}
\end{array}
$$

See the following: \\
\begin{minipage}{0.5\hsize}
$$\begin{picture}(200,100)(-100,-60)
\put(30,-35){\vector(1,0){35}} \put(43,-45){$0_{U,U}$}
\put(43,-28){$\otimes$}
\put(32,-15){\fbox{\rule[.5in]{.4in}{0in }}}
\put(-35,-30){$\bm{U}_{\up{P}} \cong$}
\put(12,-40){$U_\uparrow$}  
\put(12,-28){$\otimes$}
 \put(70,-40){$U_\uparrow$}
\put(12,18){$\bm{U}_\Delta$}    \put(70,18){$\bm{U}_\Delta$}
\put(37,0){$\Meanpre{\pi'}$}

\put(10,0){$\bm{U}_{\cal M}$} \put(70,0){$\bm{U}_{\cal M}$}
\put(12,-15){$\bm{U}_P$} \put(70,-15){$\bm{U}_P$}
\put(-80, 10){$\Meanpre{\pi}:=$}
\end{picture}$$
\end{minipage}
 \begin{minipage}{0.5\hsize}
$$\begin{picture}(50,100)(-10,-60)
\put(-5,-35){\vector(1,0){105}} \put(43,-45){$0_{1,1}$}
\put(43,-28){$\otimes$}
\put(32,-15){\fbox{\rule[.5in]{.4in}{0in }}}

\put(-50,-30){$\mathbbm{1}_{\up{P}} \cong$ }

\put(-20,-40){$1$}  
\put(-20,-30){$\otimes$}

 \put(105,-40){$1$}

\put(-20,-20){$I \stackrel{0_{I,\mathbbm{1}_P}}{\longrightarrow} $}
\put(12,15){$\mathbbm{1}_\Delta$}    
\put(70,15){$\mathbbm{1}_\Delta$}
\put(85,-20){$\stackrel{0_{\mathbbm{1}_P,I}}{\longrightarrow} I $}

\put(42,0){$f_{\pi'}$}

\put(12,0){$\mathbbm{1}_{\cal M}$} \put(70,0){$\mathbbm{1}_{\cal M}$}
\put(12,-15){$\mathbbm{1}_P$} \put(70,-15){$\mathbbm{1}_P$}

\put(-75,10){$f_\pi:=$}
\end{picture}$$
\end{minipage}

\noindent 5. ({\em Polarity Changing $\downarrow$}): \\
\vspace{2ex}

\hspace{1.75in} $\pi$ \, is  

\vspace{-8ex}
$$\infer*[\pi']{\infer[\downarrow]{\vdash [\Delta], {\cal M}, \down{N}}{
\vdash [\Delta], {\cal M}, N}}{}$$
For the interpretation of this case, 
$\mathfrak{r}_{\cal M}$ and $\mathfrak{i}_{\cal M}$ of Definition \ref{rA!A}
are used. We assume $\mathbbm{1}_{\cal M}=1^m$
for a natural number $m$.
Depending on the value of $m$,
two morphisms $g_m$ and $h_m$ 
are defined
as follows,
modulo associativity of the monoidal (resp. comonoidal)
structure of $j$ (resp. $k$) (cf. Axiom 2'):  let   
\begin{eqnarray} \label{gh}
g_m:=
\mathfrak{i} \Comp
(\mathfrak{i} \otimes 1)
\Comp
\cdots 
\Comp
(\mathfrak{i} \otimes 1^{m-2}) 
         & \mbox{and}
       & h_m:=
(\mathfrak{r} \otimes 1^{m-2}) \Comp
       \cdots \Comp (\mathfrak{r} \otimes 1) \Comp \mathfrak{r}
\end{eqnarray}
to yield a retraction $1^m \rhd_{(g_m,h_m)} 1$.
 When $m$ is $1$, 
both $g_m$ and $h_m$ are defined to be $\Id_1$.

\bigskip

For an exchange $\tau$ of the interface, the conjugate actions
$(\quad)^\tau$ and $^\tau (\quad)$ are defined as follows: $x^\tau= 
(\tau^{-} \otimes 1^m )
\, \Comp \, \,  
x
\, \, \Comp \,
\tau$ and  $^{\tau} x= 
\tau^{-}
\, \Comp \, \, 
x
\, \, \Comp \,
(\tau \otimes 1^m )$.
{In what follows in the definition,
the compositions $\Comp$'s are
modulo permutation of
$\otimes$}. \footnote{That is, in composing two maps between tensors of objects, the matching of the codomain 
of one with the domain of the other is only up to permutation of the
tensor factors.}

\smallskip

\noindent 
We define:
\begin{eqnarray*}
\Meanpre{\pi} & := 
\theta^-
\, \, \,  \Comp \, \, \, 
(h_m \otimes \Meanpre{\pi'} \otimes g_m) \, \, \,  \Comp \, \, \, 
\theta,
\end{eqnarray*}
where $\theta^- =
\alpha \otimes \,  ^{\tau} \! 
(\bm{U}_N \otimes \bm{U}_\Delta \otimes {\cal M}^\flat \otimes \mathfrak{i}_{\cal M})
\quad$ and $\quad \theta=
\alpha^* \otimes 
(\bm{U}_N \otimes \bm{U}_\Delta \otimes {\cal M}^\flat \otimes \mathfrak{r}_{\cal
 M})^{\tau}$. \\  
See the following:

$$\begin{picture}(200,130)(-70,-50)
\put(30,60){\vector(1,0){35}} \put(43,65){$h_m$}
\put(43,50){$\otimes$}

\put(36,-15){\fbox{\rule[.8in]{.34in}{0in }}}

\put(-79,35){$\bm{U}_N$} \multiput(-55,35)(4,0){9}{\line(1,0){1}}

\put(-79,18){$\bm{U}_\Delta$} \multiput(-55,18)(4,0){9}{\line(1,0){1}}

\put(12,60){$1$}   \put(70,60){$1^m$}

\put(85,52){\line(1,-3){30}} 

\put(120,60){$1$}  \put(132,60){\vector(1,0){100}} \put(172,65){$\alpha$} 
\put(237,60){$U$}

\put(-115,60){$U$} \put(-102,60){\vector(1,0){100}} \put(-52,65){$\alpha^*$}

\put(12,18){${\cal M}^\sharp$}    
\put(70,18){${\cal M}^\sharp$} \multiput(96,20)(4,0){5}{\line(1,0){1}}

\put(145,35){$\bm{U}_N$} \multiput(168,35)(4,0){10}{\line(1,0){1}}
\put(145,18){$\bm{U}_\Delta$} \multiput(168,18)(4,0){10}{\line(1,0){1}}

\put(39,10){$\Mean{\pi'}^\dagger$}
\put(12,-17){$\bm{U}_\Delta$} 
\put(70,-17){$\bm{U}_\Delta$} 
\multiput(96,-15)(4,0){5}{\line(1,0){1}}

\multiput(96,2)(4,0){5}{\line(1,0){1}}

\put(12,35){$\bm{U}_N$}
\put(70,35){$\bm{U}_N$} \multiput(96,37)(4,0){5}{\line(1,0){1}}

\put(12,-1){${\cal M}^\flat$} 
\put(70,-1){${\cal M}^\flat$}


\put(43,-30){$\otimes$}
\put(30,-35){\vector(1,0){35}} \put(43,-48){$g_m$}
\put(13,-40){$1^m$}
\put(70,-37){$1$}  \put(88,-37){\line(1,3){30}}
\put(120,-40){$1^m$}    
\multiput(132,-40)(4,0){5}{\line(1,0){1}}

\put(202,-15){\line(-1,0){35}} \put(170,-25){$\mathfrak{i}_{\cal M}$}
\put(202,-15){\line(-4,-3){37}}

\put(145,-17){${\cal M}^\sharp$}

\put(-200,0){$\Mean{\pi}^\dagger:=$}


\put(155,-40){$1^m$} 

\put(145,-1){${\cal M}^\flat$}

\multiput(168,-1)(4,0){10}{\line(1,0){1}}

\put(202,-17){${\cal M}^\sharp$}

\put(237,35){$\bm{U}_N$}
\put(237,18){${\cal M}^\sharp$}
\put(237,-17){$\bm{U}_\Delta$}
\put(237,-1){${\cal M}^\flat$}


\put(-24,-17){${\cal M}^\sharp$}

\put(-79,-17){${\cal M}^\sharp$}

\put(-61,-17){\line(1,0){35}} \put(-45,-25){$\mathfrak{r}_{\cal M}$}
\put(-61,-17){\line(3,-2){37}} 

\put(-22,-40){$\mathbbm{1}_{\cal M}$}

\put(-22,-53){$= 1^m$}

\multiput(-4,-38)(4,0){5}{\line(1,0){1}}

\put(-79,-1){${\cal M}^\flat$} 

\multiput(-55,-1)(4,0){9}{\line(1,0){1}}

\put(-162,50){$\bm{U}_{\down{N}} \cong $}

\put(-115,35){$\bm{U}_N$}  \put(-115,48){$\otimes$}

\put(-152,9){$\bm{U}_{\cal M} \cong $} 

\put(-115,18){${\cal M}^\sharp$} 

\put(-112,9){$\otimes$}

\put(-115,-1){${\cal M}^\flat$}  

\put(-115,-17){$\bm{U}_{\Delta}$} 


\put(-3,-15){\fbox{\rule[.8in]{.08in}{0in }}}

\put(-93,-15){\fbox{\rule[.8in]{.08in}{0in }}}

\put(128,-15){\fbox{\rule[.8in]{.08in}{0in }}}

\put(223,-15){\fbox{\rule[.8in]{.08in}{0in }}}

\put(-91, 11){\footnotesize $\tau$} 
\put(-1, 11){\footnotesize $\tau^-$} 
\put(132, 11){\footnotesize $\tau$}   \put(224,11){\footnotesize $\tau^{-}$}

\end{picture}$$
We define:
\begin{eqnarray*}
f_{\pi} & := 
\eta^-
\, \, \,  \Comp \, \, \, 
(h_m \otimes f_{\pi'} \otimes g_m) \, \, \,  \Comp \, \, \, 
\eta 
\end{eqnarray*}
where
$\eta^- =
 1 \otimes 0_{\mathbbm{1}_N, I} \otimes \, ^{\rho} (\mathbbm{1}_\Delta
 \otimes   \mathfrak{i}^m)$ and 
$\eta= 1 \otimes 0_{I,\mathbbm{1}_N} \otimes 
(\mathbbm{1}_\Delta  \otimes   \mathfrak{r}^m)^\rho$
for an exchange $\rho$ of the interface. See the following:

$$\begin{picture}(200,130)(-60,-70)
\put(30,45){\vector(1,0){35}} \put(43,50){$h_m$}
\put(43,35){$\otimes$}
\put(32,-15){\fbox{\rule[.6in]{.4in}{0in }}}

\put(12,45){$1$}   
\put(70,45){$1^m$}

\put(80,41){\line(1,-2){34}} 

\put(110,45){$1$}  
\multiput(122,45)(4,0){25}{\line(1,0){1}} 

\put(237,45){$1$}

\put(12,23){$\mathbbm{1}_N$}    
\put(70,23){$\mathbbm{1}_N$}

\put(85,25){\vector(1,0){150}}

\put(155,29){\footnotesize $0_{\mathbbm{1}_N,I}$}

\put(40,5){$f_{\pi'}$}
\put(12,-17){$\mathbbm{1}_\Delta$} 
\put(70,-17){$\mathbbm{1}_\Delta$} 
\multiput(90,-15)(4,0){5}{\line(1,0){1}}

\multiput(90,7)(4,0){5}{\line(1,0){1}}

\put(12,4){$\mathbbm{1}_{\cal M}$} 
\put(70,4){$\mathbbm{1}_{\cal M}$}


\put(43,-30){$\otimes$}
\put(30,-35){\vector(1,0){35}} \put(43,-48){$g_m$}
\put(12,-37){$1^m$}
\put(70,-37){$1$}  
\put(80,-30){\line(1,2){35}}
\put(110,-37){$1^m$}    \multiput(122,-35)(4,0){5}{\line(1,0){1}}

\put(202,-15){\line(-1,0){35}} \put(175,-25){$\mathfrak{i}^m$}
\put(202,-15){\line(-2,-1){37}}

\put(145,-17){$1^m$}

\put(-195,0){$f_{\pi}:=$}


\put(145,-37){$1^m$} 

\put(145,4){$\mathbbm{1}_{\Delta}$}

\multiput(167,7)(4,0){8}{\line(1,0){1}}
\put(202,-17){$1^m$}

\put(237,23){$I$}
\put(237,-17){$\mathbbm{1}_\Delta$}
\put(237,4){$1^m$}


\put(-22,-17){$1^m$}

\put(-79,-17){$1^m$}

\put(-61,-17){\line(1,0){35}} \put(-42,-26){$\mathfrak{r}^m$}
\put(-61,-17){\line(2,-1){37}} 

\put(-22,-37){$1^m$} 

\multiput(-10,-37)(4,0){5}{\line(1,0){1}}

\put(-79,4){$\mathbbm{1}_{\Delta}$} 

\multiput(-59,7)(4,0){9}{\line(1,0){1}}

\put(-114,45){$1$}  \multiput(-100,45)(4,0){27}{\line(1,0){1}} 
\put(-115,34){$\otimes$}   \put(-160,34){$\mathbbm{1}_{\down{N}} \cong $}
\put(-114,23){$I$}

\put(-100,25){\vector(1,0){110}}
\put(-40,29){\footnotesize $0_{I,\mathbbm{1}_N}$}

\put(-114,-17){$\mathbbm{1}_\Delta$} 

\put(-144,4){$\mathbbm{1}_{\cal M} =$}

\put(-114,4){$1^m$}  

\put(-4,-15){\fbox{\rule[.4in]{.1in}{0in }}}

\put(-96,-15){\fbox{\rule[.4in]{.1in}{0in }}}

\put(129,-15){\fbox{\rule[.4in]{.1in}{0in }}}

\put(222,-15){\fbox{\rule[.4in]{.1in}{0in }}}

\put(-91, 0){\footnotesize $\rho$} 
\put(-2, 0){\footnotesize $\rho^-$} 
\put(132, 0){\footnotesize $\rho$}   
\put(223,0){\footnotesize $\rho^{-}$}

\end{picture}$$

\begin{exam}[$\Mean{\pi}$ and $f_\pi$ of Definition \ref{interprf}]\label{exinter}{\em
The following two examples are interpretations of proofs
in the GoI situation $\Rel$
of Example \ref {relexpl} and Example \ref{Relex},
in which $\alpha$ is identified with 
the singleton subset $\{ n_\alpha \}$ of $\mathbb{N}=U$.
In the following, matrices are UDC representations of
$\Rel$ morphisms (see Appendix \ref{UDC}), 
the blank elements denote $0$ morphisms
(i.e., empty relations) of appropriate types
and $g_m$ and $h_m$ are from $(\ref{gh})$.
}
\end{exam}
(i) Let $\pi$ be the unique cut-free proof for
$\vdash \down{X^\bot},\up{X}$ (the $\eta$-expansion of 
$\vdash X^\bot,X$). \\
The side formula of the $\downarrow$-rule of $\pi$
is $\up{X}$ so that $\mathbbm{1}_{\up{X}}=1$.
It also holds that $\mathbbm{1}_{\down{X^\perp}}=1$.

Then 

\noindent 
$\Meanpre{\pi}=\bordermatrix{ 
   &  U_{\downarrow}     & U_{X^\perp}    & U_{X}      &  U_{\uparrow}  \cr
U_{\downarrow}       &   &  &  & \{ (n_\alpha, n_\alpha) \}  \cr
U_{X^\perp}               &   &  & \Id_U  &  \cr
U_{X}        &   &  \Id_U &  &  \cr
U_{\uparrow} & \{ (n_\alpha, n_\alpha) \} &  &  &   \cr
}$
and
$f_{\pi}=\bordermatrix{ 
& \mathbbm{1}_{\down{X^\perp}} & \mathbbm{1}_{\up{X}} \cr
\mathbbm{1}_{\down{X^\perp}} &   &   \Id_1 \cr
\mathbbm{1}_{\up{X}} &  \Id_1 &  \cr
}$. 

\bigskip

\noindent They are obtained respectively from

\bigskip

$\bordermatrix{ 
   &  U_{\downarrow}     & U_{X^\perp}    & U_{X}      &  U_{\uparrow}
   & 1  \cr
U_{\downarrow} &        &   &  &  & \alpha \Comp g_1  \cr
U_{X^\perp}   &               &  & \Id_U  &  \cr
U_{X}       &  &     \Id_U &  &  \cr
U_{\uparrow} &  &  &  &  &   \cr
1 &  h_1 \Comp \alpha^*
}$ and
$\bordermatrix{ 
& \mathbbm{1}_{\down{X^\perp}} & \mathbbm{1}_{\up{X}} & 1  \cr
\mathbbm{1}_{\down{X^\perp}} &   &   & g_1 \cr
\vspace{0.01cm} & &  \cr
\mathbbm{1}_{\up{X}} &   &  \cr
\vspace{0.01cm} & &  \cr
1 & h_1  &   \cr
}$

\bigskip

\noindent
when contracting the last two columns (resp. rows)
respectively by $\mathfrak{r}_{\up{X}}$ (resp. $\mathfrak{i}_{\up{X}}$) and by
$r$ (resp. $\mathfrak{i}$).

\smallskip

\noindent
(ii) Let $\pi$ be the unique cut-free proof for $\vdash \down{\up{(X \otimes Y)}},
Y^\perp \wp X^\perp$. \\
The side formula of the $\downarrow$-rule of $\pi$
is $Y^\perp \wp X^\perp$ so that $\mathbbm{1}_{Y^\perp \wp
     X^\perp}=1^2$.
Both $\mathbbm{1}_{\down{\up{(X \otimes Y)}}}$
and $\mathbbm{1}_{\up{(X \otimes Y)}}$ are $1$.
Note that $\bm{U}_{Y^\perp \wp X^\perp} = (Y^\perp \wp X^\perp)^\sharp=U + U$.

Then

\noindent
$\Meanpre{\pi} =
\bordermatrix{  & U_{\downarrow} & \bm{U}_{X \otimes Y} \cr
U_{\uparrow} \cr
\vspace{.3cm} &  \cr
\bm{U}_{Y^\perp \wp X^\perp} &
{\small \begin{array}{l}
\{ (n_{\alpha}, n_\alpha) \} \\
+ 
\{ (n_{\alpha}, n_\alpha) \} 
\end{array}}
 & \Id_{U + U} \cr
}
\otimes
\bordermatrix{  &  U_{\uparrow} & 
\bm{U}_{Y^\perp \wp X^\perp} \cr
U_{\downarrow} &   & {\small \begin{array}{l}
\{ (n_{\alpha}, n_\alpha) \} \\
+ 
\{ (n_{\alpha}, n_\alpha) \} 
\end{array}}
      \cr
\vspace{.3cm} &  & \cr
\bm{U}_{X \otimes Y}  &  &  \Id_{U + U}   \cr
}
$ \\

and \hfill  $f_{\pi}=
\bordermatrix{ 
&
\mathbbm{1}_{
\down{
\up{(X \otimes Y)}
}}
&
\mathbbm{1}_{Y^\perp \wp X^\perp} \cr
\mathbbm{1}_{
\down{
\up{(X \otimes Y)}
}} & & \Id_1 + \Id_1 \cr
\mathbbm{1}_{Y^\perp \wp X^\perp} & \Id_1 + \Id_1 \cr 
}$.

\bigskip

\noindent They are obtained respectively from

\bigskip

\noindent
{\small $\bordermatrix{  & U_{\downarrow} & \bm{U}_{X \otimes Y} & U_{\uparrow} & 
\bm{U}_{Y^\perp \wp X^\perp}   & 1^2 \cr 
U_{\downarrow} &  &  &  & & \alpha \Comp g_2 
   \cr
\bm{U}_{X \otimes Y}  & & & &  \Id_{U + U}   \cr
U_{\uparrow} \cr
\bm{U}_{Y^\perp \wp X^\perp} & 
 & \Id_{U + U} \cr
1^2 & h_2 \Comp \alpha^*
}$}
and
{\small $\bordermatrix{ 
&
\mathbbm{1}_{
\down{
\up{(X \otimes Y)}
}}
&
\mathbbm{1}_{Y^\perp \wp X^\perp}
& 1^2 \cr
\mathbbm{1}_{
\down{
\up{(X \otimes Y)}
}} & & &  g_2  \cr
\vspace{.03cm} & & &  \cr
\mathbbm{1}_{Y^\perp \wp X^\perp} \cr
\vspace{.03cm} & & &  \cr
1^2 & h_2 \cr 
}$}    

\bigskip

\noindent
when contracting the last two columns (resp. rows)
respectively by $\mathfrak{r}_{Y^\perp \wp X^\perp}$
(resp. 
$\mathfrak{i}_{Y^\perp \wp X^\perp}$) and by 
$\mathfrak{r}^2$  (resp. $\mathfrak{i}^2$).

\vspace{2ex}
\begin{rem}[On degenerate GoI situations] {\em
In Example \ref{pfnexpl}, we noted that the two polarized GoI situations
$\Pfn$ and $\PInj$ are degenerate.  We can now say why we chose this terminology.
The reader can check that in both {\Pfn} and {\PInj}, the interpretation above
of the polarity-changing rule $\downarrow$ has no effect; that is,
$\Mean{\pi} = \Mean{\pi'}$ for the conclusion of this rule.   This is definitely
{\em not} the case in the {\Rel} model.
}
\label{prfdegen}
\end{rem}

\vspace{2ex}

\begin{defn}[Polarized Execution formulas] \label{defex}{\em
Let $\sigma = \otimes^m s$, the $m$-fold tensor product of
the symmetry $s=s_{U,U}$,
and $\sigma_* = \otimes_{i=1}^m s_i$ where
each $s_i$ is the symmetry $s_{1^{\ell_i}, 1^{\ell_i}}$
for certain $\ell_i$.
The $s$ corresponds to the permutation between
dual cut formulas
so that $U=U_P=U_{P^\perp}$
and the $s_i$ corresponds to the permutation
induced by De Morgan duality for dual cut formulas
so that $1^{\ell_i} \cong \mathbbm{1}_{P} \cong \mathbbm{1}_{P^\perp}$.
Then {\em polarized execution formulas} are defined both on $\Mean{\pi}$ and on $f_\pi$ as follows:
\begin{eqnarray*}
{\sf Ex}(\Mean{\pi}, \sigma) :=  &
\TR{(\Id_{U_\Gamma} \otimes \sigma ) \Comp \Mean{\pi}
}{U_\Delta}{U_\Gamma}{U_\Gamma}  =
& 
\TR{(\Id_{U^n} \otimes \sigma ) \Comp \Mean{\pi}}{U^{2m}}{U^n}{U^n}  \\
 {\sf Ex}(f_\pi, \sigma_*)      := &
\TR{(\Id_{\mathbbm{1}_\Gamma} \otimes \sigma_* ) \Comp \Mean{\pi}}{\mathbbm{1}_\Delta}{\mathbbm{1}_\Gamma}{ 
\mathbbm{1}_\Gamma} =
&   
\TR{(\Id_{1^{n'}} \otimes \sigma_* ) \Comp f_\pi}{1^{2m'}}{1^{n'}}{1^{n'}} 
\end{eqnarray*}
}
\end{defn}
Note:  ${\sf Ex}(\Mean{\pi}, \sigma)\in End_{\cal C}(U_\Gamma) =
End_{\cal C}(U^n)$ \, and \, ${\sf Ex}(f_\pi, \sigma_*) \in 
End_{\cal C}(\mathbbm{1}_{\Gamma})=
End_{\cal C}(1^{n'})$, as pictured below.

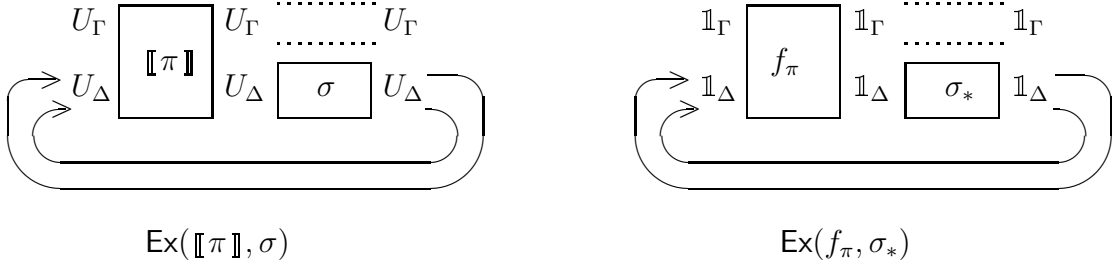
\begin{figure}[htbp]
\begin{minipage}{0.5\hsize}
$\begin{picture}(200,80)(-50,-40)
\put(30,-15){\fbox{\rule[.5in]{.4in}{0in }}}
\put(90,-15){\fbox{\rule[.2in]{.4in}{0in }}}

\put(78,-10){\oval(180,70)[b]}       
\put(147,-12){\oval(42,20)[tr]}      
\put(6,-12){\oval(36,17)[tl]}  \put(0,-7){$>$}      

\put(78,-25){\oval(160,20)[b]}                   
\put(146,-25){\oval(25,20)[tr]}                  
\put(10,-25){\oval(25,20)[tl]} \put(4,-18){$>$}  

\put(12,15){$U_\Gamma$}    \put(70,15){$U_\Gamma$}
\multiput(90,25)(4,0){10}{\line(1,0){1}}
\multiput(90,10)(4,0){10}{\line(1,0){1}}
\put(130,15){$U_\Gamma$}
\put(39,0){$\Mean{\pi}$}
\put(12,-10){$U_\Delta$} \put(70,-10){$U_\Delta$} \put(130,-10){$U_\Delta$}


\put(105, -10){$\sigma$}
\end{picture}$ \\
$${\sf Ex}(\Mean{\pi}, \sigma)$$
\end{minipage}
\begin{minipage}{0.5\hsize}
$\begin{picture}(200,80)(-50,-40)
\put(30,-15){\fbox{\rule[.5in]{.4in}{0in }}}
\put(90,-15){\fbox{\rule[.2in]{.4in}{0in }}}

\put(78,-10){\oval(180,70)[b]}       
\put(147,-12){\oval(42,20)[tr]}      
\put(6,-12){\oval(36,17)[tl]}  \put(0,-7){$>$}      

\put(78,-25){\oval(160,20)[b]}                   
\put(146,-25){\oval(25,20)[tr]}                  
\put(10,-25){\oval(25,20)[tl]} \put(4,-18){$>$}  

\put(12,15){$\mathbbm{1}_\Gamma$}    \put(70,15){$\mathbbm{1}_\Gamma$}
\multiput(90,25)(4,0){10}{\line(1,0){1}}
\multiput(90,10)(4,0){10}{\line(1,0){1}}
\put(130,15){$\mathbbm{1}_\Gamma$}
\put(39,0){$f_{\pi}$}
\put(12,-10){$\mathbbm{1}_\Delta$} 
\put(70,-10){$\mathbbm{1}_\Delta$} 
\put(130,-10){$\mathbbm{1}_\Delta$}


\put(105, -10){$\sigma_*$}
\end{picture}$ \\
$${\sf Ex}(f_{\pi}, \sigma_*)$$
\end{minipage}
\caption{Execution Formulas on $\Mean{\pi}$ and on $f_\pi$ of
Polarized Proofs of $\vdash [\Delta],\Gamma$}
 \label{polex}
\end{figure}


\subsection{Polarized execution formulas are an invariant for cut-elimination}
\label{polarex}

  In \MELL, the execution formula is only an invariant of
normalization for sequents not containing ``$?$" (see \cite{HS06, HS11}). 
But in \MLLP, $\uparrow$ and $\downarrow$
are logically weak (although still functorial) operators.
In the case of the exponentials in \cite{AHS02}, 
semantical axioms inspired directly from the syntactical rules
of  ! and ? in linear logic are imposed on top of the GoI situation.
Here, for the weaker polarized case, 
we instead use the multipoint semantical structure to simulate
the logically weak $\uparrow$ and $\downarrow$.  That is the purpose
of the rather subtle retraction structure of multipoints.
The simulation is indirect in the sense that the notion of multipoint
does not live in the syntax of $\MLLP$. But semantically
the notion is fine-grained enough to sufficiently simulate $\uparrow$ and $\downarrow$.
This allows us to prove the execution formula is a full invariant for cut-elimination,
in the sense of the following Proposition.

\bigskip

\begin{prop}[Ex is an invariant]\
If $\pi \rightarrow_{*} \pi'$ by $\MLLP$ cut-elimination,
then ${\sf Ex}(\Mean{\pi}, \sigma) = {\sf Ex}(\Mean{\pi'}, \sigma')$
and ${\sf Ex}(f_{\pi}, \sigma_*) = {\sf Ex}(f_{\pi'}, (\sigma')_*)$. 
In particular, if $\pi'$ is a (cut-free) normal form of $\pi$, then 
(since $\sigma'$ and $(\sigma')_*$ are $\Id_I$) we have:
${\sf Ex}(\Mean{\pi}, \sigma) = \Mean{\pi'}$
and ${\sf Ex}(f_{\pi}, \sigma_*) = f_{\pi'}$. 
\label{exinvariance}
\end{prop}

\begin{prf}{}
The crucial  polarized case is
where $\pi$ contains cut formulas $\down{P^\perp}$  and $\up{P}$,
which are transformed into the cut formulas $P^\perp$ and  $P$  of
$\pi'$  by cut elimination.
The other cases are similar to the non-polarized $\MLL$ case.
So the crucial case is that of a proof $\pi$, ending with a cut between
$\uparrow P$ and $\downarrow P^\bot$, which has the following form.  
A one-step reduction then gives rise to the proof
ending with a cut between $P$ in $\pi'_1$ and $P^\bot$
in $\pi'_2$:
$$
\infer[\mbox{\small cut}]{ \vdash
[ \cdots, \uparrow P, \downarrow  P^\perp ],
 \cdots, {\cal M} 
}{\infer[\mbox{\small $\uparrow$}]{\vdash [\cdots], \uparrow\! P,
\cdots }{ \infer*[\pi_1^{'}]{\vdash [\cdots], P, \cdots}{}}
 & 
\infer[\mbox{\small $\downarrow$}]{\vdash [\cdots], \downarrow\! P^\bot,
{\cal M} }{ \infer*[\pi_2^{'}]{\vdash [\cdots], P^\bot, {\cal M}}{}}
}
$$
Recall the definition of the interpretations of polarity changing
 $\uparrow$ (see 4 of Definition \ref{interprf})), where 
 $U_\uparrow \rightarrow U_\uparrow$
occurring in the construction of $\Mean{\pi}$
(for the principal $\uparrow$) is the zero morphism.
This directly means that in $(\Id \otimes s^m) \Comp \Mean{\pi}$
there arise no new loops via $U_\uparrow$ and $U_\downarrow$
for the $\up{P}$ and $\down{P}$ in the last cut.
Hence  the trace of $(\Id \otimes s^m) \Comp \Mean{\pi'}$= the trace of
 $(\Id \otimes s^m) \Comp \Mean{\pi}$.
Let us calculate this precisely using the trace axioms. 
Let $\pi_1$ (resp. $\pi_2$) denote $\pi'_1$ (resp. $\pi'_2$)
followed by a $\uparrow$ (resp. by $\downarrow$) -rule applied to $P$ (resp. to
 $P^\bot$).
Let $\Gamma_i$ (resp. $\Gamma$) be conclusions with the list of cut-formulas 
of $\pi'_i$ (resp. $\pi$).
\begin{eqnarray*}
\Meanpre{\pi_1} & = & 
\Meanpre{\pi'_1} \otimes (c: U_\uparrow \rightarrow U_\uparrow )
\, \, \, \mbox{with $c=0$}  \\
\Meanpre{\pi_2} & = &  
(\mathfrak{i}_{\cal M} \otimes \bm{U}_{\Gamma'_2 \setminus  {\cal M}})
\Comp
\{ \Mean{\pi'_2} \otimes (a: U_\downarrow \rightarrow 1^m)
\otimes (b: 1^m \rightarrow U_\downarrow )
\} \Comp
(\mathfrak{r}_{\cal M} \otimes  \bm{U}_{
\Gamma'_2 \setminus  {\cal M}}
) \\
         & & 
\hfill \mbox{where 
$\mathbbm{1}_{\cal M}=1^m$, $a=h_m \, \Comp \, \alpha^*$ and $b=\alpha \,
\Comp \, g_m$.} \\
\Meanpre{\pi}
& = &  \Meanpre{\pi_1} \otimes \Meanpre{\pi_2}
 \\
& = &
(\hat{\mathfrak{i}}_{\cal M} \otimes \bm{U}_{\Gamma \setminus  {\cal M}}
) \Comp
\{
(\Mean{\pi'_1} \otimes c)
\otimes 
(\Mean{\pi'_2} \otimes a
\otimes b)
\}
\Comp
(\hat{r}_{\cal M} \otimes 
\bm{U}_{\Gamma \setminus  {\cal M}}
) 
\end{eqnarray*}
Note that the compositions and precompositions of the middle arrows occurring in
 the above (R.H.S)
are modulo 
permutations of their
 domains and codomains. 

In the definition of $\Meanpre{\pi}$, we use the following retraction
$(\hat{\mathfrak{i}}_A, \hat{\mathfrak{r}}_A)$ derivable from 
Definition \ref{rA!A} using the canonical isomorphism (\ref{FUD}).
We define $\hat{\mathfrak{i}}_{A}:=\mathfrak{i}_A
\otimes A^\flat$ and $\hat{\mathfrak{r}}_{A} :=\mathfrak{r}_A \otimes A\flat$:
$$\begin{array}{lr}
 \bm{U}_A \otimes
 \mathbbm{1}_A \rhd_{(\hat{\mathfrak{i}}_A,\hat{\mathfrak{r}}_A)} \bm{U}_A
& \mbox{satisfying $\hat{\mathfrak{i}}_A \, \Comp \, \hat{\mathfrak{r}}_A = \bm{U}_A $}
\end{array}$$
The same definitions as above also apply
for $\hat{\mathfrak{i}}_{\cal M}$ and $\hat{\mathfrak{r}}_{\cal M}$
for a sequence ${\cal M}$.


\smallskip



\noindent
In the following proof, $\bm{U}_x$ 
is simply abbreviated by $x$: it is
either a formula $A$ or a sequence ${\cal M}$.
We abbreviate $U_\uparrow$ and $U_\downarrow$
by $\uparrow$ and $\downarrow$.
For the proof, it suffices to show
\begin{eqnarray} \label{main}
{\sf Tr}^{\uparrow \otimes \downarrow}_{\Gamma \setminus \{\downarrow, \uparrow\},
\Gamma \setminus \{\downarrow, \uparrow\}
}
(
(\Id \otimes
s_{\uparrow, \downarrow} \otimes s_{P,P^\bot}
 )
\Comp \Meanpre{\pi}
) & = & 
(\Id \otimes s_{P,P^\bot} ) \Comp
(\Meanpre{\pi'_1} 
\otimes 
\Meanpre{\pi'_2} ) 
\end{eqnarray}
where $\Id=
\Id_{\Gamma \setminus \{ P, P^\bot \}}$
and $\Meanpre{\pi'}=\Meanpre{\pi'_1} \otimes \Meanpre{\pi'_2}$. 
\bigskip

For proving (\ref{main}), we start with observing the following
identity derivable by
generalized yanking (Appendix \ref{genyank}), vanishing and dinaturality
in a traced monoidal category.

\vspace{1ex}

\noindent ({\bf Iterated generalized yanking})  

\vspace{1ex}

For any morphisms $f: X \longrightarrow U$, $g: U \longrightarrow V$ and
$h: V \longrightarrow Y$ in a traced monoidal category,
\begin{eqnarray} \label{eqitergenyank}
{\sf Tr}_{X,Y}^{V \otimes U} ( 
(s_{Y,U} \otimes V) \Comp (Y \otimes s_{U,V})
\Comp (h \otimes f \otimes g)  ) =  
h \Comp
g \Comp f
\end{eqnarray} 
Pictorially, this says;

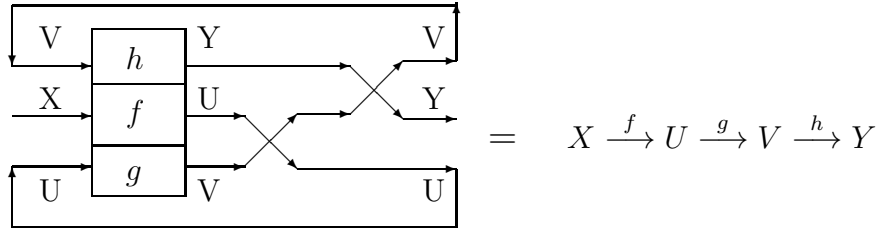
\begin{figure}[hbt]
$$\begin{picture}(20,70)(100,-30)
\put(10,15){X}
\put(43,10){$f$}
\put(43,-12){$g$}
\put(10,-20){U}
\put(70,15){U}
\put(70,-20){V}
\put(180,0){\large =}
\put(210,0){$X\stt{f}U\stt{g}V\stt{h}Y$}
\put(30,1){\line(1,0){36}}
\put(0,12){\vector(1,0){30}}
\put(0,-7){\vector(1,0){30}}
\put(66,12){\vector(1,0){22}}
\put(66,-7){\vector(1,0){22}}
\put(108,13){\vector(1,0){20}}
\put(108,-8){\vector(1,0){60}}
\put(88,12){\vector(1,-1){20}}
\put(88,-7){\vector(1,1){20}}
\put(168,-8){\line(0,-1){22}}
\put(0,-30){\vector(0,1){23}}
\put(168,-30){\line(-1,0){168}}
\put(30,-15){\fbox{\rule[.5in]{.4in}{0in }}}

\put(43,30){$h$}
\put(30, 27.4){\fbox{\rule[.2in]{.4in}{0in }}}
\put(0,31){\vector(1,0){30}}
\put(66,31){\vector(1,0){62}}
\put(128,31){\vector(1,-1){20}}
\put(128,13){\vector(1,1){20}}
\put(148,11){\vector(1,0){20}}
\put(148,33){\vector(1,0){20}}
\put(168, 33){\vector(0,1){22}}
\put(10,38){V}
\put(70,38){Y}
\put(0, 54){\vector(0,-1){23}}
\put(0,54){\line(1,0){168}}

\put(155,15){Y}
\put(155,38){V}
\put(155,-20){U}
\end{picture}
$$
\end{figure}

The (L.H.S) of (\ref{main}),
by naturality w.r.t $\hat{\mathfrak{r}}_{\cal M}$ and $\hat{\mathfrak{i}}_{\cal M}$, is
equal to
\begin{multline}
(\hat{\mathfrak{i}}_{\cal M} \otimes \Id_{\Gamma \setminus \{ {\cal M}, \downarrow,
 \uparrow \} }) \Comp \\ 
{\sf Tr}^{\downarrow}_{ 1^m \otimes \Gamma \setminus \{ \downarrow, \uparrow \},
\,  1^m \otimes \Gamma \setminus \{  \downarrow, \uparrow \}   \, \, }
\{
(\Id \otimes \sigma )
\Comp
(
\Meanpre{\pi'_1} \otimes c
\otimes 
\Meanpre{\pi'_2} \otimes a
\otimes b
)
\} 
\Comp
(\hat{\mathfrak{r}}_{\cal M} \otimes \Id_{\Gamma \setminus \{ {\cal M},
 \downarrow,\uparrow \} } )\ \label{trcdown} 
\end{multline}
The middle arrow of (\ref{trcdown}) is,
by superposing, equal to
\begin{eqnarray*}
(\Id \otimes s_{\uparrow, \downarrow} \otimes s_{P,P^\bot} ) \Comp
(\Meanpre{\pi'_1} 
\otimes 
\Meanpre{\pi'_2}) 
\, \otimes \, 
\TR{a \otimes b \otimes c}{\uparrow \otimes \downarrow}{1^m}{1^m}
\end{eqnarray*}
which by (\ref{eqitergenyank}), is equal to 
$(\Id \otimes s_{\uparrow, \downarrow}
\otimes s_{P,P^\bot} ) \Comp
(\Meanpre{\pi'_1} 
\otimes 
\Meanpre{\pi'_2}) 
\, \otimes \, 
a \Comp c \Comp b
$. \\
This amounts to showing that (\ref{trcdown}) is equal to the 
(R.H.S) of (\ref{main}) 
by Axiom 9' since
$a  \Comp c
\Comp b=0$ when $c=0$.
See  Appendix \ref{calculations} for a pictorial representation of the above calculations.

\noindent
The analogous assertion 
 for $f_{\pi_1}$, $f_{\pi_2}$ and $f_{\pi}$ is  easily checked
in the bottom layer.
\end{prf}

\vspace{1ex}

\begin{exam}[Extrusion of $\downarrow$-boxes by expanding Ex]
\label{exextru}{\em
In the dynamics of normalization,
{\em extrusion} of $\downarrow$-boxes depicted in
Example \ref{cutexextru}
is captured by running the execution formula. Let us illustrate this phenomenon
in $\Rel$, considered as a polarized GoI situation arising from Haghverdi's
notion of a UDC (Unique Decomposition Category) (see Appendix \ref{UDC}).  In this case,
the execution formulas  (defined via traces) can be written in Girard's original form of a power series, 
as in (\ref{exudc}) of Definition \ref{usualGoI} in Appendix \ref{appdx1}, and 
have a matrix representation. 

 This phenomenon is sufficiently explained
by examining the lower layer (= multipoint) interpretation
$f_{\pi_i}$ of the three proofs
($i=1,2,3$) together with permutation
$\sigma_*=
s_{\mathbbm{1}_{\uparrow_1}, \mathbbm{1}_{\downarrow_2}}
\otimes 
s_{\mathbbm{1}_{\downarrow_3}, \mathbbm{1}_{\uparrow_2}}
$ of cuts, which are common
to the three proofs: \\
$f_{\pi_i}=
{ \begin{blockarray}{ccc|cccc}
   &   \mathbbm{1}_{\downarrow_1} 
   & \mathbbm{1}_{\uparrow_3} 
   &   \mathbbm{1}_{\uparrow_1}
   &  \mathbbm{1}_{\downarrow_2}  
   &  \mathbbm{1}_{\downarrow_3}  
   &   \mathbbm{1}_{\uparrow_2} \\
\begin{block}{c(@{\hspace*{8pt}} cc|cccc@{\hspace*{10pt}})}
\mathbbm{1}_{\downarrow_1} & & \delta_{i3}  & \delta_{i1}  &  &  & \delta_{i2}  \\
 \mathbbm{1}_{\uparrow_3}  & \delta_{i3} &    &&&  1    \\
\cline{1-7}
\mathbbm{1}_{\uparrow_1}  & \delta_{i1} &     &  &  \\
\mathbbm{1}_{\downarrow_2}           &&&&&&  1  \\
\mathbbm{1}_{\downarrow_3}  && 1 & &&   &     \\
\mathbbm{1}_{\uparrow_2}     & \delta_{i2} &    && 1   &      &    \\
\end{block}
\end{blockarray}
}
$
\hfill $\sigma_* = \bordermatrix{ 
     & \mathbbm{1}_{\uparrow_1}    & \mathbbm{1}_{\downarrow_2}        \cr
\mathbbm{1}_{\uparrow_1}               &     &1    \cr
\mathbbm{1}_{\downarrow_2}           & 1 &    \cr
} \otimes
\bordermatrix{ 
   &  \mathbbm{1}_{\downarrow_3}       &  \mathbbm{1}_{\uparrow_2}   \cr
\mathbbm{1}_{\downarrow_3}        &     & 1  \cr
\mathbbm{1}_{\uparrow_2}          &  1 &       
}$ \\
In the above, $1$ abbreviates $\Id_1$
and $\delta_{ij}$ is the Kronecker delta.
Note that the normal form $\pi$ of $\pi_1$, $\pi_2$ and $\pi_3$
is the $\eta$-expansion of the axiom-link, hence
is interpreted by the $2 \times 2$ anti-diagonal identity matrix
 $f_{\pi}$
indexed with 
$\mathbbm{1}_{\downarrow_1}$ and $\mathbbm{1}_{\uparrow_3}$.

The scope extrusion of Example \ref{cutexextru}
is represented by the expansion of the Execution formula
$Ex(f_{\pi_1}, \sigma_*)$ (i.e., taking the trace of $(\operatorname{Id} \otimes \sigma_* ) \Comp 
f_{\pi_i}$).
Note that the $box_j$ with $j \in \{ 1,2,3 \}$
is represented by the two paired elements
$\delta_{ij}$ in the symmetric matrix $f_{\pi_i}$,
while the symmetric pair of $1$s
at $(\mathbbm{1}_{\downarrow_i},
 \mathbbm{1}_{\uparrow_i} )$ and
$( \mathbbm{1}_{\uparrow_i}, \mathbbm{1}_{\downarrow_i})$
 with $i \in \{ 2,3 \}$
represents the box introduced by $\downarrow_i$.
Then, the calculation passes through
$Ex(f_{\pi_2}, \sigma_*)$, then $Ex(f_{\pi_3}, \sigma_*)$ and finally
 will terminate in $f_{\pi}$.  Here,
the lower left $\delta_{i1}$ moves to the lower
left $\delta_{i2}$ by the action of
$ (f_{\pi_1})_{22} \, \, \sigma_* (f_{\pi_1})_{21}$.
Then it moves  finally to the lower left
$\delta_{i3}$ by the action of
$ (f_{\pi_1})_{12} \, \, 
\sigma_* (f_{\pi_1})_{22} \, \, \sigma_* (f_{\pi_1})_{21}$.
This says that  as  $j$ decreases, the sum for 
$Ex(f_{\pi_j}, \sigma_*)$
becomes``finer grained": that is, the information flow realized in the sum
of $Ex(f_{\pi_j}, \sigma_*)$ 
can be retrieved from that in $Ex(f_{\pi_i}, \sigma_*)$ with $i >j$.  
For the information flow arising from the Execution formula
in a UDC, the reader is referred to
Appendices \ref{usualGoI}, \ref{UDC} and Haghverdi's thesis
\cite{Hag00}.
}\end{exam}


\subsection{Polarized execution formulas characterize focusing}
\label{mainthmsec1}
 As we saw above in MLLP, the execution formula yields invariants 
 of the dynamical process of cut-elimination. In this section we 
 give a second property peculiar to the polarized execution formula.
 As far as we know it has no analogue in traditional linear logic.
 
Our main result is that  the execution formula in polarized
GoI situations is  able to
characterize the focusing property, which is the fundamental characteristic 
of polarized logics.
%
%
%
Observe that in a polarized GoI situation, a proof $\pi$ of an \MLLP \, sequent is
 interpreted as a pair $(\Mean{\pi}, f_\pi)$.  This 
interpretation does not capture the focusing property.  Instead, 
the GoI situation only provides an interpretation
of polarities in terms of multipoints arising from the retractions
$U \otimes 1^m \rhd U$ for $\Mean{\pi}$
and $1^m \otimes 1^m \rhd 1^m$ for $f_{\pi}$.
We may now ask:
how do we semantically obtain the stronger notion of positivity/negativity?
We show that this stronger
property can be characterized
in terms of {\em preservation of multipoints}
through running the execution formulas.  These results
are described in Theorem \ref{invari} below.
For this proposition, we introduce the following
definition:


\vspace{1ex}

\begin{defn}[restriction and range of morphisms] \

{\em Let ${\cal C}$ be a polarized GoI situation.
\begin{itemize} \itemsep=0pt 
\item[-]
In the presence of $0$ morphisms of ${\cal C}$, 
the two morphisms $\iota_j$ and $\rho_j$ 
are derivable, as follows, where $\check{K}:=
K \setminus \{j\}$:
$$
\xymatrix{
\iota_j: \quad   X_j \simeq X_j \otimes (\bigotimes_{i \in \check{K}} I)
\ar[rr]^(.5){X_j \otimes (\bigotimes
0_{I,X_i})}
& &
X_j \otimes (\bigotimes_{i \in \check{K} } X_i)
\simeq
\bigotimes_{i \in K} X_i  \\
\rho_j : \quad  \bigotimes_{i \in K} X_i
\simeq 
X_j \otimes (\bigotimes_{i \in \check{K} } X_i)
\ar[rr]^(.55){X_j \otimes (\bigotimes 0_{X_i,I})}
& & 
X_j \otimes (\bigotimes_{i \in \check{K} } I) \simeq X_j 
}$$
so that
$\rho_k \iota_j = X_i$ if $j=k$ and $0_{X_j,X_k}$ otherwise
(These are called quasi-injections and quasi-projections in \cite{Hag00,HS11})
\item[-] For a morphism $f$ with domain $\bigotimes_{i \in K} X_i$,
its {\em restriction} to $X_j$ is the morphism $f \Comp \iota_j$.
\item[-] For a morphism $f$ with codomain $\bigotimes_{i \in K} X_i$,
$f$ {\em ranges over $X_j$} { if 
$(\iota_j \Comp \rho_j) \Comp f = f$}.
\end{itemize}
}
\end{defn}

\smallskip

The following lemma will be used for Theorem \ref{invari} (Case 4).

\begin{lem}[Trace on zero morphisms] \label{zerotrace}\


\noindent
Zero morphisms satisfy the following property on tracing.  For
any morphism $f :  X \otimes U \longrightarrow Y \otimes U$:
\begin{multline*}
\TR{f \Comp (X \otimes 0_{U,U})}{U}{X}{Y}
=
(X \otimes 0_{U,I}) \,  \Comp \,  f \,  \Comp \,  (X \otimes 0_{I,U})
= 
\TR{(X \otimes 0_{U,U}) \Comp f}{U}{X}{Y}
\end{multline*}
\end{lem}
\begin{prf}{} We prove the first equation (a similar calculation goes for
the second). \\
First decompose $0_{U,U}=0_{I,U} \Comp \, 0_{U,I}$. \\
$\begin{aligned}
\TR{f \Comp (X \otimes 0_{I,U} \, \Comp \, 0_{U,I})}{U}{X}{Y}
=  
\TR{(X \otimes 0_{U,I}) \Comp f \Comp (X \otimes 0_{I,U})
}{I}{X}{Y}
=
(X \otimes 0_{U,I}) \Comp f \Comp (X \otimes 0_{I,U})
\end{aligned}$ \\
The first equation is by dinaturailty and the second one is by vanishing.
\end{prf}

\smallskip

\noindent
{\bf Notation:}  
for a sequence $\Gamma$ containing a formula $A$,
consider the composition\\ $\xymatrix{
\mathbbm{1}_A
\ar[r]^>(0.7){{\sf mp}(A)} &
 U_{A}
  \ar[r]^{\iota}
& U_{\Gamma}} ,$ where $\iota$ is the quasi-injection induced by the
embedding of $A$ into $\Gamma$.  This is denoted (by abuse of notation) by
${\sf mp}(A): \mathbbm{1}_A \longrightarrow U_\Gamma$, leaving the $\iota$ 
implicit.

\bigskip

\bigskip

\begin{thm}[focusing = invariance of {\sf mp}] \label{invari}
Let $\pi$ be an $\MLLP$ proof of a sequent $\vdash [\Delta], {\cal N}, P$,
where $P$ is a positive formula.
Then the execution formulas over
$\Mean{\pi} $ and over $f_\pi$ 
give rise to the following commutative diagram:

$$\xymatrix{
U_{{\cal N}, P}  \ar[rr]^{{\sf Ex}(\Mean{\pi}, \sigma)}  &      &
U_{{\cal N}, P} \\ \\
\mathbbm{1}_P \ar[uu]^{{\sf mp}(P)} \ar[rr]_{{\sf Ex}(f_{\pi}, \sigma_*)}
 &    &  \mathbbm{1}_{\cal N} \ar[uu]_{{\sf mp}({\cal N})}}$$
The bottom arrow of this diagram indicates that
${\sf Ex}(f_{\pi}, \sigma_*)$, when its domain is restricted 
to
$\mathbbm{1}_P$, ranges
over $\mathbbm{1}_{\cal N}$.
\end{thm}

\bigskip

\begin{prf}{}

We prove the proposition directly for  Case 1 (axiom) and Case 4 ($\downarrow$ rule)
and by induction on the size of $\pi$  for the other cases. {\em Note:}
symbols used in the proof are the same as those in Definition \ref{interprf}.

\smallskip

\noindent 1.({\em Axiom}:)  \\
The following diagram commutes because multipoints associated with De
Morgan dual formulas are equal by the duality-induced permutation
on their domains (which are foldings of tensors of $1$'s).
$$\xymatrix
@C=0.3in
@R=1.2pc
{
U_P \otimes U_{P^\perp} \ar[rr]^{s_{U,U}}  &     &   U_{P} \otimes U_{P^\perp} \\  \\
\mathbbm{1}_P \otimes \mathbbm{1}_{P^\perp}
 \ar[uu]^{{\sf mp}(P) \otimes 0}  \ar[rr]_{s_{1^n,1^n}}
 &    & 
\mathbbm{1}_P \otimes \mathbbm{1}_{P^\perp}
 \ar[uu]_{0 \otimes {\sf mp}(P^\perp)}}$$

\smallskip 
\noindent 2. ({\em Linear connectives:})\\
We prove it for a proof $\pi$ ending with the $\otimes$-rule 
(the result is obvious for the $\wp$-rule).\\
Suppose $\pi$ is
$$\infer[\otimes]{\vdash [\Delta_1, \Delta_2], {\cal M}_1, {\cal M}_2,  P_1 \otimes P_2  }{
\infer*[\pi_1]{\vdash  [\Delta_1], {\cal M}_1, P_1}{}  &
\infer*[\pi_2]{ \vdash [\Delta_2], {\cal M}_2, P_2}{} 
}$$
First, note that
\begin{eqnarray*}
{\sf Ex}(\Mean{\pi_1} \otimes \Mean{\pi_2}, \sigma_1 \otimes \sigma_2) & = &
{\sf Ex}(\Mean{\pi_1} , \sigma_1 )
\otimes
{\sf Ex}(\Mean{\pi_2}, \sigma_2) \\
{\sf Ex}(f_{\pi_1} \otimes f_{\pi_2}, (\sigma_1)_* \otimes (\sigma_2)_*) & = &
{\sf Ex}(f_{\pi_1} , (\sigma_1)_* )
\otimes
{\sf Ex}(f_{\pi_2}, (\sigma_2)_*)
\end{eqnarray*}
where the $\sigma_i$  are  iterated tensors of symmetries $s$, representing 
the cut formulas in $\pi_i$, for $i=1,2$.

Then commutativity of the first diagram by I.H.
yields directly
that of the second
where 
$ n= n_1 + n_2$:\\
 $\xymatrix
@C=0.3in
@R=1.2pc
{
U_{{\cal M}_i, P_i} \ar[rr]^{{\sf Ex}(\Mean{\pi_i}, \sigma_i)}  & &
 U_{{\cal M}_i, P_i} \\ \\
\mathbbm{1}_{P_i}  \ar[uu]^{{\sf mp}(P_i)} \ar[rr]_{{\sf Ex}(f_{\pi_i},
(\sigma_i)_*)}  
\ar@{}[uurr]|{ i=1,2}  
&  & \mathbbm{1}_{{\cal M}_i}
     \ar[uu]_{{\sf mp}({\cal M}_i)}
}$ \hspace{1cm} 
$\xymatrix
@C=0.3in
@R=1.2pc
{
U_{{\cal M}_1,{\cal M}_2, P_1 \otimes P_2}  \ar[rr]^{{\sf
      Ex}(\Mean{\pi}, \sigma)}  &  &  U_{{\cal M}_1,{\cal M}_2, P_1 \otimes P_2} \\ \\
\mathbbm{1}_{P_1} \otimes \mathbbm{1}_{P_2} 
 \ar[uu]^{{\sf mp}(P_1) \otimes {\sf mp}(P_2)  }
\ar[rr]_{{\sf Ex}(f_\pi, \sigma_*)    }  &   &   
\mathbbm{1}_{{\cal M}_1} \otimes \mathbbm{1}_{{\cal M}_2} 
 \ar[uu]_{{\sf mp}({\cal M}_1)
\otimes
{\sf mp}({\cal M}_2 )
}
}$

\smallskip

\noindent 3.({\em Polarity Changing $\uparrow$:}) \\
This case never happens
since a conclusion of any such  proof does not contain a positive formula.

\smallskip

\noindent 4.({\em Polarity Changing $\downarrow$:})  This case
does not use the I.H. Instead it uses
 Lemma \ref{zerotrace} directly. This lemma connects traces with compositions
of zero morphisms (see the discussion after equation (\ref{sigmazero}) below ). 
 We begin with an equation  which is entailed from the commutativity
of the following diagram (to be discussed below).     \footnote{For typographical simplicity, we rotated the diagram  90
      degrees clockwise}
\begin{eqnarray}
\alpha^m \Comp f_\pi & = 
 \Meanpre{\pi} \Comp \alpha 
& \quad \quad \text{where $f_\pi$ is restricted to
      $\mathbbm{1}_{\down{N}}=1$.}
      \label{bigdiageqn}
\end{eqnarray}
$$\xymatrix
@C=.5pc
@R=2.5pc
{ \mathbbm{1}_{\down{N}} =
1
\ar[dddd]_{f_\pi}
\ar[rrrr]^{ \alpha \, \otimes \, 0_{I,
{\cal M}^\sharp \otimes {\cal M}^\flat \otimes \bm{U}_{\Xi}} }
\ar[rrd]_{h_m}
   &     &  &   & 
  U_{\downarrow} \otimes {\cal M}^\sharp \otimes {\cal M}^\flat \otimes \bm{U}_{\Xi}
  \ar[d]_{U_{\downarrow} \otimes \mathfrak{r}_{\cal M} \otimes {\cal M}^\flat  \otimes
\bm{U}_\Xi} 
  \ar@/^10pc/[dddd]^{\Meanpre{\pi}} 
   \\     
 &  &
1^m 
\ar[d]_{\iota }
     &    &     U_{\downarrow} \otimes 1^m \otimes {\cal M}^\sharp \otimes {\cal M}^\flat
\otimes \bm{U}_\Xi
   \ar[d]^{(h_m \Comp \alpha^*) \otimes (\alpha \Comp g_m) \otimes \Meanpre{\pi'}} \\
 &   &
U_{\downarrow} \otimes  1^m \otimes 1^m \otimes {\cal M}^\flat \otimes \bm{U}_\Xi
\ar@{=}[d]^{\mbox{\tiny symmetry}} 
   &    &    1^m \otimes  U_{\downarrow} \otimes {\cal M}^\sharp \otimes {\cal M}^\flat \otimes
 \bm{U}_\Xi
   \ar@{=}[d]^{\mbox{\tiny symmetry}} \\
     &       &    1^m \otimes 1^m \otimes V
\ar[rr]^(.4){1^m \otimes \alpha^m \otimes V} 
     \ar@<1ex>[d]^{\mathfrak{i}^m \otimes V }&  &   
   1^m \otimes {\cal M}^\sharp \otimes {\cal M}^\flat \otimes U_{\downarrow}
      \otimes \bm{U}_\Xi
   \ar@<1ex>[d]^{\mathfrak{i}_{\cal M} \otimes V}\\ 
\mathbbm{1}_{\cal M} = 
1^m \ar[rr]_{\iota}    &    &    
 1^m \otimes V  
\ar[rr]_{\alpha^m \otimes V}   
  \ar@{}[urr]|(0.6){ \parbox[b]{1in}{ \tiny $V\otimes$ lifting \\
 $V={\cal M}^\flat \otimes U_{\downarrow} \otimes  \bm{U}_\Xi$} }   &  &      
{\cal M}^\sharp \otimes {\cal M}^\flat \otimes U_{\downarrow} \otimes \bm{U}_\Xi  
}$$
In the diagram, we let $\mathbbm{1}_{\cal M}:=1^m$ and
$\Xi$ denotes $\Delta, N$.  Why does this diagram commute?
The diagram consists of 4 regions from left to right:   modulo symmetry,
(i) a leftmost pentagon bordered by $f_\pi$, (ii) a middle upper
hexagon bordered by $1^m\otimes \alpha^m\otimes V$,
(iii) a middle lower square bordered by $1^m\otimes \alpha^m\otimes V$,
and (iv) a rightmost square bordered by $\Meanpre{\pi}$.
The left pentagon commutes from the definition of $f_\pi$.
The upper middle hexagon commutes because
$\alpha^* \Comp \, \alpha =\Id_1 $.
The lower middle square is $V \otimes$ lifting along $\alpha^m$
of Proposition \ref{varlifting}, hence commutes.  The rightmost
square commutes by definition of $\Meanpre{\pi}$.
These  commutativities imply the outermost hexagon commutes.

Then the following is a commutative square for equation (\ref{bigdiageqn}).
\begin{eqnarray} \label{seconddiagram}
\xymatrix
@C=0.3in
@R=1.2pc
{
\bm{U}_{\Delta, {\cal M}, \down{N}}
  \ar[rr]^{\Meanpre{\pi}}  &  & \bm{U}_{\Delta, {\cal M}, \down{N}}
\\ \\
\mathbbm{1}_{\down{N}}  \ar[rr]_{f_\pi }  \ar[uu]^{\alpha
= {\sf mp}(\down{N})   }
 &  & 1^{m} \ar[uu]_{\alpha^m= {\sf mp}({\cal M}) }
}
\end{eqnarray}
{Both
$\sigma \otimes \bm{U}_{{\cal M}, \down{N}}$ and
$0_{\bm{U}_\Delta} \otimes \bm{U}_{{\cal M}, \down{N}}$
(resp. both 
$\sigma_* \otimes 
\mathbbm{1}_{{\cal M}, \down{N}}$ and
$0_{\mathbbm{1}_\Delta} \otimes 
\mathbbm{1}_{{\cal M}, \down{N}}$)
act identically
on the component $\bm{U}_{\cal M}$ of the co-domain 
(resp. on the co-domain $1^m$) of $\alpha^m$.  Thus 
we have}
\begin{eqnarray}
(\sigma \, \otimes \bm{U}_{{\cal M}, \down{N}})
\Comp \Meanpre{\pi} \Comp \alpha  
& = & 
(0_{\bm{U}_\Delta} \otimes \bm{U}_{{\cal M}, \down{N}})
\Comp \Meanpre{\pi} \Comp \alpha
\qquad  \text{and}  \nonumber \\
 \alpha^m \Comp 
(\sigma_* \otimes 
\mathbbm{1}_{{\cal M}, \down{N}}) 
\Comp
f_\pi &  = &  \alpha^m \Comp 
(0_{\mathbbm{1}_\Delta} \otimes 
\mathbbm{1}_{{\cal M}, \down{N}}) 
\Comp
f_\pi \label{sigmazero}
\end{eqnarray}
{ By composing and precomposing the upper (resp. the lower)
horizontal arrow of (\ref{seconddiagram}) with 
$0_{\bm{U}_\Delta} \otimes \bm{U}_{{\cal M}, \down{N}}$
(resp. with 
$0_{\mathbbm{1}_\Delta} \otimes 
\mathbbm{1}_{{\cal M}, \down{N}}$), 
we have, by Lemma \ref{zerotrace},}
\begin{eqnarray*}
\TR{(\sigma \, \otimes \bm{U}_{{\cal M}, \down{N}})
\Comp \Meanpre{\pi}}{\bm{U}_\Delta}{\bm{U}_{{\cal M},
\down{N}}}{\bm{U}_{{\cal M}, \down{N}}} \Comp \, \alpha  
& = & 
 \alpha^m \Comp \,
\TR{
(\sigma_* \otimes 
\mathbbm{1}_{{\cal M}, \down{N}})}{\mathbbm{1}_\Delta}{
\mathbbm{1}_{{\cal M}, \down{N}}}{\mathbbm{1}_{{\cal M}, \down{N}}}
\Comp
f_\pi
\end{eqnarray*}
This is the assertion 
${\sf Ex}(\Meanpre{\pi}, \sigma) \Comp \, \alpha =\alpha^m \Comp
\, {\sf Ex}(f_{\pi},
\sigma_*)$.

\smallskip

\noindent 5.({\em Cut:}) \footnote{This is the case where Corollary 
\ref{stronguniform}
 (strong uniformity) is used.} \\
This case uses the following property of the associativity of cut
(cf. \cite{HS11}):
\begin{eqnarray*}
{\sf Ex}( \Mean{\pi}, \sigma \otimes s)  =  
{\sf Ex}({\sf Ex}( \Mean{\pi}, \sigma), s) & \quad \mbox{and} \quad
{\sf Ex}( f_\pi, \sigma_* \otimes s_*)  = 
{\sf Ex}({\sf Ex}( f_{\pi}, \sigma_*), s_*) 
\end{eqnarray*}
where $s$ corresponds to the last cut of $\pi$.

The sequence 
$\Xi$ of Case 2 of Definition \ref{interprf}
must be of the form $Q, {\cal M}$ with a positive $Q$: i.e.,
$\pi$ is \\  
$$
\infer[cut]{
\vdash [\Delta', \Delta'', P, P^\bot], {\cal N}, Q, {\cal M}}{
\infer*[\pi']{\vdash [\Delta'], P, {\cal N}}{}
&
\infer*[\pi'']{\vdash [\Delta''], P^\perp, Q, {\cal M}}{}
}$$
Note first that $\sigma=\sigma' \otimes \sigma''$
{with $\sigma'$ (resp. $\sigma''$) representing
all the cuts of $\pi'$ (resp. $\pi''$)}.

The I.H. implies the following two diagrams commute:

$\xymatrix
@C=0.3in
@R=1.2pc
{
U_{P, {\cal N}} \ar[rr]^{{\sf Ex}(\Mean{\pi'},\sigma')} &   & U_{P, {\cal N}} \\ \\
\mathbbm{1}_P \ar[rr]_{{\sf Ex}(f_{\pi'},\sigma'_*)}  \ar[uu]^{{\sf
      mp}(P)}    &     &  \mathbbm{1}_{\cal N} \ar[uu]_{{\sf mp}({\cal N})}
}$ $\hspace{1cm}$
$\xymatrix
@C=0.3in
@R=1.2pc
{
U_{P^\perp, Q, {\cal M}}  \ar[rr]^{{\sf Ex}(\Mean{\pi''},\sigma'')}  &
      & U_{P^\perp, Q, {\cal M}} \\ \\
\mathbbm{1}_Q  \ar[rr]_{{\sf Ex}(f_{\pi''},\sigma''_*)}  \ar[uu]^{{\sf
      mp}(Q)} &    &  \mathbbm{1}_{P^\perp} \otimes \mathbbm{1}_{\cal M}
  \ar[uu]_{{\sf mp}(P^\bot, {\cal M})}
}$

Since $\Mean{\pi} \cong \Mean{\pi'} \otimes \Mean{\pi''}$
modulo permutation of interface,
$${\sf Ex}(\Mean{\pi},\sigma) \cong {\sf Ex}(\Mean{\pi'},\sigma') \otimes
{\sf Ex}(\Mean{\pi''},\sigma'')$$
Then
$$
( U_{{\cal N}, Q, {\cal M}} \otimes s_{U_{P},U_{P^\perp}}) \Comp {\sf
Ex}(\Mean{\pi},\sigma)  \, \cong \, 
(U_{{\cal N}, Q, {\cal M}} \otimes s_{U_{P},U_{P^\perp}}
) \Comp
({\sf Ex}(\Mean{\pi'},\sigma') \otimes {\sf Ex}(\Mean{\pi''},\sigma'')) 
$$
This composing morphisms is realized by
the upper horizontal morphisms of the following composition
of the two commutative squares whose left one is obtained
by tensoring the above two squares: \\
$\xymatrix@C=0.6in
{
U_{P,{\cal N}} \otimes U_{P^\perp, Q, {\cal M}} \ar[rr]^{
{\sf Ex}(\Mean{\pi'},\sigma') \otimes {\sf Ex}(\Mean{\pi''},\sigma'')} 
&    &   U_{P,{\cal N}} \otimes U_{P^\perp, Q, {\cal M}}
\ar[rr]^{U_{{\cal N}, Q, {\cal M}} \otimes 
s_{U_P,U_{P^\perp}}}& & 
U_{{\cal N}, Q, {\cal M}} \otimes U_{P, P^\perp}
 \\  \\
\mathbbm{1}_{P} \otimes \mathbbm{1}_{Q}  \ar[uu]_{{\sf mp}(P) \otimes {\sf mp}(Q)}  
\ar[rr]_{{\sf Ex}(f(\pi'),\sigma'_*) \otimes {\sf Ex}(f(\pi''),\sigma''_*)}
 & & 
\mathbbm{1}_{\cal N} \otimes  
\mathbbm{1}_{P^\perp}
 \otimes
      \mathbbm{1}_{\cal M} 
\ar[uu]^{{\sf mp}(P^\bot) \otimes {\sf mp}({\cal N, M})}  
\ar[rr]_{
\mathbbm{1}_{{\cal N}, Q, {\cal M}}
\otimes s_{\mathbbm{1}_{P}, \mathbbm{1}_{P^\perp}}}
 &   &  \mathbbm{1}_{\cal N} \otimes
      \mathbbm{1}_{\cal M}  \otimes \mathbbm{1}_{P}  
\ar[uu]^{{\sf mp}(P)
\otimes {\sf mp}({\cal N, M})   }
}$

In the outermost square, 
taking trace along $U_P \otimes U_{P^\perp}$
(resp. along $\mathbbm{1}_P \otimes
\mathbbm{1}_{P^\perp}$)
of the upper (resp. the lower) horizontal morphism,
by Corollary \ref{stronguniform}
 (strong uniformity) we have the following diagram,
      which is the assertion
by virtue of the associativity mentioned at the beginning of this case. 

$$\xymatrix@C=1in@R=1.6pc
{ U_{\cal N} \otimes U_{Q, {\cal M}}  \ar[rr]^{
{\sf Tr}^{U_P \otimes U_{P^\bot}} ((
U_{{\cal N}, Q, {\cal M}}
 \otimes s) \Comp {\sf Ex}(\Mean{\pi},\sigma)
)}  &  & 
U_{\cal N} \otimes U_{Q, {\cal M}}
\\  \\
\mathbbm{1}_Q \ar[uu]^{{\sf mp}(Q)}   \ar[rr]_{
{\sf Tr}^{\mathbbm{1}_P \otimes \mathbbm{1}_{P^\bot} }((
\mathbbm{1}_{{\cal N}, Q, {\cal M}} \otimes s_{ \mathbbm{1}_P,
\mathbbm{1}_{P^\perp} }) \Comp {\sf Ex}(f_\pi,\sigma_*))
} &  & 
\mathbbm{1}_{\cal N} \otimes \mathbbm{1}_{\cal M} 
 \ar[uu]_{{\sf mp}(
{\cal N}) \otimes  {\sf mp}({\cal M})} 
}$$
\end{prf}

\bigskip


\smallskip

\begin{exam}[invariance of Theorem \ref{invari}]{\em
We give two examples of Theorem \ref{invari}
for the interpretations shown in Example \ref{exinter}. In these examples
${\sf Ex}(\Mean{\pi}, \sigma)$ and 
${\sf Ex}(f_{\pi}, \sigma_*)$ are $\Mean{\pi}$
and $f_\pi$ respectively,  since $\pi$ is cut-free.
} \end{exam}\label{exinvari}
\smallskip

\noindent (i) In this case, $P=\down{X^\perp}$
and ${\cal M}=\up{X}$. \\ Since $f_\pi$ restricted to 
$\mathbbm{1}_{\down{X^\perp}} :=1$
ranges over $\mathbbm{1}_{\up{X}}$,
the Proposition follows from
$$\Meanpre{\pi} \Comp
(\alpha: \mathbbm{1}_{\down{X^\perp}}\longrightarrow
U_{\downarrow})=
(\alpha: \mathbbm{1}_{\up{X}} \longrightarrow U_{\uparrow})
\Comp f_\pi \quad \mbox{under this restriction}.$$
Recall that in the $\Rel$ examples, the equated morphisms
are identified with $\{ n_\alpha \}$.

\noindent (ii) In this case, $P=\down{\up{(X \otimes Y)}}$
and ${\cal M}=Y^\perp \wp X^\perp$. \\
Since $f_\pi$ restricted to
$\mathbbm{1}_{\down{\up{(X \otimes Y)}}}$
ranges over $\mathbbm{1}_{Y^\perp} \otimes \mathbbm{1}_{X^\perp}$,
the commutativity follows from:
$$
\Meanpre{\pi} \Comp
(\alpha: \mathbbm{1}_{\down{\up{(X \otimes Y)}}}
\longrightarrow U_\downarrow)
=
( \alpha^2 : \mathbbm{1}_{Y^\perp \wp X^\perp}
 \longrightarrow 
\bm{U} _{Y^\perp \wp X^\perp}) \Comp f_{\pi} \quad
\mbox{under the restriction.}$$
The equated morphisms are identified with
$\{ n_\alpha \} + \{ n_\alpha \}$.

\smallskip

We now show there is a kind of   ``converse":  any \MLLP-provable sequent $\vdash [\Delta],{\cal N},A$ with $A$ invariant in
the sense of the diagram  in Theorem \ref{invari} must have $A$ positive.  More precisely:


\begin{prop}[converse of focusing]\label{converse_focus}
Let $\vdash [\Delta],{\cal M},A$ be a sequent provable in $\MLLP$
such that $A$ contains a polarity changing connective
and ${\cal M}$ is a sequence of negative formulas.  If the following diagram 
nontrivially commutes in all models\footnote{
We say a commutative diagram {\em trivially commutes} if the unique arrow in it is the zero morphism,
and {\em nontrivially commutes} if it is not the zero morphism.} 
$$\xymatrix{
U_{{\cal M}, A} \ar[rr]^{{\sf Ex}(\Mean{\pi}, \sigma)}  &      & U_{{\cal M}, A} \\ \\
\mathbbm{1}_{A} \ar[uu]^{{\sf mp}(A)} \ar[rr]_{{\sf Ex}(f_{\pi}, \sigma_*)}
&    &  \mathbbm{1}_{\cal M} \ar[uu]_{{\sf mp}({\cal M})}}$$
then $A$ is a positive formula.
\end{prop}
\begin{prf}{} 
It suffices to prove the assertion for cut-free proofs
for $\vdash {\cal M},A$
so that
${\sf Ex}(\Mean{\pi}, \sigma)$ and
${\sf Ex}(f_{\pi}, \sigma_*)$ become respectively
$\Mean{\pi}$ and $f_\pi$
because of the invariance of cut-elimination of
Proposition \ref{exinvariance}.
For the proof, we make the following observation, which is
proved for $f_\pi$ by induction on the construction of
proofs $\pi$: \\[1ex]
{\em For a positive formula $Q$ and any negative formula $M$
in a conclusion of a cut-free focused proof $\pi$,
the following inequality holds in any $\Rel$ polarized GoI situation:}
\begin{eqnarray}
[\mathbbm{1}_M] f_\pi \cap \mathbbm{1}_{Q}  \not = \emptyset \label{noemp}
\end{eqnarray}
(cf. Notation  \ref{notrel} for $\Rel$ in Section \ref{rel-mp} below.)

\noindent
{\em Proof of (\ref{noemp}):}
Note first that the last rule of $\pi$ must not be an $\uparrow$-rule.
Since the induction is straightforward for the axiom and
linear connectives $\otimes$ and $\wp$, we prove the case
when the last rule is $\downarrow$. The corresponding case of
Definition \ref{interprf} says that 
$[\mathbbm{1}_{\cal M} ] f_\pi \supseteq \mathbbm{1}_{\down{N}}$,
where $\mathbbm{1}_{\cal M} = 1^m$ and $\mathbbm{1}_{\down{N}}=1$,
which directly implies (\ref{noemp}).
({See the last I/O diagram
of Definition \ref{interprf} for the inclusion, where
$\mathbbm{1}_{\cal M}$ of the input on the left
first goes via $\rho$ to $1^m$, which splits via
$\mathfrak{r}^m$, and the bottom output
goes to the $1$ of the (right hand) side, by $g_m$ }.)
\hfill {\em End of Proof of (\ref{noemp})}

\bigskip
We prove  Proposition  \ref{converse_focus}
by contradiction; suppose that $A$ is negative
and consider the polarized GoI situation $\Rel$.
From the assumption, the
given cut-free proof $\pi$ has a bottom most $\uparrow$-rule, say $I$,
whose principal formula is denoted by $\up{Q}$.  
So  any inference (if it exists) below $I$ 
is either the $\wp$-rule or exchange.
Let $\pi'$ be the subproof of $\pi$
ending at the premise of $I$. See the proof figure below
for $\pi'$, where ${\cal N}$ is a certain sequence (including the empty one)
of negative formulas.
$\pi'$ is focused with the formula $Q$ positive. \\
\begin{minipage}{0.05\hsize}
$\pi= $
\end{minipage}
\begin{minipage}{0.3\hsize}
$\infer{\infer{\vdash {\cal M}, A}{\txt{\footnotesize $\wp$'s and exchanges}
}}{
 \infer[I]{\vdash {\cal N}, \up{Q}}{
 \infer*[\pi']{
 \vdash {\cal N},  Q}{}}}$ \\
\end{minipage} \\
We have the following equation: 
\begin{eqnarray}
f_\pi \cong f_{\pi'} \otimes 0_{\mathbbm{1}_{\up{Q}},\mathbbm{1}_{\up{Q}}}
\hfill \mbox{\hspace{3ex} modulo exchange rules below $I$.} \label{f'f}
\end{eqnarray}

\smallskip

First, we observe the following claim used in (Case 1) and (Case 2.2)
 below:
 
 \vspace{1ex}

\noindent ({\bf Claim})
{\em The inequality (\ref{noemp}) for $M=A$ and the condition
$\mathbbm{1}_{Q} \cap \mathbbm{1}_{\cal M} = \emptyset$
yield a contradiction. }\\
The contradiction is that $f_\pi$ restricted to $\mathbbm{1}_A$
ranges over $\mathbbm{1}_{\cal M}$ (the commutative diagram in the
 assertion).

\smallskip

\smallskip

\noindent (Case 1)
The case where the distinguished occurrence $\up{Q}$ occurs in ${\cal M}$: \\
In this case, ${\cal M}$ is constructed
from negative formulas and the $\up{Q}$
via $\wp$-connectives and commas.
The conclusion of $\pi'$ has occurrences
$Q$ and a factorization $A'$ of $A$ by means of $\wp$-rules
below $I$: That is $A$ is either $A'$ or $A' \wp \check{A}$
for some subformula $\check{A}$.
Note that $\mathbbm{1}_{A'}
\subseteq \mathbbm{1}_A$ because of the factorization.
Because $\pi'$ is focused, we have
$[\mathbbm{1}_{A'}] f_{\pi'} \cap \mathbbm{1}_{Q}  \not = \emptyset$
by (\ref{noemp}).
This directly implies
$[\mathbbm{1}_{A}] f_{\pi} \cap \mathbbm{1}_{Q}  \not = \emptyset$
by (\ref{f'f}). On the other hand, since $\up{Q}$ occurs in ${\cal M}$
but $\mathbbm{1}_{\up{Q}} \cap \mathbbm{1}_{Q} = \emptyset$,
we have $\mathbbm{1}_{Q} \cap \mathbbm{1}_{\cal M} = \emptyset$.
Thus we have a contradiction by the above claim.

\smallskip

\noindent (Case 2)
The case where the distinguished occurrence $\up{Q}$ occurs in $A$: \\
In this case $A$ is constructed
from negative formulas and the $\up{Q}$
via $\wp$-connectives.

\smallskip
\noindent (Case 2.1) The case where $A=\up{Q}$. \\
From the definition of the interpretation of the $\uparrow$-rule,
$f_\pi$ restricted to $\mathbbm{1}_A=1$ is $0$,
hence the diagram trivially commutes in the sense of footnote 7.

\smallskip
\noindent (Case 2.2) Otherwise. In this case
$Q$ and another factor $A'$
both occur in the conclusion of $\pi'$,
where $A'$ as well as $\up{Q}$
is a factorization of $A$ by means of $\wp$-rules
below $I$: That is $A$ is (modulo commutativity of $\wp$)
either $\up{Q} \wp A'$
or $\up{Q} \wp A' \wp \check{A}$
for some subformula $\check{A}$.
By the same argument for $Q$ and $A'$ as Case 1,
we have
$[\mathbbm{1}_{A'}] f_{\pi'} \cap \mathbbm{1}_{Q}  \not = \emptyset$
by (\ref{noemp}).
This directly implies
$[\mathbbm{1}_{A}] f_{\pi} \cap \mathbbm{1}_{Q}  \not = \emptyset$
by (\ref{f'f}).
On the other hand,
since this $Q$ does not occur in ${\cal M}$,
$\mathbbm{1}_{Q} \cap \mathbbm{1}_{\cal M} = \emptyset$.
Thus we have a contradiction by the above claim.

\end{prf}


\section{Constructing a compact polarized category via GoI}
\label{polcomp-int}
In this section we describe certain kinds of polarized categories arising from 
a different view of GoI; namely, from Joyal-Street-Verity's \Int-construction in the multipointed setting.
Although the material of this section is formally independent of
the previous Section \ref{polgoisit}, it highlights different aspects of GoI
arising from \Int-constructions, in particular the construction of compact polarized models 
in the sense of our previous paper \cite{HamSc07}.

\subsection{Polarized logic and polarized categories}
In Section \ref{MLLP} we described O. Laurent's polarized multiplicative linear logic \MLLP.  
For references to polarized categories, 
the reader is 
referred to  our own paper \cite{HamSc07}, as well as the original sources referred to there. 
We begin with a very general definition of a polarized categorical model for  \MLLP.  Our models are  included among Cockett-Seely polarized categories \cite{CS07}, although theirs are considerably more general.  
The definition below is  simpler than assumed in our previous paper \cite{HamSc07},
which emphasized full completeness theorems.  As also mentioned there, 
our previous categorical models (as well as the categorical models below) are related to the dialogue categories and dialogue chiralities of Paul-Andr\'e Melli\`es
\cite{PAM2,PAM3}, although our motivation arises from the proof theory of \MLLP.  

\vspace{1ex}
  
\begin{defn}[Polarized Categories]
\label{polcat}
{ \em
A {\em polarized   category  }  \ 
$( \langle \Cplus,  \Cminus \rangle, \widehat{\cC} )$
consists of the following data:
\begin{itemize}

\item A pair of monoidal categories   $(\Cplus, \otimes)$  and $(\Cminus, \wp)$ called 
 {\em positive} (resp. {\em negative})  . 
\item
A contravariant monoidal equivalence $(\ )^{\perp}$  of the two categories: 
$$(-)^{\perp}: (\Cplus)^{op} \stt{\cong}\Cminus$$

\item ({\em Polarity changing functors})
A pair of adjoint functors $\uparrow \ \dashv \ \downarrow $, where
$\uparrow : \Cplus \longrightarrow \Cminus$ and $\downarrow: \Cminus\longrightarrow \Cplus$. Diagrammatically:
$$\xymatrix{\Cplus  \ar@/^/[rr]^{\uparrow }
\ar@{}[rr]|{\bot} &    &    \ar@/^/[ll]^{\downarrow}
 \Cminus}$$

\item De Morgan duality for $\downarrow$ and $\uparrow$ (wrt the monoidal equivalence):
\begin{eqnarray*}
(\down{X})^\perp  \cong  \up{X^\perp}  & \mbox{and} & 
(\up{X})^\perp \cong  \down{X^\perp}  
\end{eqnarray*}
\item A bimodule \footnote{Intuitively, families of {\em nonfocused
proofs}} $\widehat{\cC}:   \Cplus^{op}\times \Cminus \rightarrow \Set $  satisfying that there is
a natural equivalence:  
\begin{eqnarray*}
\widehat{\cC}(P,N) &  \cong  & \Cplus(P,\downarrow N)
\hspace{2ex} \mbox{for all $P\in \Cplus, N\in\Cminus$.}  
\end{eqnarray*}
 This may
be written as a reversible ``rule": 
$$\begin{array}{c}
\widehat{\cC}(P,N)\\
\hline
\hline
\Cplus(P,\down{N})   
\end{array}\downarrow
$$
\end{itemize}
}
\end{defn}
\begin{rem}{\em  By duality, there is a dual  rule:
$$\begin{array}{c}
\Cminus(\up{P}, N) \\
\hline
\hline
  \widehat{\cC}(P,N)\\
\end{array}\uparrow
$$
In the language of distributors (profunctors), we are demanding that
$\widehat{\cC}$ be left and right representable \cite{joyal-catlab} and compatible
with $( \ )^\perp$.

}
\end{rem}
At the *-autonomous category level ($\otimes,\wp,(\ )^\perp$), the various coherence theorems in \cite{CHS06} guarantee  negation can be taken to be strictly
 involutive (along with strict monoidal structure for each monoidal product).  We have not, however, examined the more general question of strictness for polarized categories.

 \vspace{2ex}

 \begin{rem}[ The case where the
profunctor $\widehat{\cC}$ is  a hom functor $\cC(-,-)$ ] \label{prohom}{\em

A polarized category arises from
the following framework of adjoint functors
 $L \dashv \; \downarrow$ and $\uparrow \; \dashv R$
 with contravariant equivalence $(~)^\bot$, as shown in the following diagram:
 
  $$
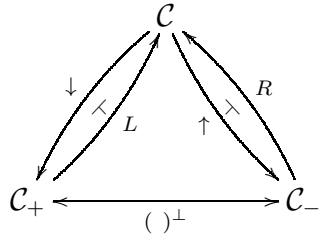
\xymatrix@C=3pc{                                                       
 &  \cC                                                               
 \ar@/_/[ddl]_-{\downarrow}               
 \ar@{}[ddl]|-{\rightthreetimes}   
 \ar@/_/[ddr]_-{\uparrow}               
 \ar@{}[ddr]|-{\leftthreetimes}                                         
 \\ \\  \Cplus                                                           
 \ar@/_/[uur]_-{L}  \ar@{<->}[rr]_{(~)^\bot}                                                  
 &     &  \Cminus                                                    
 \ar@/_/[uul]_-{R}    
}$$
These functors are subject to a natural isomorphism
$ ( \uparrow L (~) )^\bot \simeq \downarrow R (~)^\bot$
between two endofunctors on $\Cplus$.
In this framework, the polarity changing functors for Definition \ref{polcat} are defined 
by $\uparrow \Comp L$ and
$\downarrow \Comp R$:

 }\end{rem}
 
\vspace{2ex}

The following definition of compactness is sufficient for the purposes of this paper, although more general
 frameworks are possible.  
\begin{defn}[Compact Polarized  Category]{\em 
We say a polarized category is {\em compact}  if 
\begin{itemize}
\item $\Cplus^{op} = \Cminus $.
\item $(A\otimes B)^\perp =  B^\perp \otimes A^\perp$
for all objects $A, B$.
\end{itemize}
A compact polarized category is {\em degenerate}  if $\downarrow$ (equivalently $\uparrow$)
is the identity functor.   
}
\label{comp-pol}
\end{defn}

\vspace{1ex}

Examples of compact polarized categories include various categories of multipointed relations 
(arising in work of Hamano and Takemura \cite{HamTak08}), as well as various polarized Int-categories
arising from GoI, to be discussed below. The reader is referred to
Section \ref{rel-mp} for the above-mentioned category of multipointed
relations. An analogous approach to pointed relations is seen
in Ehrhard's category {\sf PpL} of preorders with projections \cite{Ehr11}. 

\begin{rem} 
{\em Although the notion of compact polarized category may appear to be ``degenerate"
in some informal  sense, nevertheless the notion is sufficiently robust to distinguish the two key
proofs in our paper  \cite{HamSc07} (Example 2.2). 
In other words,  compact polarized categories are adequate
to account for $\downarrow$-boxes in \MLLP.
These boxes are intrinsic for \MLLP, but not for the weaker
logic ${\sf MLL}^{\downarrow \uparrow}$ of \cite{OLaur02}.
}
\end{rem}

\smallskip
\subsection{The {\sf Int} construction}
\label{Int-construct}
The original connection of GoI to categories was realized by several researchers (e.g. M. Hyland and S. Abramsky)
as being related to the so-called \Int-construction in the original paper of Joyal-Street Verity \cite{JSV96}. 
 Further history and related notions are discussed
in  the paper of Abramsky, Haghverdi, and Scott \cite{AHS02}. 

Given a traced monoidal
category ${\cal C}$, we can define a compact closed category $\Int(\cal C)$
as follows: an object is a pair $(A^+,A^-)$ of ${\cal C}$-objects
and a morphism $f : (A^+,A^-) \longrightarrow (B^+,B^-)$ is a ${\cal C}$-morphism
$f: A^+ \otimes B^- \longrightarrow B^+ \otimes A^-$.
The composition of $\Int(\cal C)$ is defined by 
\begin{eqnarray}
g \Comp f & = & 
{\sf Tr}_{A^+ \otimes C^-,  \; C^+ \otimes A^-}^{B^-}
(
(C^+ \otimes s_{B^-, A^-}) \Comp (g \otimes A^-) \Comp
(B^+ \otimes s_{A^-, C^-}) \Comp (f \otimes C^-) \Comp
(A^+ \otimes s_{C^-, B^-})
) \qquad 
\label{comp-int}
\end{eqnarray}
This composition is shown in Figure 3 below, after Proposition \ref{int-P-cat}.   
An arrow $(A^+,A^-)\stt{f}(B^+,B^-) \in \Int(\cC)$ is  really an arrow $A^+ \otimes B^- \stt{f} B^+ \otimes A^-\in \cC$.
We picture the arrow  pointing upwards and denote it by a box, with the four objects at the corners. The domain $A^+ \otimes B^-$ is denoted by the lower edge,   the  codomain $B^+ \otimes A^-$  denoted by the upper edge.

\noindent
We would like to think of an $\Int({\cal C})$ morphism $f : (A^+,A^-) \longrightarrow (B^+,B^-)$ 
intuitively as a bidirectional data flow: a {\em pair} of arrows, one
from $A^+$ to $B^+$  and the other ``backwards" from $B^-$ to $A^-$.
Unfortunately, this view of $f$ is only heuristic; officially, $f$ is not a tensor of two maps
going in opposite directions, i.e. $f \not = g\otimes h $,
 where $g: A^+\rightarrow B^+$ and $h:B^-\rightarrow A^-$.
However, in the following subsections, we shall explicitly model this bidirectional
dataflow by using multipoints. 

\subsection{A polarized {\sf Int} construction} \label{PolInt}

In what follows in this subsection, let $\cC$ be a traced monoidal category 
$$({\cal C}, \otimes, I, s, 1, (\mathfrak{i},\mathfrak{r}), 0)$$
with $0$ morphisms (Axiom 8) and a 
distinguished object $1$  satisfying the retraction
$1 \otimes 1 \rhd_{(\mathfrak{i},r)} 1$ (Axiom 3)
of Definition \ref{GoIsitu} of the previous Section.
Moreover $\cC$ is supposed to satisfy
the following variations of the Lifting properties
(Axiom 6') 
and (Axiom 9) 
in which
$U$ can be taken to be {\em any} object $A$ of $\cC$
and $\beta$  can be {\em any}
morphism $1^m \longrightarrow A$:
\footnote{Note:  unlike in the previous Section, here $A$
is not assumed to have any retraction structure and we also assume
 $A \otimes 1^m \rhd_{(\mathfrak{i}_\beta, \mathfrak{r}_\beta)} A$ lifts
the $m$-fold retraction structure of $1 \otimes 1 \rhd_{(\mathfrak{i},r)} 1$ on
$1$.}

\smallskip

\noindent {\bf Axiom 6''\!' 
:(Lifting Property along $\beta$)} \\
For any object $A$ of $\cC$, any $m\in\Z^+$ and any morphism $\beta:1^m \longrightarrow A$,
there exists a retraction pair $A \otimes 1^m \rhd_{(\mathfrak{i}_\beta, \mathfrak{r}_\beta)} A$ lifting the
retraction $1^m \otimes 1^m \rhd_{(\mathfrak{i}^m, \mathfrak{r}^m)} 1^m$
along $\beta$:
$$
\xymatrix
@R=1.2pc{
A \otimes  1^m \ar@<1ex>[rr]^{\mathfrak{i}_{\beta}} &  &  \ar@<1ex>[ll]^{\mathfrak{r}_{\beta}}
 A\\ \\
1^m \otimes 1^m \ar[uu]^{\beta \otimes 1^m} \ar@<1ex>[rr]^{\mathfrak{i}^m}   &            &  
  \ar[uu]_{\beta} \ar@<1ex>[ll]^{\mathfrak{r}^m}  1^m  \\
}$$

\smallskip

\noindent {\bf Axiom 9''
} \\
{For any morphism
$f: V \otimes X_1 \longrightarrow W \otimes X_2$ 
with $X_i=A_i$ or $X_i= 1^{m_i}$ ($i=1,2$) for any non-zero
natural number $m_i$,
and any morphism $\beta_i : 1^{m_i} \longrightarrow A_i$,}
\begin{eqnarray*}
(W \otimes \mathfrak{i}_{\beta_2} ) \Comp (f \otimes 0_{1^{m_1},1^{m_2} }
) \Comp
(V \otimes \mathfrak{r}_{\beta_1})  = f  &\mbox{and} &   
(W \otimes \mathfrak{i} ) \Comp ( f \otimes 0_{1^{m_1},1^{m_2} } ) \Comp
(V \otimes \mathfrak{r} ) = f 
\end{eqnarray*}
See the following figure for the respective equations
when $X=A$ and $X=1$:


\begin{minipage}{0.45\hsize} 
$\begin{picture}(180,95)(-40,-60)
\put(30,-15){\fbox{\rule[.5in]{.4in}{0in }}}
\put(15,-38){$1^{m_1} $}   \put(69,-38){$1^{m_2} $}

\put(30,-40){\vector(1,0){35}} \put(43,-50){$0_{1^{m_1},1^{m_2}}$}

\put(12,18){$V$}    \put(70,18){$W$}
\put(41,0){$f$}
\put(12,-5){$A_1$} \put(70,-5){$A_2$}
\put(130, 0){$= \ \ f$}

\put(-15,-25){\vector(2,1){25}}  
\put(-15,-25){\vector(2,-1){25}}

\put(43,-30){$\otimes$}
\put(-30,-30){$A_1$}

\put(-5,-28){$\mathfrak{r}_{\beta_1}$}
\put(85,-27){$\mathfrak{i}_{\beta_2}$}
\put(12,-10){\vector(1,0){18}}
\put(66,12){\vector(1,0){25}}
\put(66,-10){\vector(1,0){10}}
\put(5,12){\vector(1,0){25}}


\put(78,-10){\vector(2,-1){25}}
\put(78,-40){\vector(2,1){25}}

\put(110,-30){$A_2$}
\end{picture}$
\end{minipage}
\begin{minipage}{0.1\hsize}
and 
\end{minipage}\hspace{-6ex}
\begin{minipage}{0.4\hsize} 
$\begin{picture}(180,95)(-40,-60)
\put(30,-15){\fbox{\rule[.5in]{.4in}{0in }}}
\put(15,-38){$1^{m_1}$}   \put(67,-38){$1^{m_2} $}

\put(30,-40){\vector(1,0){35}} \put(43,-50){$
0_{1^{m_1},1^{m_2}}
$}

\put(12,18){$V$}    \put(70,18){$W$}
\put(41,0){$f$}
\put(12,-5){$1^{m_1}$} \put(70,-5){$1^{m_2}$}
\put(130, 0){$= \ \ f$}

\put(-15,-25){\vector(2,1){25}}  
\put(-15,-25){\vector(2,-1){25}}

\put(43,-30){$\otimes$}
\put(-30,-30){$1^{m_1}$}

\put(-5,-28){$\mathfrak{r}^{m_1}$}
\put(85,-29){$\mathfrak{i}^{m_2}$}
\put(12,-10){\vector(1,0){18}}
\put(66,12){\vector(1,0){25}}
\put(66,-10){\vector(1,0){10}}
\put(5,12){\vector(1,0){25}}

\put(78,-10){\vector(2,-1){25}}
\put(78,-40){\vector(2,1){25}}

\put(110,-30){$1^{m_2}$}
\end{picture}$
\end{minipage}

\smallskip

\begin{notn}[morphism $f_{X_i, Y_j}$]{\em
For a morphism $f: X_0 \otimes X_1 \longrightarrow Y_0 \otimes Y_1$,
we denote by $f_{X_i, Y_j}: X_i \longrightarrow Y_j$
the following composition (where $(~)+1$ is mod $2$)):
$$\xymatrix{   
X_i \simeq X_i \otimes I \ar[rr]^(.4){X_i \otimes 0_{I,X_{i+1}}}
& &  X_{i} \otimes X_{i+1} \simeq X_0 \otimes X_1 \ar[r]^f & 
Y_0 \otimes Y_1 \simeq Y_{j} \otimes Y_{j+1}
\ar[rr]^(.6){Y_j \otimes 0_{Y_{j+1},I}} 
& &  Y_j \otimes I \simeq Y_j
}$$
\label{mor-fxy}
}\end{notn}

\smallskip
\begin{defn}[Positive category ${\sf Int}_{P}({\cal C})$]
\ {\em
The positive category  ${\sf Int}_{P}({\cal C})$ consists of the following data:
\begin{itemize}
\item objects are multipointed objects of ${\sf Int}_{P}({\cal C})$:
$$(A^{+}_{\alpha^+}, A^{-}_{\alpha^-})$$
where
$(A^{+}, A^{-})$ is an object of ${\sf Int}({\cal C})$
and for $* \in \{+,- \}$,
$\alpha^*$ is a  morphism
$\xymatrix{
1^{m^*}  \ar[r]^{\alpha^*}    & A^*
}$, whose domain is the $m^*$-ary tensor-folding of $1$.
Here  $m^*$ is a natural number associated to $\alpha^*$. 
We call $\alpha^*$ {\em a multipoint of} $A^*$.

\item morphisms are 3-tuples:
$$\xymatrix{
(A^{+}_{\alpha^+}, A^{-}_{\alpha^-}) \ar[rr]^{(f,f_+,f_-)}
&   & (B^{+}_{\beta^+}, B^{-}_{\beta^-})
}$$
where
\begin{itemize}
\item[-]
$f: (A^+, A^{-}) \longrightarrow (B^{+}, B^{-})$
is a morphism in ${\sf Int}({\cal C})$.

\item[-]
$f_+$ and $f_-$ are morphisms in ${\cal C}$
making the following respective diagrams commute 
(see Notation  \ref{mor-fxy}   ):

\begin{center}
\begin{minipage}{0.4\hsize} 
$\xymatrix
@C=0.3in
@R=1.2pc
{
A^+  \ar[rr]^{f_{A^+,B^+}}  &    &  B^+  \\  \\
1^{m^+}  \ar[uu]^{\alpha^+} \ar@[-][rr]_{f_{+}} 
 &  &  1^{n^+}  \ar[uu]_{\beta^+}  
}$
\end{minipage}
\begin{minipage}{0.4\hsize} 
$\xymatrix
@C=0.3in
@R=1.2pc
{
B^- \ar[rr]^{f_{B^-,A^-}}  &    &  A^- \\  \\
1^{n^-}  \ar[uu]^{\beta^-} \ar@[-][rr]_{f_{-}} 
 &  &  1^{m^-}  \ar[uu]_{\alpha^-}  
}$
\end{minipage}
\end{center}
\item[-]
$f$ makes the following diagram commute:
$$\xymatrix
@C=0.3in
@R=1pc
{
B^- \ar[rr]^{f_{B^-,B^+}}  &    &  B^+ \\  \\
1^{n^-}  \ar[uu]^{\beta^-} \ar@[-][rr]_{0} 
 &  &  1^{n^+}  \ar[uu]_{\beta^+}  
}$$
\end{itemize}
Diagrammatically a morphism $(f,f_+,f_{-})$ of ${\sf Int}_{P}({\cal C})$
is described as follows:
$$\xymatrix@C=0.4in@R=1pc
{
    B^+  \ar@{-}[rr]  &                                            &  A^- &   \\
                              & A^+ \ar@{-}[ul]  \ar@{-}'[rr]  \ar@{}[ru]|{f} 
                              &    
                               &  \ar@{-}[ul] B^- \\
    1^{n^+} \ar[uu]^{\beta^+}  &                                             & \ar[uu]_(0.3){\alpha^-} 1^{m^-}          \\
                              &  \ar[uu]_{\alpha^+} 1^{m^+}
      \ar[ul]^{f_+}   
                                &                 & 
                                \ar@{.>}[ulll]^0
                                1^{n^-} \ar[ul]_{f_-}  \ar[uu]_{\beta^-}  
}$$
\end{itemize}

}
\label{intposcat}
\end{defn}
In the above diagram, we say $f_+$ and $f_-$ {\em represent} the bidirectional
dataflow implicit in the upper arrow $f$.

\vspace{1ex}

\begin{prop}
 ${\sf Int}_{P}({\cal C})$ forms a monoidal category.
\label{int-P-cat}
\end{prop}

\begin{prf}{}
$\Id_{(A^{+}_{\alpha^+}, A^{-}_{\alpha^-})}$ is defined
by $(\Id_{(A^+,A^-)}, \Id_+, \Id_-)$ where $\Id_*$ is $\operatorname{Id}$ on the domain of $\alpha^*$
with $* \in \{ +,- \}$. 
Since the first element of the tuple is $\Id_{A^+} \otimes \Id_{A^-}$ in ${\cal C}$,
$\Id_{(A^{+}_{\alpha^+}, A^{-}_{\alpha^-})}$ belongs to ${\sf
 Int}_{P}({\cal C})$. 

It is immediate that the tensor product of ${\sf Int}({\cal C})$
restricts to a tensor product in ${\sf Int}_{P}({\cal C})$,
forming a monoidal subcategory.
We show that ${\sf Int}({\cal C})$ composition preserves
the positivity of morphisms. Recall the composition of two ${\sf Int}({\cal C})$ morphisms 
$(A^+, A^-) \stackrel{f}{\longrightarrow} (B^+, B^-)$
and $(B^+, B^-) \stackrel{g}{\longrightarrow} (C^+, C^-)$
given in Equation (\ref{comp-int}) and Figure 3 below.

The general ${\sf Int}_{P}({\cal C})$ composition with multipoints is shown in  Figure 5 below.
It represents the morphism  $(g,g_+,g_-)\Comp (f,f_+,f_-)$.
Here the top plane corresponds to the ordinary ${\sf Int}({\cal C})$ composition in Figure 3 below.
The bottom plane represents the analogous composition at the level of
multipoints,  where the composition coincides more specifically
with $(g_+ \Comp f_+, f_- \Comp g_-)$ by generalized 
yanking,
and is
illustrated separately in Figure 4 below.


\begin{figure}[!htbp] 
\begin{minipage}{0.5\hsize}
\label{compint}
\begin{center}
$$
{\tiny
\xymatrix@C=0.1in@R=0.4pc{ &&&
                                      &    &  A^{-}       &    &   B^- \ar@/^4pc/[dddddddddd]_{Tr^{B^-}(~)}   \\ \\
(C^+,C^-)  &&& C^+ \ar@{-}[rr] \ar@{-}[dd]     &  & B^- \ar[rruu]  &    & A^- \ar[lluu]\\ 
                  &&&                                               &    g & \\
(B^+,B^-) \ar[uu]^g     &&&  B^+ \ar@{-}[rr]     &     &   C^- \ar@{-}[uu]   &   &  A^- \ar[uu]  \\ \\
(B^+,B^-)  &&& B^+ \ar@{-}[rr] \ar@{-}[dd]     &  & A^- \ar[rruu]  &    & C^- \ar[lluu]\\ 
          &&&                                                         &     f& \\
(A^+,A^-) \ar[uu]^f           &&& A^+ \ar@{-}[rr]     &     &   B^- \ar@{-}[uu]   &   &  C^- \ar[uu]      \\ \\
                 &&&                           &      &   C^-   \ar[rruu]  &      &  B^- \ar[lluu]}
 } 
 $$
\end{center}
\caption{Composition  in $\Int({\cC})$}
\end{minipage}
\begin{minipage}{0.6\hsize}
\label{compintmp}
\begin{center}
$$
{\tiny
\xymatrix@C=0.1in@R=0.4pc{ &&&
                                      &    &  1^{m^-}      &    &
 1^{n^-} \ar@/^4pc/[dddddddddd]_{Tr^{1^{n^-}}(~)}   \\ \\
(1^{l^+},1^{l^-})  &&& 1^{l^+} \ar@{-}[rr]      &  & 1^{n^-} \ar[rruu]  &    & 1^{m^-} \ar[lluu]\\ 
                  &&&                                               &   & \\
(1^{n^+},1^{n^-}) \ar[uu]^{g_+\otimes g_{-}}     &&&  1^{n^+}\ar[uu]^{g_+} \ar@{-}[rr]     &     &   1^{l^-} \ar[uu]_{g_{-}}  &   &  1^{m^-} \ar[uu]  \\ \\
(1^{n^+},1^{n^-})  &&& 1^{n^+} \ar@{-}[rr]     &  & 1^{m^-} \ar[rruu]  &    & 1^{l^-} \ar[lluu]\\ 
          &&&                                                         & & \\
(1^{m^+},1^{m^-}) \ar[uu]^{  f_+\otimes f_{-}}           &&& 1^{m^+}\ar[uu]^{f_+} \ar@{-}[rr]     &     &   1^{n^-} \ar[uu]_{f_{-} }  &   &  1^{l^-} \ar[uu]      \\ \\
                 &&&                           &      &   1^{l^-}
 \ar[rruu]  &      &  1^{n^-} \ar[lluu] \\ 
&&  \ar@{}[r] |{\mbox{which is equal to}}      &  \\
& &
   \ar@{}[r] |{(g_+ \Comp f_+, \quad   f_- \Comp g_- )}   &
}
}
 $$
\end{center}
\caption{Composition in ${\sf Int}_{P}({\cal C})$ for multipoints}
\end{minipage}
\end{figure}
\noindent {Figures 3 and 4 are the upper and lower surfaces
of a 3-dimensional diagram pictured in Figure 5 below.
Consider the two central cubes in Figure 5. The top and bottom 
squares of these cubes compose because of Figures 3 and 4.  
The main question is the composition of the two left and, respectively, the two right
vertical faces of the two central cubes. The left vertical two squares 
obviously compose to form a commutative square.
The right (rearmost) vertical two squares in the cubes 
compose to be a commutative square
because of
generalized yanking in Appendix \ref{genyank}.}
\end{prf}


\vspace{2ex}

Obviously there is a forgetful functor
$|~~|:   {\sf Int}_{P}({\cal C})   \longrightarrow {\sf Int}({\cal C})$.
\bigskip
\begin{defn}[functor $\downarrow$]{\em
The functor \ $\downarrow :   {\sf Int}({\cal C}) \longrightarrow 
{\sf Int}_{P}({\cal C})$
is defined as follows:
}
\label{funcdown}
\end{defn}

\begin{itemize}
\item[-] On objects: \ $\down{(A^+, A^-)}:= ((A^+ \otimes 1)_1, (A^- \otimes 1)_1)$, \ 
where $(A^* \otimes 1)_1$ denotes \\ {\it adjoining the point }
$1\cong I\otimes 1\stt{0_{I,A^*}\otimes 1}A^*\otimes 1$, $*\in\{+,-\}$.

\item[-] On morphisms: \ for $f:(A^+, A^-) \longrightarrow (B^+, B^-)$, define
$$\down{f} :((A^+ \otimes 1)_1, (A^- \otimes 1)_1) \longrightarrow
((B^+ \otimes 1)_1, (B^- \otimes 1)_1)$$ as 
$\down{f}  := (s_{1,B^+} \otimes A^- \otimes 1) \Comp
(1 \otimes f \otimes 1) \Comp 
(s_{A^+,1} \otimes B^- \otimes 1)$ and $f_+$ and $f_-$ are
$\operatorname{Id}$'s
on $1$.

\noindent
 Diagrammatically, 
$$
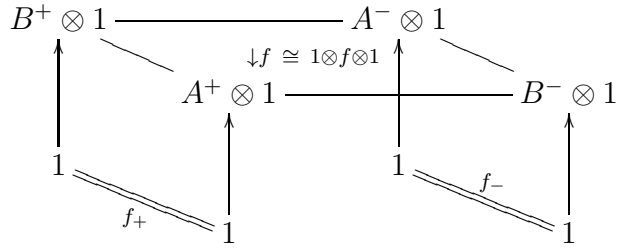
\xymatrix
@C=0.3in@R=1pc
{
    B^+ \otimes 1
\ar@{-}[rr]  &                                            &  A^-
	 \otimes 1
&   \\
                              & A^+ \otimes 1
 \ar@{-}[ul]
	 \ar@{-}'[rr]  
\ar@{}[ru]|{\downarrow f~\cong~ 1
\otimes f	 \otimes 1
}
                              &    
                               &  \ar@{-}[ul] B^- \otimes 1
\\
    1
\ar[uu]  &                                             & \ar[uu] 1          \\
                              &  \ar[uu]  1
  \ar@{=}^{f_+}[ul] 
                                &                 & 1
 \ar@{=}_{f_-}[ul]  \ar[uu]   
}$$

\end{itemize}


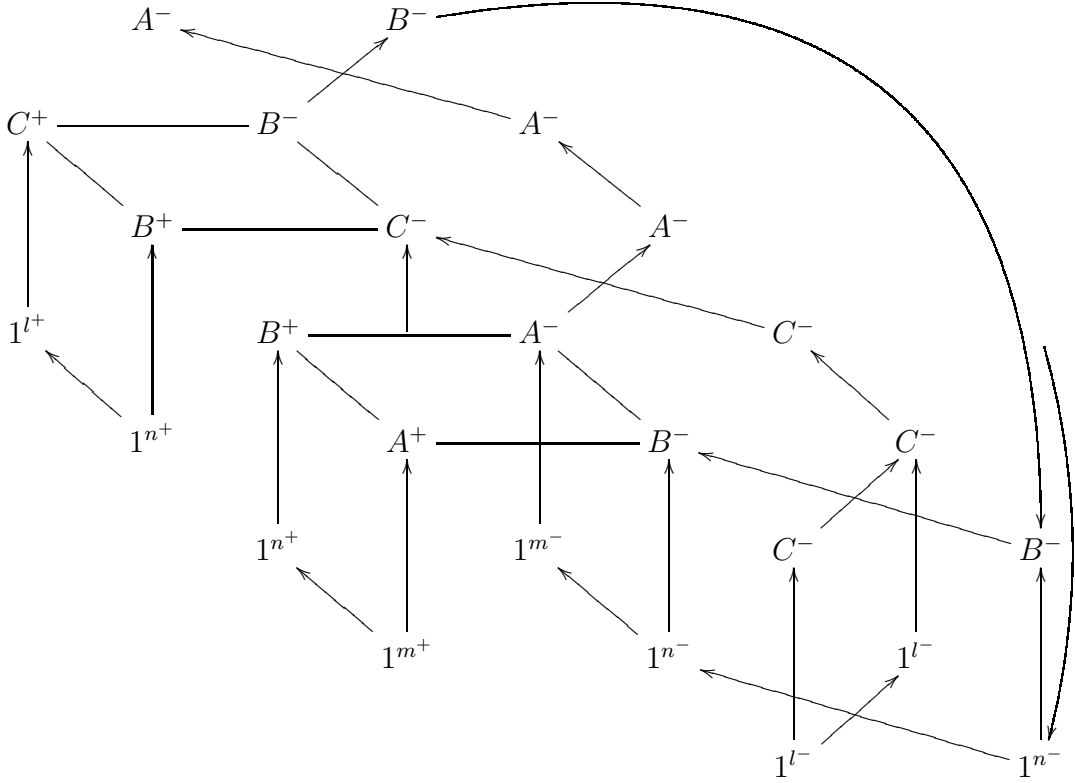
\begin{figure}[!tb]
\label{intPcomp}
$\xymatrix
{
                 &                     &   A^-  &   &   B^- \ar@/^8pc/[dddddrrrrr] \\
                     &  C^+ \ar@{-}[rr] 
 &         & B^- \ar[ur]  
&   & 
                      A^- \ar[ulll]          \\
    &   &   B^+ \ar@{-}[rr] \ar@{-}[ul]  &          &   C^- \ar@{-}[ul]
     &   &  A^- \ar[lu]  \\
    &        \ar[uu] 1^{l^+}     &     &   B^+  
\ar@{-}[rr]  
 & \ar[u]  &   A^-  \ar[ur]    
&   &  
    C^- \ar[ulll]  &    &   \ar@/^1pc/[dddd] \\
      &                                 &  1^{n^+} \ar[uu]  \ar[ul] &
 &   A^+  
\ar@{-}[ul]  
         \ar@{-}[rr]    &      &  B^- \ar@{-}[ul]   &   &  C^-  \ar[lu]\\
         &   &   &  \ar[uu] 1^{n^+}     &   &  \ar[uu]1^{m^-}  &   & C^- \ar[ur]  &   & B^- \ar[ulll] \\
         &   &   & &   \ar[lu] \ar[uu] 1^{m^+} &  &  \ar[lu] \ar[uu] 1^{n^-} &   & \ar[uu]  1^{l^-} \\
          & & & & & & & \ar[uu] 1^{l^-} \ar[ur]  &   &  1^{n^-} \ar[uu] 
\ar[ulll]
   	     		 }  
   $  
\begin{minipage}{\textwidth}
\caption[]{
Composition $(g,g_+,g_{-})\circ(f,f_+,f_{-})$ in ${\sf
 Int}_{P}({\cal C})$ 
 (the top plane is from Figure 3 and the
bottom plane is from Figure 4.) \footnotemark[13]
} 
\end{minipage}
\end{figure}
\footnotetext[13]{The rightmost feedback arrow on the lower level
maps $1^{n^-}$ (below the upper $B^-$) to $1^{n^-}$.}

\begin{prop}[adjunction] ~\\
\label{adjunction}
$\downarrow$ is right adjoint to the forgetful functor $|~~|$, i.e.
\begin{eqnarray}
{\sf Int} ({\cal C})( (A^+, A^-) , (B^+, B^-) ) & 
\cong & {\sf Int}_{P}({\cal C})
( (A^+_{\alpha^+}, A^-_{\alpha^-}) , ((B^+ \otimes 1)_1, 
(B^- \otimes 1)_1) ) 
\label{adj}
\end{eqnarray}
\end{prop}

\begin{prf}{}
Note first (\ref{adj}) gives the required adjunction,
because
$ (A^+, A^-) = |  (A^+_{\alpha^+}, A^-_{\alpha^-})  | $ and
$\downarrow \!\! (B^+, B^-) =  ((B^+ \otimes 1)_{1}, (B^- \otimes 1)_{1})$.
The natural bijection in (\ref{adj}) is given by: 
\smallskip

\begin{itemize}
\item {\bf (From right to left)}: \\  define
$\epsilon_{(B^+, B^-)}: \,        
\abs{\down{(B^+,B^-)}}=(B^+ \otimes 1, B^- \otimes 1 ) \longrightarrow 
(B^+,B^- )$~by:
$$\xymatrix{ \epsilon_{(B^+, B^-)}:
B^+ \otimes 1 \otimes B^- 
\ar[rr]^(.65){B^+ \otimes 0_1 \otimes  B^-}
& & 
B^+ \otimes 1 \otimes B^- 
\ar[rr]^(.5){B^+ \otimes s_{1,B^-}} & & 
B^+ \otimes B^- \otimes 1 }$$
modulo canonical associativity isomorphisms of $\otimes$.
Then $\epsilon$ is the  co-unit of the adjunction.
Given a morphism $g$ of the R.H.S,
we obtain a morphism of the  L.H.S. by composing with $\epsilon_{(B^+,
      B^-)}$. That is, by the composition of ${\sf Int} ({\cal C})$,
\begin{eqnarray}
\epsilon_{(B^+, B^-)} \Comp \abs{g} 
:= &
\TR{
\begin{array}{l}
(B^+ \otimes s_{B^- \otimes 1, A^-}) \Comp (\epsilon_{(B^+, B^-)}
\otimes A^-) \Comp \\
\quad \quad \quad (B^+ \otimes 1 \otimes s_{A^-,B^-})
\Comp (\abs{g} \otimes B^-)
\Comp (A^+ \otimes s_{B^-,B^- \otimes 1})
\end{array}
}{B^- \otimes 1}{A^+ \otimes B^-}{\, B^+ \otimes A^-} \nonumber \\
 = &  \TR{
((B^+ \otimes s_{1,A^-}) \Comp
(B^+ \otimes A^- \otimes 0_1) \Comp \abs{g})
\otimes B^-}{B^- \otimes 1}{A^+ \otimes B^-}{\, B^+ \otimes A^-}
\nonumber \\
 \stackrel{\mbox{\footnotesize vanishing}}{=}  & \TR{ (B^+ \otimes s_{1,A^-})
\Comp
(B^+ \otimes 0_1 \otimes A^-) \Comp \abs{g}}
{1}{A^+ \otimes B^-}{\, B^+ \otimes A^-} \label{rtol}
\end{eqnarray}


\smallskip

\item {\bf (From left to right):}  This is the part where certain commutativity
conditions   will be used (to compare the two layers).
 
Given a morphism $f : (A^+, A^-) \longrightarrow (B^+, B^-)$
in ${\sf Int} ({\cal C})( (A^+, A^-) , (B^+, B^-) )$, i.e.
$$\xymatrix@C=1pc@R=1pc
{B^+ \ar@{-}[rr]    &    &   A^- \\
                                  &    \ar@{-}[lu] A^+ \ar@{-}[rr] 
                                  \ar@{} [ru] |{f}
                                  &  &   B^- \ar@{-}[lu],
} $$
$f' : (A^+_{\alpha^+}, A^-_{\alpha^-}) \longrightarrow \, 
\downarrow \!\!(B^+, B^-) :=( (B^+ \otimes 1)_1, 
(B^- \otimes 1)_1 )$ 
is constructed by the following diagram:
$$\xymatrix
@C=0.2in@R=0.9pc
{ \tiny
B^+\otimes 1     \ar@{-}[rr]  &                                            &  A^- &   \\
                             &  A^+\otimes 1^{m^+} \ar[ul]   &  
                              &  \ar[ul]_{\mathfrak{i}_{\alpha^-}} 
                               A^- \otimes 1^{m^-}
        			 &                                              \\                          
   1  \ar[uu]^{1}   &                                             & \ar@{.>}[uu]_{\alpha^-} A^+ \ar[ul]_{\mathfrak{r}_{\alpha^+}}      &       
   &  \ar@{-}[ll]  B^- \ar[ul]\otimes 1  &         \\
                             &  \ar[uu]_{  \alpha^+ \otimes 1^{m^{+}} }   1^{m^+}\otimes 1^{m^+} \ar[ul]^{g_{2m^+}}      &                & 
                     1^{m^-}\otimes 1^{m^-}   \ar@{.>}[uu]_(0.3){ \alpha^- \otimes 1^{m^{-}} }   \ar[ul]^{\mathfrak{i}^{m^-}} 
     &         \\                
                    &    &   1^{m^+} \ar[uu]_(.4){\alpha^+} 
\ar[ul]^{\mathfrak{r}^{m^+}} 
&    
                    &  \ar[uu]_{1} 1  \ar[ul]^{h_{2m^-}}    }$$
{Note that in the diagram the domain $1$ of $\alpha^-$
is hidden, situated behind $A^+$.} \\[2ex]
In the diagram,
the upper (outer) square denotes the morphism $f'$ being constructed
and the parallelogram (inside the square) with the vertices\\
$A^+\otimes 1^{m^{+}}, B^-\otimes 1 , A^-\otimes 1^{m^-}, B^+\otimes1
$ denotes the following morphism \\
with $g_{m^+}: 1^{m^+} \longrightarrow 1 $ and $h_{m^-} : 1 \longrightarrow
      1^{m^-}$ of (\ref{gh}):
\begin{multline}
\lefteqn{(s_{1,B^+} \otimes A^- \otimes 1^{m^-}) \Comp
(g_{m^+} \otimes f \otimes h_{m^-}) \Comp 
(s_{A^+,1^{m^+}} \otimes B^- \otimes 1):} \\
 (A^+ \otimes 1^{m^+} ) \otimes (B^- \otimes 1  ) 
\longrightarrow
(B^+ \otimes 1 ) \otimes (A^- \otimes 1^{m^-} ) \label{parallelogram}
\end{multline}
The vertical square on the right front face and the left rear face
are lifting properties (Axiom 6''\!') over $\alpha^+$ and $\alpha^-$,  respectively. 
 Hence, as the upper surface of the diagram depicts,
\begin{eqnarray}
f':= &  
(B^+ \otimes 1 \otimes \, \mathfrak{i}_{\alpha^-}) \Comp \, (\ref{parallelogram}) \Comp
(\mathfrak{r}_{\alpha^+} \otimes B^- \otimes 1) \nonumber \\
= & (s_{1,B^+} \otimes \mathfrak{i}_{\alpha^-}) \Comp (g_{m^+} \otimes f \otimes
 h_{m^-}) \Comp (\mathfrak{r}_{\alpha^+} \otimes B^- \otimes 1) \label{ltor}
\end{eqnarray} 
In the bottom surface, $f'_+$ and $f'_-$ are constructed by 
composing the bottom arrows in the diagram.
The right and left faces are shown to be commutative
by virtue of the fact that the two morphisms $g_{2m} \Comp \,  \mathfrak{r}^{m}$  and 
$\mathfrak{i}^{m} \Comp \, h_{2m}$ give the retraction structure
$1^{m} \rhd 1$.

\end{itemize}

\noindent
Finally we show that when the $f'$ of (\ref{ltor})
is applied to the above ``from right to left" construction,
then the original $f$ is recovered.
$$\begin{array}{l}
\epsilon_{(B^+,B^-)} \Comp f'
\stackrel{(\ref{rtol})}{=}
\TR{ 
(B^+ \otimes s_{1,A^-}) \Comp
(B^+ \otimes 0_1 \otimes A^-) \Comp f'}{1}{A^+ \otimes B^-}{B^+
 \otimes A^-} \\
 = \hfill \mbox{modulo symmetry} \\
\TR{
\begin{array}{r} 
(B^+ \otimes \mathfrak{i}_{\alpha^-} \otimes 1)
\Comp 
(B^+ \otimes A^- \otimes s_{1,1^{m^-}})
\Comp
(s_{1,B^+ \otimes A^-} \otimes 1^{m^-})
\Comp  \\
\quad \quad \quad \quad (0_1 \Comp g_{m^+} \otimes f \otimes h_{m^-})
\Comp
(\mathfrak{r}_{\alpha^+} \otimes B^- \otimes 1)
\end{array}
}
{1}{A^+ \otimes B^-}{B^+ \otimes A^-}
\\
= \hfill \mbox{naturalities} 
\\
\begin{array}{l}
(B^+ \otimes \mathfrak{i}_{\alpha^-}) \Comp \\
\TR{
(B^+ \otimes A^- \otimes s_{1,1^{m^-}})
\Comp
(s_{1,B^+ \otimes A^-} \otimes 1^{m^-})
\Comp
(0_{1^{m^+},1} \otimes f \otimes h_{m^-})
}
{1}{ 1^{m^+}  \otimes A^+ \otimes B^-}{B^+
 \otimes A^- \otimes 1^{m^-}}  \\ 
\hfill \Comp  (\mathfrak{r}_{\alpha^+} \otimes B^-),
\end{array}
\end{array}$$
whose ${\sf Tr}$ part is
$$\begin{array}{l}
 \TR{
(B^+ \otimes A^- \otimes s_{1,1^{m^-}})
\Comp
( f \otimes 0_{1^{m^+},1} \otimes h_{m^-})
\Comp
(s_{1^{m^+}, A^+ \otimes B^-} \otimes 1)
}
{1}{ 1^{m^+}  \otimes A^+ \otimes B^-}{B^+
 \otimes A^- \otimes 1^{m^-}} \\  
= \hfill \mbox{naturality}  \\
 \TR{
(B^+ \otimes A^- \otimes s_{1,1^{m^-}})
\Comp
( f \otimes 0_{1^{m^+},1} \otimes h_{m^-})
}
{1}{ A^+ \otimes B^- \otimes 1^{m^+}}{B^+
 \otimes A^- \otimes 1^{m^-}}  
\Comp
s_{1^{m^+}, A^+ \otimes B^-}
\\
 = \hfill \mbox{superposing} \\  
 f \otimes 
\TR{
s_{1, 1^{m^-}} 
\Comp
(0_{1^{m^+},1} \otimes h_{m^-})}
{1}{1^{m^+}}{1^{m^-}}
\Comp s_{1^{m^+}, A^+ \otimes B^-}
\\
 = \hfill  \text{generalized yanking}  \\ 
 f \otimes 
((h_{m^-} \Comp \, 0_{1^{m^+},1})
\Comp \,  s_{1^{m^+}, A^+ \otimes B^-})
\quad = \quad  
f \otimes 
(0_{1^{m^+},1^{m^-}}
\Comp  \, s_{1^{m^+}, A^+ \otimes B^-} )
\hfill  \text{zero absorbing}  \\ 
\end{array}$$

\noindent
We conclude: \
$\epsilon_{(B^+,B^-)} \Comp f' = 
(B^+ \otimes \mathfrak{i}_{\alpha^-})  \Comp
( f \otimes 0_{1^{m^+},1^{m^-}} )
\Comp \, s_{1^{m^+}, A^+ \otimes B^-} \Comp
\, (\mathfrak{r}_{\alpha^+} \otimes B^-)$ 
 = $f$, \\ by Axiom 9''.
 \end{prf}

Recall that the duality $(~)^\bot$ of ${\sf Int}({\cC})$
is a contravariant endofunctor such that $(A^+,A^-)^\bot:=  (A^-,A^+)$
and $f^\bot:= s_{B^+, A^-} \Comp f \Comp s_{B^-, A^+}$ for
$f: (A^+,A^-) \longrightarrow (B^+,B^-)$.
This duality $(~)^\bot$ acts on ${\sf Int}_{P}({\cal C})$ to yield the following dual category
${\sf Int}_{N}({\cal C})$.

\vspace{2ex}

Recall that the duality $(~)^\bot$ of ${\sf Int}({\cC})$
is a contravariant endofunctor such that $(A^+,A^-)^\bot:=  (A^-,A^+)$
and $f^\bot:= s_{B^+, A^-} \Comp f \Comp s_{B^-, A^+}$ for
$f: (A^+,A^-) \longrightarrow (B^+,B^-)$.
This duality $(~)^\bot$ acts on ${\sf Int}_{P}({\cal C})$ to yield the following dual category
${\sf Int}_{N}({\cal C})$.

\vspace{1ex}

\begin{defn}[Negative category ${\sf Int}_{N}({\cal C})$] \ {\em
The negative category  ${\sf Int}_{N}({\cal C})$ consists of the following data:
\begin{itemize}
\item objects:  
those of ${\sf Int}_{P}({\cal C})$.
\item morphisms: 
those of ${\sf Int}_{P}({\cal C})$
but the last condition on $f$ is replaced by:
$$\xymatrix
@C=0.3in
@R=1.2pc
{
A^+  \ar[rr]^{f_{A^+,A^-}}  &    &   A^- \\  \\
1^{m^+}  \ar[uu]^{\alpha^+  } \ar@[-][rr]_{0} 
 &  &  1^{m^-}  \ar[uu]_{  \alpha^-}  
}$$
\end{itemize}
}
\end{defn}
Diagrammatically a morphism $(f,f_+,f_{-})$ of ${\sf Int}_{N}({\cal C})$
is described as follows:
$$\xymatrix@C=0.5in@R=1.3pc
{
    B^+  \ar@{-}[rr]  &                                            &  A^- &   \\
                              & A^+ \ar@{-}[ul]  \ar@{-}'[rr]  \ar@{}[ru]|{f} 
                              &    
                               &  \ar@{-}[ul] B^- \\
    1^{n^+} \ar[uu]^{\beta^+}  &                                             & \ar[uu]_(0.3){\alpha^-} 1^{m^-}          \\
                              &  \ar[uu]_{\alpha^+} 1^{m^+}
    \ar[ul]^{f_+}    
                                \ar@{.>}[ur]_0
                                &                 & 
                      1^{n^-} \ar[ul]_{f_-}  \ar[uu]_{\beta^-}  
}$$
Note that the 0 morphism occurring in the bottom level 
is antidiagonal to that of the $0$ morphism of ${\sf Int}_{P}({\cal C})$.


\smallskip
Hence the positive and the negative categories are contravariantly equivalent.
The functor $\uparrow : {\sf Int}({\cal C}) \longrightarrow
{\sf Int}_{P}({\cal C})$ is defined by
de Morgan duality $\uparrow \! (~):= (\downarrow (~~)^\bot)^\bot$.   

\bigskip


\noindent
Thus we obtain a compact polarized category (Definition \ref{comp-pol}),
in the style of  Remark \ref{prohom}:

\vspace{2ex}

\begin{thm}[A compact polarized category] \

\hspace{-4ex} $(\langle {\sf Int}_{P}({\cal C}), {\sf Int}_{N}({\cal C})\rangle, \widehat{{\sf Int}({\cal C})})$ is
a polarized category such that
$\downarrow$ (resp. $\uparrow$) is right (resp. left) adjoint to the forgetful
functor $| \ \ |$. The polarized category is compact so that
$({\sf Int}_{P}({\cal C}))^{op}={\sf Int}_{N}({\cal C})$. In diagrammatic form:


$$\xymatrix@C=3pc{                                                       
 & {\sf Int}({\cal C})                                                               
 \ar@/_/[ddl]_-{\downarrow}               
 \ar@{}[ddl]|-{\rightthreetimes}   
 \ar@/_/[ddr]_-{\uparrow}               
 \ar@{}[ddr]|-{\leftthreetimes}                                         
 \\ \\  {\sf Int}_{P}({\cal C})                                                             
 \ar@/_/[uur]_-{|~~|}  \ar@{<->}[rr]_{(~)^\bot}                                                  
 &     &  {\sf Int}_{N}({\cal C})                                                        
 \ar@/_/[uul]_-{|~~|}    
}$$
\end{thm}

\section{A polarized {\sf Int} construction using $\Rel$
with multipoints}
\label{rel-mp}

In this section we show how to build a concrete instance of the 
previous  polarized \Int-construction using $\Rel$
with multipoints to 
construct the associated commutativity conditions compatible with these multipoints.
  Thus we obtain a 
concrete compact polarized model of \MLLP.   

Let us make the following observations.
First, this section is a relational instance of the previous Section \ref{polcomp-int}
and the reader can read it independently of that.  Second,  in Sections \ref{MLLP} and \ref{polgoisit}
of this paper,  we often use the matrix formalism of Haghverdi's UDC's (see Appendix \ref{UDC}). This agrees with the usual matrix calculus in linear algebra.  
 In what follows, we adopt instead the matrix notation of Joyal-Street-Verity \cite{JSV96}
 for {\sf Int}(\Rel), since these authors do similar calculations to those below.   We introduce the 
following standard notions (\cf \cite{AHS02,HamSc07}).

\vspace{1ex}

\begin{notn} \label{notrel}  \
{\em
\begin{itemize}
\item For a relation $R: A \rightarrow B$, and subsets $X \subseteq A$
and $Y \subseteq B$,\\
$$\begin{array}{c}
[Y] R \, : = \, \{ x \mid \exists b \in Y . (x,b) \in R  \}  \subseteq A
 \qquad
R[X] \, : = \, \{ y \mid \exists a \in X . (a, y) \in R  \} \subseteq B
\end{array}$$
{We write $R^*$  for the smallest reflexive and transitive relation
      containing $R$. }

\item { (The category $\Rel$) \\
$\Rel$ denotes the category of sets and relations.
Relational composition of $R: A \longrightarrow B$
and $S: B \longrightarrow C$ is written from right to left.
 We write $SR: A\longrightarrow C$ where
$S R =\{  (x,z)\in A\times C \mid \exists y \in B . \ (x,y) \in R \ \, \mbox{and} \ \,
(y,z) \in S \}$
and omit the $\Comp$ symbol.
 $\Rel$ becomes monoidal with disjoint union $A + B$
of sets as the tensor product. The empty set $\emptyset$ is the tensor
      unit.}

\item
A morphism $R: (A^+, A^-) \rightarrow (B^+, B^-)$ of
{\sf Int(Rel)} is represented by the following matrix (where
the border objects represent appropriate domains and codomains
of the entries, as shown below:)
$$
\bordermatrix{ 
& A^+ & B^- \cr
B^+ & R_{12} & R_{22}\cr
A^{-} & R_{11} & R_{21}\cr
}
$$
\noindent The entries are relations
$R_{11}: A^+ \rightarrow A^-$, $R_{12}: A^+ \rightarrow B^+$,
$R_{21}: B^- \rightarrow A^-$, \\ $R_{22}: B^- \rightarrow B^+$.

E.g.  $\Id_{(A^+, A^-)}$ is represented by
$\begin{pmatrix}
1 & 0 \\
0 & 1 \\
\end{pmatrix}$
{where $1$ denotes the singleton $\{ * \}$.}

\item
The dual morphism $R^\perp: (B^-, B^+) \rightarrow (A^-, A^+)$
in ${\sf Int(Rel)}$ is represented by
\begin{eqnarray*}
R^\perp := 
\begin{pmatrix}
0 & 1 \\
1 & 0 \\
\end{pmatrix}
R 
\begin{pmatrix}
0 & 1 \\
1 & 0 \\
\end{pmatrix}
=
\begin{pmatrix}
R_{21} & R_{11} \\
R_{22} &  R_{12} \\
\end{pmatrix}
\end{eqnarray*}

\end{itemize}

}
\end{notn}

\begin{fact}[composition in ${\sf Int(Rel)}$] 
Given morphisms $R: (A^+, A^-) \rightarrow (B^+, B^-)$
and $S: (B^+, B^-) \rightarrow (C^+, C^-)$ of ${\sf Int(Rel)}$ represented by
\bordermatrix{ 
& A^+ & B^- \cr
B^+ & R_{12} & R_{22}\cr
A^{-} & R_{11} & R_{21}\cr} \ and \ 
\bordermatrix{ 
& B^+ & C^- \cr
C^+ & S_{12} & S_{22}\cr
B^{-} & S_{11} & S_{21}\cr
}, \\
the composition $S\Comp R$, in ${\sf Int(Rel)}$ is given by the
 following relation: 

\begin{minipage}{0.6\hsize}
\begin{multline*}
\bordermatrix{
 & A^+  & C ^- \cr
C^+ & \emptyset & S_{22} \cr
A^-  & R_{11} & \emptyset} \quad \cup \quad
\bordermatrix{
& A^+  & C ^- \cr
C^+ &  S_{12} (R_{22}S_{11})^* R_{12} &
S_{12} R_{22} (S_{11}R_{22})^* S_{21} \cr
A^-  & 
R_{21} S_{11}  (R_{22}S_{11})^* R_{12}
&  
R_{21} (S_{11} R_{22})^* S_{21}
\cr
}
\end{multline*}
\end{minipage}

\noindent
\mbox{\em which we can write as:}
\begin{minipage}[c]{0.4\hsize}
$$
\xymatrix
@C=0.2in
@R=1.2pc
{
\ar[dd]_{R_{11}}  A^+      \ar[rr]^{R_{12}}  &   &  B^+ \ar[rr]^{S_{12}}  \ar@/^/[dd]^{S_{11}}  &  &  C^+ \\ \\
A^-       &   &   \ar[ll]^{R_{21}}  B^- 
\ar@/^/[uu]^{R_{22}} 
&   &  \ar[ll]^{S_{21}}   C^- 
\ar[uu]_{S_{22}}
\ \ }
$$
\end{minipage}



\end{fact}

\bigskip

{In general in $\Rel$,
a multipoint $1^m\stt{} A$ is an $m$-indexed
family of subsets of $A$ (cf. Example \ref{mprelexpl}). 
In what follows, 
we denote a multipoint of a set $A$ by ${\sf mp}(A)$ and 
think of it
as a subset ${\sf mp}(A)\subseteq A$.}


\begin{defn}[Positive category {\sf Pos}] \
\begin{itemize}{\em 
\item Objects are $(A^+_{{\sf mp}(A^+)}, A^-_{{\sf mp}(A^-)} )$
where $(A^+, A^-)$ is an object of ${\sf Int(Rel)}$
and ${\sf mp}(A^+) \subseteq A^+$ and ${\sf mp}(A^-) \subseteq A^-$
are multipoints respectively of  $A^+$ and $A^-$.
\item Arrows (also called {\em positive maps}) are morphisms $R: (A^+,A^-)\rightarrow (B^+,B^-)$ in $\Int(\Rel)$ satisfying the following three conditions:
\begin{enumerate}
\item[1)]
$\begin{array}{ccc}
[{\sf mp}(B^+)] R_{12}  & =  & {\sf mp}(A^+)  
\end{array}$
\item[2)]
$
\begin{array}{ccc}
R_{21} [{\sf mp}(B^-)]  & =  &  {\sf mp}(A^-)   \\
\end{array}
$
\item[3)]
$\begin{array}{ccc}
[{\sf mp}(B^+)] R_{22}  & = \emptyset =  &  R_{22} [{\sf mp}(B^-)]  \\
\end{array}$
\end{enumerate}
}
\end{itemize}
 
\end{defn}

\noindent
{The above three conditions are instances of the three conditions
on morphisms of Definition \ref{intposcat}.
The following figure explains how 
the conditions 1) and 2) correspond to
the left and right vertical squares, respectively of that definition, and that 
the condition 3) corresponds to the diagonal $0$ morphism.}
$$
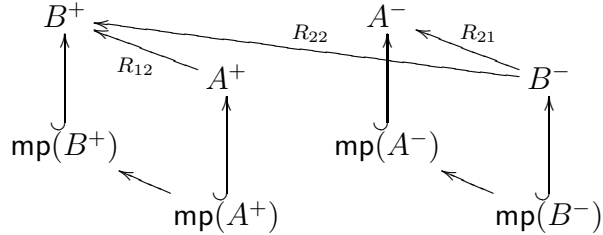
\xymatrix@C=0.2in@R=0.7pc{
   B^+  
    &                                            &  A^- &   \\
                             & A^+ \ar[ul]^{R_{12}}  
                             &    
                              &  \ar[ul]_{R_{21}} B^- \ar[ulll]_{R_{22}} \\
   {\sf mp}(B^+)  \ar@{^{(}->}[uu]  &                                             & 
   \ar@{^{(}->}[uu] 
   {\sf mp}(A^-) 
             \\
                             &  \ar@{^{(}->}[uu] {\sf mp}(A^+)  \ar[ul]
                               &                 &  {\sf mp}(B^-)  \ar[ul]
                                 \ar@{^{(}->}[uu]  
}$$


\begin{prop} 
  {\sf Pos} forms a category.
\end{prop}

\begin{prf}{}  \
$\operatorname{Id}$'s are {\sf Pos} maps since $\operatorname{Id}$ on $(A^+, A^-)$ is given by
the matrix s.t. $\Id_{12}=\Id_{A^+}$, $id_{21}=\Id_{A^-}$
and $\Id_{11}=\Id_{22}=\emptyset$. 
We check that the
composition of {\sf Int(Rel)} preserves {\sf Pos} maps. The
computation below is essentially a concrete instance 
of Figure 5.

Given two morphisms $R: (A^+, A^-) \rightarrow (B^+, B^-)$
and $S: (B^+, B^-) \rightarrow (C^+, C^-)$ of ${\sf Int(Rel)}$.
\begin{enumerate}
\item[1)] 
$\begin{array}[t]{ccl}
[{\sf mp}(C^+)] (SR)_{12} & = &
[{\sf mp}(C^+)] (S_{12} (R_{22}S_{11})^* R_{12}) \\
   & = & [ {\sf mp}(B^+)] ((R_{22}S_{11})^* R_{12}) \\
   & = & [ {\sf mp}(B^+) ] R_{12} \cup  [ {\sf mp}(B^+) ] ((R_{22}S_{11})
(R_{22}S_{11})^* R_{12}) \\
  & = &  {\sf mp}(A^+)  \cup \emptyset
\end{array}$

\item[2)]
$\begin{array}[t]{ccl}
(SR)_{21} [{\sf mp}(C^-)]  &  = & (R_{21} (S_{11} R_{22})^* S_{21})  [{\sf mp}(C^-)] \\
                                        & =  & (R_{21} (S_{11} R_{22})^* )  [{\sf mp}(B^-)] \\
                                        &= &  R_{21} [{\sf mp}(B^-)] 
                                        \cup (R_{21} (S_{11} R_{22})^* S_{11} ) (R_{22} [{\sf mp}(B^-)]) \\
                                        & = &  {\sf mp}(A^-) \cup \emptyset
\end{array}$

\item[3)]
$\begin{array}[t]{ccl}
[ {\sf mp}(C^+)] (SR)_{22} & = & [ {\sf mp}(C^+) ] S_{22} \cup
      [ {\sf mp}(C^+) ] (S_{12} R_{22} (S_{11}R_{22})^* S_{21} ) \\
     & = & \emptyset \cup
         [ {\sf mp}(B^+)] (R_{22} (S_{11}R_{22})^* S_{21} ) \\
     & = & \emptyset \cup \emptyset
\end{array}$

$\begin{array}{ccl}
(SR)_{22} [ {\sf mp}(C^-)] & = &  
S_{22} [ {\sf mp}(C^-)]  \cup
      (S_{12} R_{22} (S_{11}R_{22})^* S_{21} ) [ {\sf mp}(C^-) ] \\
  & = & \emptyset \cup (S_{12} R_{22} (S_{11}R_{22})^* ) [ {\sf mp}(B^-)]  \\
    & = & \emptyset \cup
(S_{12} R_{22}) [{\sf mp}(B^-) ]  \cup
(S_{12} R_{22} (S_{11}R_{22})^* S_{11}R_{22} ) [{\sf mp}(B^-)]  \\
  & = & \emptyset \cup \emptyset \cup \emptyset 
\end{array}$
\end{enumerate}
\end{prf}

\begin{prop}
The category {\sf Pos}   is monoidal
with respect to the tensor product of ${\sf Int(Rel)}$.
\end{prop}

\begin{prf}{}
Given {\sf Pos} maps $R: (A^+_{{\sf mp}(A^+)}, A^-_{{\sf mp}(A^-)} )
\rightarrow (B^+_{{\sf mp}(B^+)}, B^-_{{\sf mp}(B^-)} )$
and $S: 
(C^+_{{\sf mp}(C^+)}, C^-_{{\sf mp}(C^-)} )
\rightarrow (D^+_{{\sf mp}(D^+)}, D^-_{{\sf mp}(D^-)} )$,
the tensor product
$R \otimes S : ((A^+ + C^+)_{{\sf mp}(A^+) + {\sf mp}(C^+)},
(A^- + C^-)_{{\sf mp}(A^-) + {\sf mp}(C^-)})
\rightarrow 
( (B^+ + D^+)_{{\sf mp}(B^+) + {\sf mp}(D^+)},
(B^- + D^-)_{{\sf mp}(B^-) + {\sf mp}(D^-)})$
is given by:

$$\bordermatrix{ 
   &  A^+  & C^+ & B^- & D^- \cr
D^+ & \emptyset & S_{12} & \emptyset & S_{22} \cr
B^+ &   R_{12} & \emptyset & R_{22} & \emptyset \cr
C^- & \emptyset & S_{11} & \emptyset & S_{21} \cr
A^+ & R_{11} & \emptyset & R_{21} & \emptyset \cr
} $$It is straightforward that $\otimes$ preserves positivity.
\end{prf}

\vspace{1ex}

Dually in {\sf Int(Rel)},
negative maps are defined
so that they
form a monoidal category {\sf Neg}.

\vspace{1ex}

\begin{defn}[Negative category {\sf Neg} ] \
{\em
The objects of {\sf Neg} are the same as those of {\sf Pos} and
the morphisms ({\em negative maps}) are dual; that is, they satisfy the
 following three conditions:}
\end{defn}
\begin{minipage}{0.5\hsize}
\begin{enumerate}
\item[1)]
$\begin{array}{ccc}
R_{12} [{\sf mp}(A^+)]  & =  &  {\sf mp}(B^+)   
\end{array}$
\item[2)]
$
\begin{array}{ccc}
[{\sf mp}(A^-) ] R_{21}  & =  & {\sf mp}(B^-) \\
\end{array}
$
\item[3)]
$\begin{array}{ccc}
[{\sf mp}(A^-)] R_{11}  & = \emptyset =  &  R_{11} [{\sf mp}(A^+)]  \\
\end{array}$
\end{enumerate}
\end{minipage}
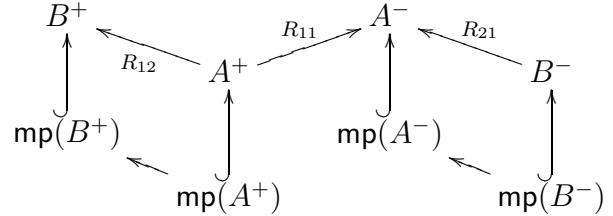
\begin{minipage}{0.5\hsize}
$$\xymatrix@C=0.2in@R=0.5pc{
   B^+  
    &                                            &  A^- &   \\
                             & A^+ \ar[ul]^{R_{12}}  
                             \ar[ru]^{R_{11}} 
                             &    
                              &  \ar[ul]_{R_{21}} B^- 
                              \\
   {\sf mp}(B^+)  \ar@{^{(}->}[uu]  &                                             & 
   \ar@{^{(}->}[uu] 
   {\sf mp}(A^-) 
             \\
                             &  \ar@{^{(}->}[uu] {\sf mp}(A^+)  \ar[ul]
                               &                 &  {\sf mp}(B^-)  \ar[ul]
                                 \ar@{^{(}->}[uu]  
}$$
\end{minipage}

\begin{prop}[negative category]
{\sf Neg} forms a monoidal category.
\end{prop}

\bigskip

\begin{rem}[positive $\not =$ negative] \
{\em {\sf Pos} maps and {\sf Neg} maps are different,
so the categories are different and the model is in this sense
non-degenerate.
 }
\end{rem}

\smallskip

The following functor  $\downarrow$ is a special instance  of the previously-defined
functor  in Definition \ref{funcdown}   and dually for $\uparrow$.

\begin{defn}[functors $\downarrow$ and $\uparrow$] \
Functors 
$\begin{array}{ccc}
\downarrow : {\sf Int(Rel)} & \longrightarrow & {\sf Pos} 
\end{array}$,
$\begin{array}{ccc}
\uparrow : {\sf Int(Rel)} & \longrightarrow & {\sf Neg} 
\end{array}$
are defined as follows:
\end{defn}
\begin{itemize}
\item On objects:
$\downarrow (A^+, A^-) := \uparrow (A^+, A^-):= 
((A^++1)_{1}, (A^-+1)_{1})$
\item On morphisms: For $R: (A^+, A^-) \rightarrow (B^+, B^-)$,
$\downarrow R$ and $\uparrow R$ are defined by
\begin{eqnarray*}
\down R  & := \up R & :=  
\bordermatrix{
  &  1 &  A^+   &   B^-  & 1 \cr
1  &  (\star , \star)  &  \emptyset & \emptyset & \emptyset &  \cr
B^+ & \emptyset   & R_{12} & R_{22} &
\emptyset \cr
A^- & \emptyset   & R_{11} & R_{21} & \emptyset \cr
1  &  \emptyset  &  \emptyset & \emptyset & (\star, \star) \cr}
\end{eqnarray*}
\end{itemize}
Note that the functors $\uparrow$ and $\downarrow$ are not full since
${\sf Pos}\not = {\sf Neg}$.

\begin{prop}[adjunctions]~~\\ 
$\downarrow$ (resp. $\uparrow$) is right (resp. left) adjoint to the forgetful
functor $| \, ~ \, |$:
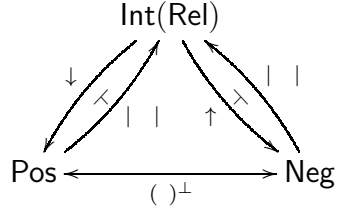
$$\xymatrix@C=1.5pc@R=1.5pc
{                                                       
& {\sf Int}({\sf Rel})                                                               
\ar@/_/[ddl]_-{\downarrow}               
\ar@{}[ddl]|-{\rightthreetimes}   
\ar@/_/[ddr]_-{\uparrow}               
\ar@{}[ddr]|-{\leftthreetimes}                                         
\\ \\  {\sf Pos} 
\ar@/_/[uur]_-{|~~|}  \ar@{<->}[rr]_{(~)^\bot}                                                  
&     & {\sf Neg}  
\ar@/_/[uul]_-{|~~|}    
}$$


\end{prop}

\begin{prf}{}
We show the following: 
\begin{eqnarray*}
\Int({\Rel})((A^+, A^-),
(B^+ , B^- )
) & \cong & {\sf Pos} ((A^+_{{\sf mp}({A^+})}, A^-_{{\sf mp}({A^-})} ),
((B^++1)_{1}, (B^-+1)_{1}) \, ) \\
& \cong & {\sf Neg} ( ((A^++1)_{1}, (A^-+1)_{1}),
(B^+_{{\sf mp}({B^+})}, B^-_{{\sf mp}({B^-})} ) )
\end{eqnarray*}
where
$((B^++1)_{1}, (B^-+1)_{1}) \, = \, \downarrow \! (B^+, B^- )$
and
$((A^++1)_{1}, (A^-+1)_{1}) \, = \,  \uparrow\! (A^+, A^- )$.


\noindent
Every positive map $R'$ from the R.H.S is
of the form 
$$
\bordermatrix{
  & A^+ & B^- &   1 \cr
1  &  {\sf mp}(A^+) \times 1     &  \emptyset    & \emptyset  \cr
B^{+} &  R_{12}   &  R_{22}   & \emptyset  \cr
A^{-} &  R_{11}  &  R_{21}   &   1 \times {\sf mp}(A^-)  \cr
} $$so that 
$\begin{array}{ccc}
R' & = & R \cup \;  {\sf mp}(A^+) \times 1 \; \cup  \; 1 \times {\sf mp}(A^-)
\end{array}$ with $R$ from the L.H.S.
This gives a natural bijective correspondence.
\end{prf}

\section{Conclusion and Future Work}In this paper we presented two independent studies of GoI for multiplicative polarized linear logic (\MLLP): one based
on the notion of GoI situations \cite{AHS02} and the other based on a
direct application of Joyal-Street-Verity's  {\sf Int} construction
\cite{JSV96}.  Both modellings use the idea of adjoining multipoints  to
account for polarities, hence focusing. In polarized GoI situations,
preservation of multipoints via the execution formula allows us to
characterize focusing semantically.  In the case of the {\sf Int}
construction, the goal was instead to construct compact polarized
denotational models.  This involved adding multipoints to the {\sf Int}
construction so as to be  compatible with those commutativity
conditions previously discussed.
 
 Finally, in the last section, we construct a concrete example of such a polarized category, based on the {\sf Int} construction applied to the category of multipointed relations.
For future studies, we leave open the following questions.
 
  (1)  What is the logical status of multipoints?   For example, multipoints have no counterpart in the syntax: they are an additional structure added to a nonpolarized (although somewhat ``degenerate") compact closed model. Yet multipoints allow us to characterize syntactic questions of polarized logic, for example, characterizing focusing.

(2)  One question of interest is how Sections 3 and 4 of this paper are
related.  We note that our main theorem characterizing focusing
(Theorem \ref{invari}) involves commutative squares
which can also be shown to be weak pullbacks. Thus weak pullbacks arise from the
termination of the execution formula given by traces. In Section 4, we
start from weak pullbacks in the definition of morphisms in
${\sf Int}_P({\cal}C)$  (below Definition \ref{intposcat}),
where we see the property that the two side
vertical faces are actually weak pullbacks.  The main results of Section
4 show these weak pullbacks are preserved not only under composition but
more strongly under tracing. We have used the fact that the squares
that arise in both sections are analogous.  We hope to give a
categorical characterization of such analogies.

   (3) This paper is restricted to the multiplicative fragment. It
   would be interesting to extend this to the full MALLP level, which is
   the language of Girard's Ludics, as studied in our paper
   \cite{HamSc07}. This seems more promising compared to nonpolarized
   additive models because the additive connectives are less complicated
   in the polarized setting, as mentioned in the work of O. Laurent (for
   example, \cite{OLaur02}). This future work may relate our work to
   Laurent's GoI model for additives \cite{OLaur01}.
 
 
 \vspace{2ex}

 \noindent
 {\bf Acknowledgement}  The authors would like to thank the referee 
 for detailed and helpful comments that have clarified many fine points
 of our exposition.

\section{Appendices}

\subsection{Appendix 1: GoI situations and Execution formulas for \MELL}
\label{appdx1}

In \cite{AHS02} the authors introduced a general algebraic framework for studying Girard's Geometry of Interaction (GoI) (\cite{Gi89,Gi95}) for  
multiplicative-exponential linear logic, \MELL\!\! .  This framework, called a  {\em GoI Situation}, contains an underlying traced monoidal category $\cC$, along with a reflexive object 
$U$  and an endofunctor $T$ used to represent the exponentials of linear logic. It is shown in that paper how to interpret GoI as yielding linear combinatory algebras on the endomorphism monoids  $End_{\cC}(U)$.

In later work (\cite{HS06,HS11}) the underlying traced category $\cC$
of a GoI situation was specialized to a traced Unique Decomposition Category (e.g. \Rel, \Pfn, \PInj) 
(see Appendix \ref{UDC} below) which is equipped with a standard {\em particle-style} trace, 
 as developed in Haghverdi's thesis \cite{Hag00}, together with an abstract  categorical execution formula given in terms of that trace (see below).   The categorical GoI interpretation of Haghverdi-Scott (which captures Girard's GoI\,1 \cite{Gi89})  contains three components:
 (i) an interpretation of proofs in $End_{\cC}(U)$, (ii) an interpretation of formulas
 as types (= bi-orthogonally closed subsets of $End_{\cC}(U)$, with respect to an appropriate
 Girard-Hyland-Schalk orthogonality), and (iii) an analysis of the dynamics of cut-elimination
 via the execution formula.  For the polarized system $\MLLP$ studied in this paper, we only discuss (i)
 and (iii).

\vspace{1ex}

\begin{defn} \label{usualGoI}
{\em
A {\em GoI Situation} is a triple $(\cC,T,U)$ where:
\begin{enumerate} \itemsep =2pt
\item $\cC$ is a traced symmetric monoidal category and
 $T : \cC \to \cC$ is a traced symmetric monoidal functor
with
the following {\em monoidal} retractions (i.e. the retraction pairs are monoidal natural transformations):
\begin{enumerate} \itemsep =2pt
\item   $e': T \rhd TT : e$ 
\,\,(Comultiplication)
\item $d': T \rhd \operatorname{Id} : d $  
\,\,(Dereliction)
\item $c': T \rhd T \otimes T : c$   
 \,\,(Contraction)
\item $w': T \rhd \cK_I : w$ 
\,\,(Weakening). Here $\cK_I$ is the
constant $I$
functor.
\end{enumerate}

\item $U$ is an object of $\cC$, called {\em a reflexive object}, with
retractions:

(a) $k: U \rhd U \otimes U : j$ 
\ \
(b) $U \rhd I$ 
, \ \ and
(c) $v: U \rhd TU : u$ 
\end{enumerate}
}
\end{defn}
\vspace{1ex}

Here \, $ e': T \rhd TT : e $  
means that  $e_X: TTX \to TX$ and $e'_X: TX \to TTX$ are monoidal natural  transformations such that
$e'e = \Id_{TT}$.  We say that  $TT$ {\em is a (monoidal) retract of} $T$. Similarly for the other items.

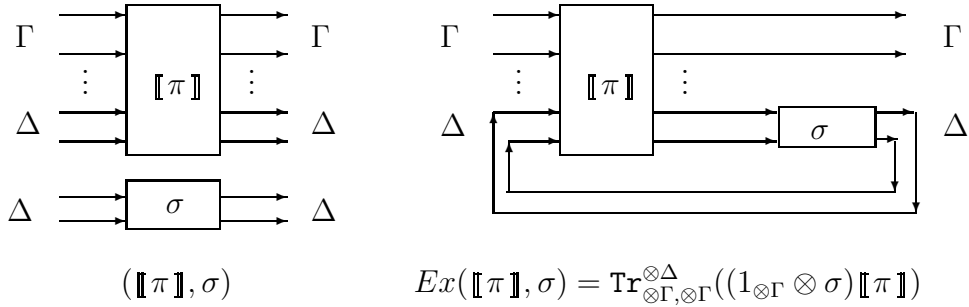
\begin{figure}[h]
\begin{picture}(50,130)(-60,-90)
\put(-12,18){$\Gamma$}
\put(-12,-15){$\Delta$}
\put(39,0){$\Mean{\pi}$}
\put(28,-75){$(\Mean{\pi},\sigma)$}
 \put(100,18){$\Gamma$}
\put(100,-15){$\Delta$}
\put(-15,-48){$\Delta$}
\put(100,-48){$\Delta$}
\put(13,0){$\vdots$}
\put(76,0){$\vdots$}
\put(5,15){\vector(1,0){25}}
\put(5,30){\vector(1,0){25}}
 \put(5,-7){\vector(1,0){25}}  
 \put(5,-18){\vector(1,0){25}}  
  \put(66,15){\vector(1,0){25}}
 \put(66,30){\vector(1,0){25}}
 \put(66,-7){\vector(1,0){25}}
\put(66,-18){\vector(1,0){25}}
 \put(30,-20){\fbox{\rule[.7in]{.4in}{0in }}}
\put(30,-48){\fbox{\rule[12pt]{.4in}{0in }}}
\put(66,-39){\vector(1,0){25}}
\put(66,-48){\vector(1,0){25}}
\put(5,-39){\vector(1,0){25}}  
 \put(5,-48){\vector(1,0){25}}  
\put(45,-45){$\sigma$}
\end{picture}
 \begin{picture}(50,70)(-170,-90)
\put(-15,18){$\Gamma$}
\put(39,0){$\Mean{\pi}$}
 \put(-15,-16){$\Delta$}
\put(175,18){$\Gamma$}
 \put(175,-16){$\Delta$}
 \put(125,-17){$\sigma$}
\put(-25,-75){$Ex(\Mean{\pi},\sigma) = \mathtt{Tr}^{\otimes\Delta}_{\otimes \Gamma, \otimes\Gamma}((1_{\otimes\Gamma}\otimes \sigma)\Mean{\pi})$}
\put(13,0){$\vdots$}
\put(76,0){$\vdots$}
\put(5,15){\vector(1,0){25}}
\put(5,30){\vector(1,0){25}}
 \put(5,-7){\vector(1,0){25}}  
 \put(11,-18){\vector(1,0){19}}  
  \put(66,15){\vector(1,0){95}}
\put(150,-7){\vector(1,0){14}}
\put(150,-17){\vector(1,0){8}}
 \put(66,30){\vector(1,0){95}}
 \put(66,-7){\vector(1,0){46}}
 \put(11,-37){\vector(0,1){19}}
 \put(165,-45){\line(-1,0){160}}
\put(5,-45){\vector(0,1){38}}
\put(157,-37){\line(-1,0){147}}
 \put(165,-7){\vector(0,-1){38}}
\put(66,-18){\vector(1,0){46}}
\put(157,-18){\vector(0,-1){19}}
 \put(30,-20){\fbox{\rule[.7in]{.4in}{0in }}}
\put(113,-17){\fbox{\rule[9pt]{30pt}{0in }}}
\end{picture}
\caption{Proofs of \ $\vdash [\Delta],\Gamma $ as I/O Boxes and the Execution Formula}
 \label{morgraph}
\end{figure}
 Following the presentation in \cite{HS06,HS11}, given a GoI situation $(\cC, T, U)$, the GoI interpretation of a proof $\pi$ of an \MELL\, sequent (with explicit cuts) 
 $\vdash [\Delta],\Gamma$ (where $\Delta$
 denotes the set of all pairs of cut formulas $A, A^\perp$ used in $\pi$) is determined
by a pair of morphisms $(\Mean{\pi},\sigma)$ as in Figure \ref{morgraph}, where $\sigma$
represents the cuts $\Delta$.  If $|\Delta| = 2m$ and $|\Gamma| = n$, these data are given by $\cC$-arrows, $\sigma: U^{2m}\rightarrow U^{2m}$, $\Mean{\pi}: U^{n+2m}\rightarrow U^{n+2m}$.  Finally,
Girard's {\em Execution Formula} determines  an arrow
${\sf Ex}(\Mean{\pi},\sigma): U^n\rightarrow U^n$, where $U^k = U\otimes \cdots \otimes U$ ($k$ times).  If $\cC$ is a Haghverdi traced Unique Decomposition Category (UDC) with a standard (particle-style) trace
(as in the $\Rel$-based models in this paper:  see Appendix \ref{UDC} below)
we can write the Execution Formula in the more familiar form
\begin{eqnarray}
{\sf Ex}(\Mean{\pi},\sigma) & = & \pi_{11} + \sum_{n\geq
 0}\pi_{12}(\sigma\pi_{22})^n(\sigma\pi_{21})  \label{exudc}
\end{eqnarray}
where $[\pi_{ij}]$ is the matrix representation of $\Mean{\pi}$; 
this was shown in \cite{HS06} to agree with Girard's original execution
formula \cite{Gi89} in his model {\sf Hilb}$_2$ (= $\ell_2[\PInj]$).  Such UDC models
also support a robust matrix calculus to represent morphisms, which agrees with the 
usual matrix representation of relations used in this paper (see Proposition \ref{matrix} below). 

\subsection{Appendix 2: Unique Decomposition Categories (UDCs)}
\label{UDC}   E. Haghverdi, in his thesis \cite{Hag00}, introduced {\em Unique Decomposition
Categories} (UDCs). These were specifically developed for modelling ``particle-style" GoI as in GoI\,1 \cite{Gi89,HS06}.

Briefly, UDCs are symmetric monoidal categories with the following additional structure:
\begin{itemize}
\item The homsets
are enriched with a $\Sigma$-monoid additive structure, such that composition distributes over addition, both from the left and the right.  For the precise
$\Sigma$-monoid axioms, we refer to Haghverdi's thesis, Chapter 4.   In particular, there are zero morphisms
$0_{XY}: X\rightarrow Y$ between any two objects $X,Y$.
\item  For a finite set $I$ and for each $j\in I$, there are {\em quasi injections}
$\iota_j: X_j\rightarrow \otimes_I X_i$ , and {\em quasi projections} $\rho_j: \otimes_IX_i \rightarrow X_j$
such that:
\begin{enumerate}\itemsep=1ex
\item $\rho_k \iota_j = \Id_{X_j}$ if $j = k$   and  $0_{X_j X_k}$ otherwise. 
\item $\sum_{i\in I} \iota_i\rho_i = \Id_{\otimes_I X_i}$.
\end{enumerate}
\end{itemize}

\noindent
Examples of UDC's (for Geometry of Interaction) include variations of $\Rel_+$, for example:
the categories {\bf \sf Pfn} and {\bf \sf PInj}  of {\em partial functions} (resp.  {\em partial injective functions}).

The main theorem on UDC's, which is used in various places in this paper, is the representation of morphisms
as matrices, with an associated full matrix calculus for computations.  This can be summarized as follows
(see Haghverdi \cite{Hag00}, Prop. 4.0.6):

\begin{prop}[Matricial Representation]  \label{matrix}Given a morphism
$f: \otimes_JX_j\rightarrow \otimes_IY_i$ in a UDC, with $|I| = m, |J| = n$, there exists a unique family
$\{f_{ij}\}_{i\in I, j\in J}:  X_j\rightarrow Y_i$
with $f = \sum_{i\in I,j\in J}\iota_i f_{ij} \rho_j$, where $f_{ij} = \rho_i f \iota_j .$  Moreover, composition of 
morphisms in a UDC corresponds to matrix multiplication of their  associated matrices.

\end{prop}

\subsection{Appendix 3: Generalized Yanking
for a traced monoidal category} 
\label{genyank}

The following identity is frequently used in calculating traces (see Proposition 2.4, in
\cite{AHS02}),  $${\sf Tr}_{X,Y}^U ( s \Comp (f \otimes g)  ) =  g \Comp f$$   Pictorially,
this says:
$$\begin{picture}(20,70)(80,-30)
\put(10,20){X}
\put(43,10){$f$}
\put(43, -12){$g$}
\put(10,-20){U}
\put(70,20){U}
\put(115,20){Y}
\put(70,-20){Y}
\put(150,0){\large =}
\put(180,0){$X\stt{f}U\stt{g}Y$}
\put(30,1){\line(1,0){36}}
\put(0,12){\vector(1,0){30}}
\put(0,-7){\vector(1,0){30}}
\put(66,12){\vector(1,0){22}}
\put(66,-7){\vector(1,0){22}}
\put(108,13){\vector(1,0){20}}
\put(108,-8){\vector(1,0){20}}
\put(88,12){\vector(1,-1){20}}
\put(88,-7){\vector(1,1){20}}
\put(128,-8){\line(0,-1){22}}
\put(0,-30){\vector(0,1){23}}
\put(128,-30){\line(-1,0){128}}
\put(30,-15){\fbox{\rule[.5in]{.4in}{0in }}}
\end{picture}
$$

\subsection{Appendix 4: Omitted Proofs} \label{REP}

\smallskip

\smallskip
\noindent {\bf Proof of Proposition \ref{mplifting}}  \\
By induction on the construction of a multipoint $\bm{p}$. \\
(Base Case): This is when $\bm{p}$ is the distinguished point $\alpha$,
hence the property is the original lifting property 6'.

\noindent (Induction Case): $\bm{p}$ is constructed either by 2 or 3
of Definition \ref{mpclass} (Note in 2, the $\bm{p}$ is a point.):
We define
$$\begin{array}{c|c}
\mbox{$\bm{p}$ constructed by 3} & 
\mbox{$\bm{p}$ constructed by 2} \\ \hline
\begin{array}{c}
\mathfrak{r}_{\bm{p}}:= ((j_m \, \Comp \, \tau) \otimes  1^{m} ) \, \Comp \,  
\bigotimes_i \mathfrak{r}_{\bm{p}_i}  \, \Comp \,  \tau^{-} \, \Comp \,  k_m \\
\mathfrak{i}_{\bm{p}}:=j_m \, \Comp \, \tau \, \Comp \,
  \bigotimes_i \mathfrak{i}_{\bm{p}_i}  \, \Comp
 \,  
(k_m  \otimes 1^{m} )
\end{array}
&
\begin{array}{c}
\mathfrak{r}_{\bm{p}}:= (j \otimes  1 ) \, \Comp \,  
(U \otimes \mathfrak{r}_{\bm{p'}})  \, \Comp \, k \\
\mathfrak{i}_{\bm{p}}:=j \, \Comp \,  (U \otimes \mathfrak{i}_{\bm{p'}})  \, \Comp
 \,  
(k  \otimes 1 )
\end{array}
\end{array}$$
Then the commutativity  property follows from
I.H.'s (the lower square of the following diagram)
and the retractions $U \rhd U^{m}$ and $U \otimes 1 \rhd U^m \otimes
1^{m}$ with $k$  either $m_1+m_2$ or $1$
by Axiom 2  (the upper right and left vertical arrows, respectively).
 Note especially that the lower square of the left case is the $m$-fold tensoring
of Axiom 6:

\smallskip

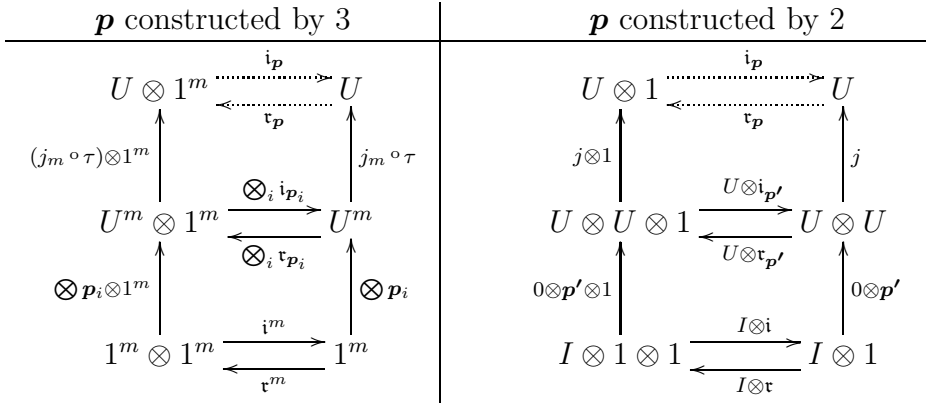
\begin{tabular}{c|cc}
$\bm{p}$ constructed by 3 &  &  $\bm{p}$ constructed by 2  \\ \hline
$ \xymatrix
@C=0.2in 
@R=0.2in
{
U \otimes 1^{m} 
\ar@{.>}@<1ex>[rr]^(0.6){\mathfrak{i}_{\bm{p}}}
&  &
\ar@{.>}@<1ex>[ll]^(0.4){\mathfrak{r}_{\bm{p}}}
U 
\\ \\
\ar[uu]^{(j_m \, \Comp \, \tau) \otimes 1^{m}}
U^m \otimes  1^{m} \ar@<1ex>[rr]^(0.6){\bigotimes_i \mathfrak{i}_{\bm{p}_i}} &  &
 \ar@<1ex>[ll]^(0.4){ \bigotimes_i \mathfrak{r}_{\bm{p}_i}  }
U^m
\ar[uu]_{j_m \, \Comp \, \tau}
\\ \\
1^{m}
 \otimes 1^{m}
\ar[uu]^{\bigotimes \bm{p}_i \otimes 1^{m}
} \ar@<1ex>[rr]^(0.6){\mathfrak{i}^{m}  }   &            &  
  \ar[uu]_{\bigotimes \bm{p}_i} \ar@<1ex>[ll]^(0.4){\mathfrak{r}^{m}
} 1^{m}
  \\ 
}
$
& \hspace{0.5cm} &
$
\xymatrix
@C=0.2in 
@R=0.2in
{U \otimes 1 
\ar@{.>}@<1ex>[rr]^(0.6){\mathfrak{i}_{\bm{p}}}
&   &
\ar@{.>}@<1ex>[ll]^(0.4){\mathfrak{r}_{\bm{p}}}
U 
\\ \\
\ar[uu]^{j \otimes 1}
U \otimes U \otimes  1 \ar@<1ex>[rr]^(0.6){
U \otimes \mathfrak{i}_{\bm{p'}}  } &  &
 \ar@<1ex>[ll]^(0.4){U \otimes \mathfrak{r}_{\bm{p'}}  }
U \otimes U 
\ar[uu]_{j}
\\ \\
I \otimes 1
 \otimes 1
\ar[uu]^{0 \otimes \bm{p'}\otimes 1
} \ar@<1ex>[rr]^(0.6){I \otimes \mathfrak{i}    }   &            &  
  \ar[uu]_{0 \otimes \bm{p'}} \ar@<1ex>[ll]^(0.4){I \otimes \mathfrak{r} 
} I \otimes 1
  \\
}
$ 
\end{tabular}

\smallskip
We check one commutativity for each of the constructions 2 and 3 of
 $\bm{p}$.   
 $$\begin{array}{c|c|c}
\mbox{$\bm{p}$ constructed by 3} & \mbox{$\bm{p}$ constructed by 2} \\
   \hline  & \\
\footnotesize
\begin{array}{lr}
\mathfrak{r}_{\bm{p}} \Comp j_m \Comp \tau \, \Comp \otimes \bm{p}_i \Comp \, \mathfrak{i}^{m} \\ = 
(j_m \otimes  1^{m} ) \, \Comp \,  
\otimes \mathfrak{r}_{\bm{p}_i}  \, \Comp \, \tau^{-} \, \Comp \,  k_m 
\, \Comp 
j_m \Comp \tau \\
\hfill \Comp \,  \otimes \bm{p}_i \, \Comp \,  \mathfrak{i}^{m} \\
 =
(j_m \otimes  1^{m} ) \, \Comp \,  
\otimes \mathfrak{r}_{\bm{p}_i} 
\Comp \,  \otimes \bm{p}_i \, \Comp \,  \mathfrak{i}^{m}  
& 
 \\ \\
 =
(j_m \otimes  1^{m} ) \, \Comp \,  
(\otimes \bm{p}_i \otimes 1^{m} )
 & 
\\
 = \bm{p} \otimes 1^{m} & 
\end{array}
& 
\footnotesize
\begin{array}{lr}
\mathfrak{r}_{\bm{p}} \Comp j \Comp (0 \otimes \bm{p'} ) \Comp (I \otimes \mathfrak{i} ) \\
= 
(j \otimes  1 ) \, \Comp \,  
(U \otimes \mathfrak{r}_{\bm{p'}})  \, \Comp \, k 
\, \Comp 
j  \\
\hfill
 \Comp \,  (0 \otimes \bm{p'}) \, \Comp \,  (I \otimes \mathfrak{i}) \\
 =
(j \otimes  1 ) \, \Comp \,  
(U \otimes \mathfrak{r}_{\bm{p'}})  \\
\hfill
 \Comp \,  (0 \otimes \bm{p'}) \, \Comp \,  (I
\otimes \mathfrak{i}) 
&  
 \\
 =
(j \otimes  1 ) \, \Comp \,  
(0 \otimes \bm{p'} \otimes 1) 
 &  
\\
 = \bm{p} \otimes 1 & 
\end{array}
& 
\footnotesize
\begin{array}{l}
  \\ \\  
  \mbox{We use the retractions:}
  \\
\mbox{In 3: $U \rhd_{(k_m,j_m)} U^m$} 
\\
\mbox{In 2: $U \rhd U^2$} \\  
\mbox{by commutativity of lower square}    \\ 
\mbox{by definition of $\bm{p}$} 
\end{array}
\end{array}$$

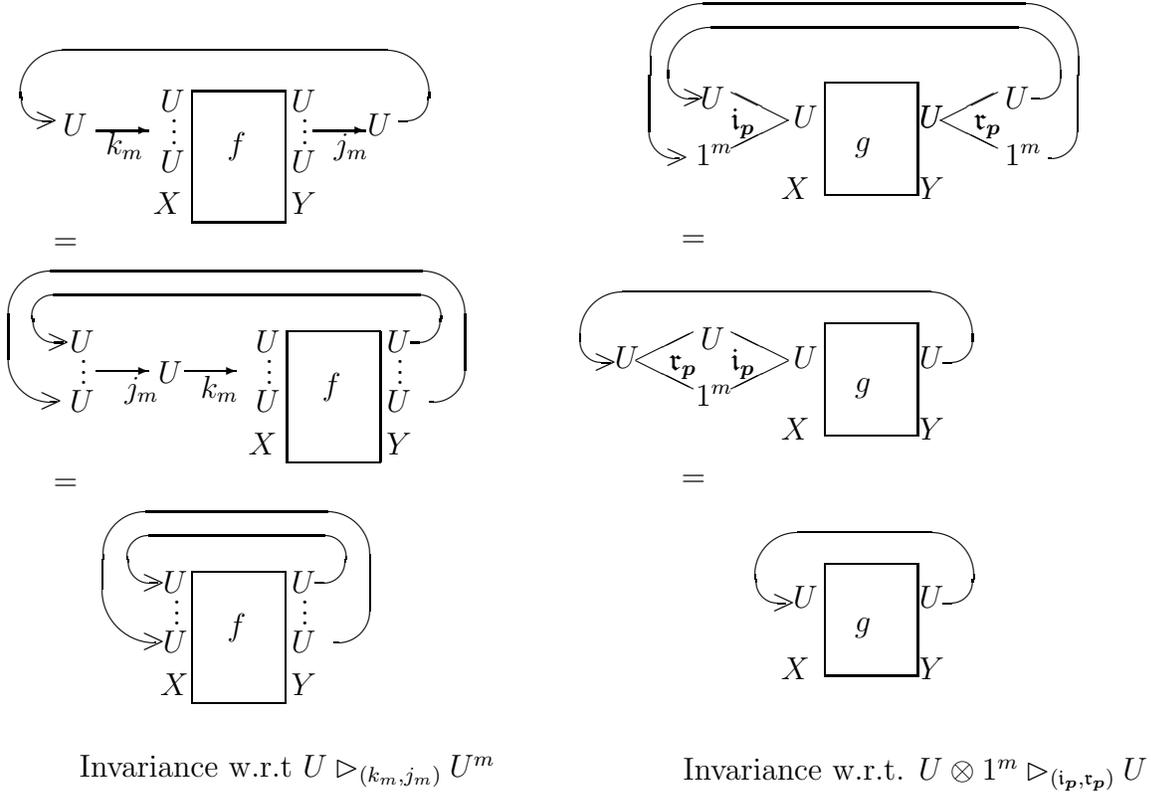
\begin{figure}[!hbt]
\begin{minipage}{0.5\hsize} 
\setlength{\unitlength}{0.9pt} 
\begin{picture}(200,100)(-60,-50)
\put(28,-25){\fbox{\rule[.6in]{.4in}{0in }}}
\put(15,18){$U$}    \put(70,18){$U$}
\put(15,-6){$\stackrel{\vdots}{U}$}    \put(70,-6){$\stackrel{\vdots}{U}$}
\put(43,0){$f$}
\put(12,-25){$X$} \put(70,-25){$Y$}
\put(-26,8){$U$}
\put(-12,10){\vector(1,0){22}} \put(-8,0){$k_m$}
\put(79,10){\vector(1,0){22}}  \put(88,0){$j_m$}
\put(102,8){$U$}

\put(42,24){\oval(172,40)[t]}   
\put(115,24){\oval(25,20)[rb]}  
\put(-31,24){\oval(25,20)[lb]}  
\put(-38,11){$>$}                 

\put(-30,-40){$=$}

\end{picture}
\begin{picture}(300,100)(-100,-50)
\put(28,-25){\fbox{\rule[.6in]{.4in}{0in }}}
\put(15,18){$U$}    \put(70,18){$U$}
\put(15,-6){$\stackrel{\vdots}{U}$}  \put(70,-6){$\stackrel{\vdots}{U}$}
\put(43,0){$f$}
\put(12,-25){$X$} \put(70,-25){$Y$}
\put(-26,5){$U$}
\put(-15,10){\vector(1,0){22}} \put(-8,-1){$k_m$}



 \put(-63,18){$U$}
  \put(-63,-6){$\stackrel{\vdots}{U}$}

\put(-52,10){\vector(1,0){22}}  \put(-40,-1){$j_m$}

\put(7,32){\oval(192,40)[t]}   
\put(88,32){\oval(30,70)[rb]}  
\put(-69,32){\oval(40,70)[lb]}  
\put(-76,-6){$>$}                 

\put(7,32){\oval(172,20)[t]}   
\put(80,32){\oval(25,20)[rb]}  
\put(-66,32){\oval(25,20)[lb]}  
\put(-73,19){$>$}                 

\put(-70,-40){$=$}

\end{picture}
\begin{picture}(300,80)(-60,-30)
\put(28,-25){\fbox{\rule[.6in]{.4in}{0in }}}
\put(16,18){$U$}    \put(70,18){$U$}
\put(16,-6){$\stackrel{\vdots}{U}$}    \put(70,-6){$\stackrel{\vdots}{U}$}
\put(43,0){$f$}
\put(15,-25){$X$} \put(70,-25){$Y$}

\put(47,32){\oval(112,40)[t]}   
\put(88,32){\oval(30,70)[rb]}  
\put(13,32){\oval(45,70)[lb]}  
\put(7,-6){$>$}                 

\put(47,32){\oval(92,20)[t]}   
\put(80,32){\oval(25,20)[rb]}  
\put(14,32){\oval(25,20)[lb]}  
\put(7,19){$>$}                 
\end{picture}
$$\mbox{Invariance w.r.t $U \rhd_{(k_m, j_m)}
U^m$
}$$
\end{minipage}
\begin{minipage}{0.5\hsize}
\setlength{\unitlength}{0.9pt} 
\begin{picture}(200,100)(-60,-50)
\put(30,-15){\fbox{\rule[.5in]{.4in}{0in }}}

\put(-22,18){$U$}   
\put(-24,-6){$1^m$}

\put(43,0){$g$}
\put(12,-20){$X$} \put(70,-20){$Y$}

\put(17,10){$U$}

\put(15,13){\line(-2,1){24}}
\put(15,13){\line(-2,-1){24}}

\put(-9,9){$\mathfrak{i}_{\bm{p}}$}

\put(70,10){$U$}

\put(79,13){\line(2,1){24}}
\put(79,13){\line(2,-1){24}}
\put(93,9){$\mathfrak{r}_{\bm{p}}$}
\put(70,10){$U$}
\put(106,18){$U$}   
\put(106,-6){$1^m$}   

\put(-30,-40){$=$}

\put(47,32){\oval(165,40)[t]}      
\put(117,32){\oval(25,20)[rb]}      
\put(-23,32){\oval(25,20)[lb]}     
\put(-30,19){$>$}                   

\put(47,32){\oval(180,60)[t]}      
\put(124,32){\oval(25,70)[rb]}      
\put(-31,32){\oval(25,70)[lb]}      
\put(-37,-6){$>$}                   
\end{picture}
\begin{picture}(200,100)(-60,-50)
\put(30,-15){\fbox{\rule[.5in]{.4in}{0in }}}

\put(-22,18){$U$}    
\put(-24,-6){$1^m$}    

\put(43,0){$g$}
\put(12,-20){$X$} \put(70,-20){$Y$}

\put(17,10){$U$}

\put(15,13){\line(-2,1){24}}
\put(15,13){\line(-2,-1){24}}

\put(-9,9){$\mathfrak{i}_{\bm{p}}$}

\put(70,10){$U$}

\put(-49,13){\line(2,1){24}}
\put(-49,13){\line(2,-1){24}}

\put(-35,9){$\mathfrak{r}_{\bm{p}}$}
\put(-58,10){$U$}

\put(-30,-40){$=$}

\put(10,22){\oval(165,40)[t]}

\put(80,22){\oval(25,20)[rb]}

\put(-60,22){\oval(25,20)[lb]}

\put(-67,9){$>$}

\end{picture}

\begin{picture}(300,80)(-60,-30)
\put(30,-15){\fbox{\rule[.5in]{.4in}{0in }}}


\put(43,0){$g$}
\put(12,-20){$X$} \put(70,-20){$Y$}
\put(17,10){$U$}

\put(70,10){$U$}


\put(47,22){\oval(92,40)[t]}    
\put(80,22){\oval(25,20)[rb]}
\put(14,22){\oval(25,20)[lb]}
\put(8,9){$>$}



\end{picture} \\
$$\mbox{Invariance w.r.t. $U \otimes 1^m \rhd_{(\mathfrak{i}_{\bm{p}}, \mathfrak{r}_{\bm{p}})}
U $ 
}$$
\end{minipage}
\caption{Invariance of traces under conjugate actions} \label{invunderconj}
\end{figure}

\noindent {\bf Proof of Lemma \ref{leminvunderconj}} \\
These retractions are compatible with 
traced monoidal categories \cite{AHS02}
by virtue of dinaturality and the directions of
the retractions $(k_m, j_m)$ and
$(\mathfrak{i}_{\bm{p}}, \mathfrak{r}_{\bm{p}})$, respectively, as
follows: 
$$\begin{array}{l|r}
\begin{array}{ll}
\mbox{
For ($\ref{2'}$);
} & 
{\sf Tr}^{U}_{X,Y}(
(j \otimes Y) \Comp f \Comp (k \otimes X))   \\ & 
= \hfill \mbox{dinaturality} \\ & 
{\sf Tr}^{U^m}_{X,Y}(
f \Comp (k \otimes X)) \Comp (j \otimes X))   \\ & 
= \\ & 
{\sf Tr}^{U^m}_{X,Y}(
f \Comp (k \Comp j \otimes X) )   \\ &
=  \hfill \mbox{$k \Comp j = \Id_{U^m}$}  \\ &
{\sf Tr}^{U^m}_{X,Y}(f)
\end{array} & 
\begin{array}{ll}
\mbox{For ($\ref{5'}$);} & 
{\sf Tr}^{U \otimes 1^m}_{X,Y}(
(\mathfrak{r}_{\bm{p}} \otimes Y) \Comp g \Comp (\mathfrak{i}_{\bm{p}} \otimes X)) \\ &
=  \hfill \mbox{dinaturality} \\ & 
{\sf Tr}^{U}_{X,Y}(
g \Comp (\mathfrak{i}_{\bm{p}} \otimes X) \Comp (\mathfrak{r}_{\bm{p}} \otimes X) 
)
 \\ & 
=  \\ & 
{\sf Tr}^{U}_{X,Y}(
g \Comp (\mathfrak{i}_{\bm{p}} \Comp \mathfrak{r}_{\bm{p}} \otimes X) 
) \\ & 
= \hfill \mbox{$\mathfrak{i}_{\bm{p}} \Comp \mathfrak{r}_{\bm{p}} = \Id_{U}$} \\ &
{\sf Tr}^{U}_{X,Y}(g)
\end{array}
\end{array}$$
See the following pictures in Figure \ref{invunderconj},
{in which ${\sf Tr}^{U^m}$ is described via Vanishing II}.


\subsection{Appendix 5: Remarks on  Retractions $U \rhd_{(k,j)} U \otimes U$ and
$ U \otimes 1 \rhd_{(\mathfrak{i}_\alpha, \mathfrak{r}_\alpha )} U$}
\label{retract.remks}
 

\noindent
 The reader may wonder about the opposite directions of the retractions  
$U \rhd_{(k,j)} U \otimes U$ and $U
\otimes
1 \rhd_{(\mathfrak{i}_\alpha, \mathfrak{r}_\alpha)} U$
in the two-layered GoI interpretation $\Mean{\pi}$ and $f_\pi$.

\vspace{1ex}

\noindent {\bf (i)} (On the retraction $U \rhd_{(k,j)}  U \otimes U$ ) \\
The retract
$U \otimes U$ (of $U$) follows the form of the logical constructions.
In (untyped) GoI, where there is a reflexive object $U$,
the interpretations of $\otimes$ and $\wp$
are indistinguishable.  In general, for any formula $A$, $U_A$ is identified with $U$.  So to
make sense of the logical connectives, via reflexivity of $U$, we use the retraction.   Thus
$U_A \otimes U_B$, which is $U\otimes U$  is faithfully
projected to $U$ which is defined to be both
$U_{A \otimes B}$ as well as $U_{A \wp B}$.   Similarly, letting
$U_\downarrow = U_\uparrow = U$, we have
 $U_\downarrow \otimes U_A$ (resp. $U_A \otimes U_\uparrow$ 
) is faithfully projected to $U$, which itself
is defined to be $U_{\down{A}}$ (resp. $U_{\up{A}}$).
The faithfulness is guaranteed by $k \Comp j=\Id_{U \otimes U}$.
Note that at the level of the $U$'s, the dual logical connectives
$\downarrow$ and $\uparrow$ are not distinguishable, just as
$\otimes$ and $\wp$ are not distinguishable.  
However to account for the asymmetry in logical rules for the polarities,
we will need  to employ a new ingredient, the object 1.

\smallskip

\noindent
{\bf (ii)} On the retraction $U \otimes 1 \rhd_{(\mathfrak{i}_\alpha,  \mathfrak{r}_\alpha)} U$   \\
The retract $1$ (of $1 \otimes 1$)
is for the sake of realizing
the retract $U$ (of $U \otimes 1$)
of the lifting property (along $\alpha$) of
Definition \ref{GoIsitu}.
Then what is the meaning of 
the retraction $U \otimes 1 \rhd_{(\mathfrak{i}_\alpha, \mathfrak{r}_\alpha)} U$ ?
Tensoring $1$ with $U$ in the construction of $\mathfrak{r}_\alpha: U \longrightarrow U \otimes 1$\
corresponds to making the point $\alpha: 1 \longrightarrow U$ explicitly appear.
Conversely $\mathfrak{i}_\alpha: U \otimes 1 \longrightarrow U$ hides the
	point $\alpha$    (i.e. makes it implicit).
The faithfulness of making  the point $\alpha$ explicit
is guaranteed by $\mathfrak{i}_\alpha \Comp \mathfrak{r}_\alpha = \Id_{U}$, i.e. intuitively,
making $\alpha$ explicit, then hiding it gives the identity. \\
For
a multipoint ${\sf mp}(A) : 1^m \stackrel{\bm{p}}{\longrightarrow} U$
for any given polarized formula $A$ so that $\mathbbm{1}_A \cong 1^m$,
the retraction $1 \otimes 1 \rhd_{(\mathfrak{i}, \mathfrak{r})} 1$ and its lifting
$U \otimes 1 \rhd_{(\mathfrak{i}_\alpha, \mathfrak{r}_\alpha)} U$
are correspondingly generalized into
the retraction $1^m \otimes 1^m \rhd_{(\mathfrak{i}^m, \mathfrak{r}^m)} 1^m$
and the lifting 
$U \otimes 1^m \rhd_{(\mathfrak{i}_{\bm{p}}, \mathfrak{r}_{\bm{p}})} U$
of Axiom 6' of Proposition \ref{mplifting}.
These retractions are in order to accommodate polarities in $U$
and in $1^m$. 

\smallskip

\noindent
{\bf (iii)}  Note also
the opposite directions of retractions between
(i) and (ii) above
are compatible with the conjugate actions of Lemma
\ref{leminvunderconj}.



\subsection{Appendix 6: Pictorial Proof for Prop \ref{exinvariance} (Ex
  is an Invariant )}
\label{calculations} 
{\tiny
\begin{longtable}[b]{ll}
   \mbox{\footnotesize L.H.S. of (\ref{main})} &  
\begin{minipage}[t]{.05in}  
 {\footnotesize 
 naturality \\ = }
 \end{minipage} 
\\ \\ 
$\xymatrix
@C=0.08in 
@R=0.25pc
{
\save [].[ddddddddddddddd].[dddddddddddddddrrrrrrrr].[rrrrrrr]*[F--]\frm{}
\restore &  &  &  & & & &  & &    \\
& & U_{\downarrow} \ar[rr]^a   &   & 1^m  & & &    \\
& &                           & \otimes &  & &&&&  \\
& &     1^m \ar[rr]_b   &   & U_\downarrow  \ar[rrdddddddddd]
  &  & U_\downarrow
\ar `^u[uuulllll] `[uuulllll] `^d[uuulllll] `^r[uuulllll] [uullll] 
& &  
\\
& \bm{U}_{\cal M} 
\ar@{-}[ru]^<<{}="n"  \ar@{-}[rddd]^<<{}="m"  
\ar@{-}^{\hat{\mathfrak{r}}_{\cal M}}@/^/"n";"m"
&            & \otimes  &  &  & & 
\bm{U}_{\cal M}  
\ar@{-}@/^1pc/[dddlll]^<<{}="v"     
\ar@{-}@/_1pc/[uuulll]^<<{}="w" 
\ar@{-}^{\hat{\mathfrak{i}}_{\cal M}}@/^/"v";"w"
\\
& & \bm{U}_{P^\bot} \ar@{-}[rr] \ar@{-}[dd] 
\ar@{}[ddrr]|{\Meanpre{\pi'_2}}
&   & \bm{U}_{P^\bot} \ar[rrdddddd]  &  &  \bm{U}_{P^\bot} \\
& &         &    &   &      \\
&  & \bm{U}_{\cal M}     &   &  \bm{U}_{\cal M} 
\ar@{-}[ll]  \ar@{-}[uu] 
& &  \\
& &  &  \otimes &  & &   \\
&  &  \ar@{-}[rr] \ar@{-}[dd] 
\ar@{}[ddrr]|{\Meanpre{\pi'_1}} &   &  \\
  &  & & &         &    &       \\
 &  & \bm{U}_P     &   &  \bm{U}_P \ar@{-}[ll] \ar@{-}[uu] \ar[rruuuuuu]  
& 
&  \bm{U}_P   \\
& &                            & \otimes &  &   &   \\
& & U_{\uparrow} \ar[rr]_c   &   &  U_{\uparrow} \ar[rruuuuuuuuuu]  &  &
U_{\uparrow}  
\ar `_d[lllll] `[lllll] `_u[lllll] `_r[lllll] [llll]
& \\
& &  &  &  &  &  &  & &  \\
& &  &  &  &  &  &  & &  \\
& &  &  &  &  &  &  & & 
\\ \\ \\ 
}$
& 
$\xymatrix
@C=0.08in 
@R=0.25pc
{ & 
\save [].[ddddddddddddddd].[dddddddddddddddrrrrrr].[rrrrrr]*[F--]\frm{}
\restore   &  &  & & & &  & &  \\
& & U_{\downarrow} \ar[rr]^{a}   &   & 1^m  & & &    \\
& &                           & \otimes &  & &&&&  \\
& &     1^m \ar[rr]_b   &   & U_\downarrow  \ar[rrdddddddddd]
  &  & U_\downarrow
\ar `^u[uuulllll] `[uuulllll] `^d[uuulllll] `^r[uuulllll] [uullll] 
& &  
\\
\bm{U}_{\cal M} \ar@{-}[rru]^<{}="n"  \ar@{-}[rrddd]_<{}="m"  &  &            & \otimes  &  &
& &  &  
\bm{U}_{\cal M}  \ar@{-}@/^1pc/[dddllll]^<<{}="v"    
\ar@{-}@/_1pc/[uuullll]^<<{}="w"         
\ar@{-}^{\hat{\mathfrak{r}}_{\cal M}}@/^/"n";"m"
\ar@{-}^{\hat{\mathfrak{i}}_{\cal M}}@/^/"v";"w" 
\\
& & \bm{U}_{P^\bot} \ar@{-}[rr] \ar@{-}[dd] 
\ar@{}[ddrr]|{\Meanpre{\pi'_2}}
&   & \bm{U}_{P^\bot} \ar[rrdddddd]  &  &  \bm{U}_{P^\bot} \\
& &         &    &   &      \\
&  & \bm{U}_{\cal M}     &   &  \bm{U}_{\cal M} 
\ar@{-}[ll]  \ar@{-}[uu] 
& &  \\
& &  &  \otimes &  & &   \\
&  &  \ar@{-}[rr] \ar@{-}[dd] 
\ar@{}[ddrr]|{\Meanpre{\pi'_1}} &   &  \\
  &  & & &         &    &       \\
 &  & \bm{U}_P     &   &  \bm{U}_P \ar@{-}[ll] \ar@{-}[uu] \ar[rruuuuuu]  
& 
&  \bm{U}_P   \\
& &                            & \otimes &  &   &   \\
& & U_{\uparrow} \ar[rr]_c   &   &  U_{\uparrow} \ar[rruuuuuuuuuu]  &  &
U_{\uparrow}  
\ar `_d[lllll] `[lllll] `_u[lllll] `_r[lllll] [llll]
& \\
& &  &  &  &  &  &  & &  \\
& &  &  &  &  &  &  & &  \\
}$ \\ 
\begin{minipage}[t]{.05in}  
 {\footnotesize 
 superposing \\ = }
 \end{minipage} 
  \\  
\begin{tabular}{c}
$\xymatrix
@C=0.1in 
@R=0.3pc
{ 
& & \bm{U}_{P^\bot} \ar@{-}[rr] \ar@{-}[dd] 
\ar@{}[ddrr]|{\Meanpre{\pi'_2}}
&   & \bm{U}_{P^\bot} \ar[rrdddddd]  &  &  \bm{U}_{P^\bot} \\
& &         &    &   &      \\
&  & \bm{U}_{\cal M}     &   &  \bm{U}_{\cal M} 
\ar@{-}[ll]  \ar@{-}[uu] 
& &  \\
\bm{U}_{\cal M} \ar@{-}[rru]^(.3){}="n" \ar@{-}@/_1pc/[rrdddddd]^(.1){}="m"
\ar@{-}^{\hat{\mathfrak{r}}_{\cal M}}@/^/"n";"m" 
& &  &  \otimes &  & & &  &  \bm{U}_{\cal M} \ar@{-}@/^1pc/[ddddddllll]^<<{}="v"       
\ar@{-}@/_1pc/[ullll]^<<{}="w"         
\ar@{-}^{\hat{\mathfrak{i}}_{\cal M}}@/^/"v";"w" 
\\
&  &  \ar@{-}[rr] \ar@{-}[dd] 
\ar@{}[ddrr]|{\Meanpre{\pi'_1}} &   &  \\
  &  & & &         &    &       \\
 &  & \bm{U}_P     &   &  \bm{U}_P \ar@{-}[ll] \ar@{-}[uu] \ar[rruuuuuu]  
& 
&  \bm{U}_P   \\
& &                            & \otimes &       &   &   \\
 &   &  &        &    &    &  \\
& & \quad 1^m      &  \mbox{ \footnotesize $\star$}  
 &  \hspace{-1em}
 1^m  
&  &
& \\
& 
 & \save [].[uu].[uurr].[rr]*[F--]\frm{}  \restore
 &  &  
&  &  &  & & \\
}$  \\ \\ \\
\begin{minipage}[b]{1.5in} 
{\footnotesize
$\star$ is L.H.S of Fig for (\ref{eqitergenyank})
when $f=b,g=c$ and $h=a$.}
 \end{minipage}
\end{tabular}
& 
\begin{tabular}{l}
\begin{minipage}[t]{1in} 
{\footnotesize
(\ref{eqitergenyank})
and $c=0$} \\ =
\end{minipage} \\
\begin{tabular}{c}
$\xymatrix
@C=0.1in 
@R=0.2pc
{ 
& & \bm{U}_{P^\bot} \ar@{-}[rr] \ar@{-}[dd] 
\ar@{}[ddrr]|{\Meanpre{\pi'_2}}
&   & \bm{U}_{P^\bot} \ar[rrdddddd]  &  &  \bm{U}_{P^\bot} \\
& &         &    &   &      \\
&  & \bm{U}_{\cal M}     &   &  \bm{U}_{\cal M} 
\ar@{-}[ll]  \ar@{-}[uu] 
& &  \\
\bm{U}_{\cal M} \ar@{-}[rru]^(.3){}="n" \ar@{-}@/_1pc/[rrddddd]^(.1){}="m"
\ar@{-}^{\hat{\mathfrak{r}}_{\cal M}}@/^/"n";"m" 
& &  &  \otimes &  & & &  &  \bm{U}_{\cal M}
\ar@{-}@/^1pc/[dddddllll]^<<{}="v"
\ar@{-}@/_1pc/[ullll]^<<{}="w"
\ar@{-}^{\hat{\mathfrak{i}}_{\cal M}}@/^/"v";"w" 
\\
&  &  \ar@{-}[rr] \ar@{-}[dd] 
\ar@{}[ddrr]|{\Meanpre{\pi'_1}} &   &  \\
  &  & & &         &    &       \\
 &  & \bm{U}_P     &   &  \bm{U}_P \ar@{-}[ll] \ar@{-}[uu] \ar[rruuuuuu]  
& 
&  \bm{U}_P   \\
& &                            & \otimes &       &   &   \\
 &   & 1^m \ar[rr]_{a  c  b =0} &        &  1^m  &    &  }$
\end{tabular}
 \\ \\   
\begin{minipage}[t]{.6in} 
{\footnotesize  Ax 9'} \\ =
\end{minipage} \hspace{-2.5cm}
\\
\hspace{-4ex}$\xymatrix
@C=0.1in 
@R=0.2pc
{ & & \bm{U}_{P^\bot} \ar@{-}[rr] \ar@{-}[dd] 
\ar@{}[ddrr]|{\Meanpre{\pi'_2}}
&   & \bm{U}_{P^\bot} \ar[rrdddddd]  &  &  \bm{U}_{P^\bot} \\
& &         &    &   &      \\
&  & \bm{U}_{\cal M}     &   &  \bm{U}_{\cal M} 
\ar@{-}[ll]  \ar@{-}[uu] 
& &  \\
& &  &  \otimes &  & & &  &  
\\
&  &  \ar@{-}[rr] \ar@{-}[dd] 
\ar@{}[ddrr]|{\Meanpre{\pi'_1}} &   &  \\
  &  & & &         &    &       \\
 &  & \bm{U}_P     &   &  \bm{U}_P \ar@{-}[ll] \ar@{-}[uu] \ar[rruuuuuu]  
& 
&  \bm{U}_P 
}$ \\ \\
 {\footnotesize R.H.S. of~(\ref{main})}
\end{tabular} 
 \end{longtable}
}

 
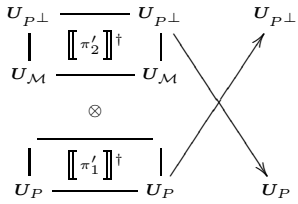
\end{document}